\documentclass[a4paper,fleqn]{cas-dc}

\usepackage[authoryear]{natbib}

\def\tsc#1{\csdef{#1}{\textsc{\lowercase{#1}}\xspace}}
\tsc{WGM}
\tsc{QE}

\newcommand{\widfig}{0.9}
\newcommand{\halffig}{0.51}
\newcommand{\fullfig}{0.99}
\newcommand{\overfig}{1.5}

\usepackage{amsmath}	
\usepackage{amssymb}	
\usepackage{tabularx}
\usepackage{graphicx}	
\usepackage{subfigure}	
\usepackage{hyperref}

\newcommand\aj{AJ}        
\newcommand\apj{ApJ}    
\newcommand\apjl{ApJL}     
\newcommand\aap{A\&A}     
\newcommand\baas{BAAS}     
\newcommand\icarus{Icarus} 
\newcommand\mnras{MNRAS}   
\newcommand\pasp{PASP}     
\newcommand\nat{Nature}  

\newcommand{\textbackslashmathfrak}{\cal}
\newcommand{\xy}{$x-y$}
\newcommand{\equ}{\{equ\}}
\newcommand{\ecl}{\{ecl\}}
\newcommand{\obs}{\{obs\}}
\newcommand{\hor}{\{hor\}}
\renewcommand{\o}{^\circ}
\newcommand{\obsr}{\mathrm{obs}}
\newcommand{\planet}{\mathrm{p}}
\newcommand{\ring}{\mathrm{r}}

\newcommand{\sub}[1]{_\mathrm{#1}}
\renewcommand{\sup}[1]{^\mathrm{#1}}
\newcommand{\beq}[1]{\begin{equation}\label{#1}}
\newcommand{\eeq}{\end{equation}}
\newcommand{\beqa}{\begin{eqnarray}}
\newcommand{\eeqa}{\end{eqnarray}}

\newcommand{\af}{a\sub{s}}
\newcommand{\nf}{\hat n\sub{f}}
\newcommand{\Rs}{R\sub{\star}}
\newcommand{\Rp}{R\sub{p}}

\newcommand{\AS}{A\sub{\nu}\sup{(sph)}}

\newcommand{\AL}{A\sup{(Lamb)}}
\newcommand{\ALp}{A\sub{p}\sup{(Lamb)}}
\newcommand{\ALr}{A\sub{\nu,r}\sup{(Lamb)}}
\newcommand{\ALo}{A\sub{\nu,0}\sup{(Lamb)}}

\newcommand{\model}{Pryngles}
\newcommand{\comp}[1]{{\tt #1}}
\newcommand{\github}{{\tt GitHub}}

\newcommand{\fibpy}[1]{{\tt fibpy}\footnote{Repository in \github: \url{https://github.com/matt77hias/fibpy}. For notes and references \url{http://extremelearning.com.au/evenly-distributing-points-on-a-sphere/} (visited Jan. 21st, 2020).}}
\newcommand{\BrightRings}[1]{{\tt BrightRings}\footnote{\url{http://github.com}}}

\newcommand{\beqn}{\begin{equation}}
\newcommand{\eeqn}{\end{equation}}

\definecolor{mygreen}{RGB}{9,1,122}
\definecolor{myblack}{RGB}{80,30,20}

\ExplSyntaxOn
\keys_set:nn { stm / mktitle } { nologo }
\ExplSyntaxOff

\begin{document}
\let\WriteBookmarks\relax
\def\floatpagepagefraction{1}
\def\textpagefraction{.001}

\shorttitle{The bright side of the light curve}    

\shortauthors{Zuluaga, J.I., Sucerquia, M. \& Alvarado-Montes J. A.}  

\title [mode = title]{The bright side of the light curve: a general photometric model of non-transiting exorings}  

%

\author[1]{Jorge I. Zuluaga}
\cormark[1]
\ead{jorge.zuluaga@udea.edu.co}
\credit{Conceptualization, physics of light scattering and software development}

\author[2,3]{Mario Sucerquia}
\ead{mario.sucerquia@uv.cl}
\credit{Light curve analysis and exoplanetary science}

\author[4,5]{Jaime A. Alvarado-Montes}
\ead{jaime-andres.alvarado-montes@hdr.mq.edu.au}
\credit{Light curve analysis and exoplanetary science}

\affiliation[1]
            {organization={SEAP/FACom, Instituto de F\'{\i}sica - FCEN, Universidad de Antioquia},
            addressline={Calle 70 No. 52-21}, 
            city={Medell\'in},
            citysep={}, 
            postcode={050026}, 
            country={Colombia}}

\affiliation[2]
            {organization={Universidad de Valpara\'iso},
            addressline={Av. Gran Breta\~na 1111}, 
            city={Valpara\'iso},
            citysep={}, 
            postcode={}, 
            country={Chile}}

\affiliation[3]
            {organization={N\'ucleo Milenio Formaci\'on Planetaria - NPF},
            city={Valpara\'iso},
            citysep={}, 
            postcode={}, 
            country={Chile}}

\affiliation[4]
            {organization={School of Mathematical and Physical Sciences, Macquarie University},
            city={Sydney},
            citysep={}, 
            postcode={NSW 2109}, 
            country={Australia}}

\affiliation[5]
            {organization={Research Centre for Astronomy, Astrophysics and Astrophotonics, Macquarie University},
            city={Sydney},
            citysep={}, 
            postcode={NSW 2109}, 
            country={Australia}}

\cortext[1]{Corresponding author}



\begin{abstract}
Rings around exoplanets (exorings) are one of the most expected discoveries in exoplanetary research. There is an increasing number of theoretical and observational efforts for detecting exorings, but none of them have succeeded yet. Most of those methods focus on the photometric signatures of exorings during transits, whereas less attention has been paid to light diffusely reflected: what we denote here as the bright side of the light curve. This is particularly important when we cannot detect the typical stellar flux drop produced by transiting exoplanets. Here, we endeavour to develop a general method to model the variations on the light curves of both ringed non-transiting and transiting exoplanets. Our model (dubbed as \comp{\model}) simulates the complex interaction of luminous, opaque, and semitransparent objects in planetary systems, discretizing their surface with small circular plane discs that resemble sequins or spangles. We perform several numerical experiments with this model, and show its incredible potential to describe the light curve of complex systems under various orbital, planetary, and observational configurations of planets, moons, rings, or discs. As our model uses a very general approach, we can capture effects like shadows or planetary/ring shine, and since the model is also modular we can easily integrate arbitrarily complex physics of planetary light scattering. A comparison against existing tools and analytical models of reflected light reveals that our model, despite its novel features, reliably reproduces light curves under common circumstances. \comp{\model} source code is written in {\tt Python} and made publicly available.
\end{abstract}



\maketitle


\section{Introduction}

Exoplanetary search has revealed a diversity of planets whose sizes range from Mercury to Jupiter, in unusual and extreme orbital configurations \citep[see e.g.][]{Wittenmyer2020,Kanodia2021,Wong2021,Singh2022,Subjak2022}. These discoveries have been accomplished with the combined power of ground- and space-based telescopes and spectrographs (e.g. \textit{Kepler}, {\it TESS}, \textit{HARPS}, etc).  New instruments and incoming missions like \textit{SPHERE/ZIMPOL}, \textit{E-ELT}, \textit{PLATO}, \textit{CHEOPS}, \textit{GMT} and \textit{JWST} (see e.g. \citealt{Lopez-Morales2019}) have increased the hope for new discoveries and improved observations of already discovered exoplanets.

To date, the discovery of two specific planetary phenomena remains elusive: exomoons and exorings.  Both are fairly common in the Solar System and there is no reason to think they will be rare in extrasolar systems. The de facto consensus on why key discoveries in this area are still missing, is that the current instrumental sensitivities for the most common techniques (transit photometry and radial velocity - RV) are probably not enough. Although it was early estimated that rings around Saturn-like exoplanets could be detected with a photometric sensitivity down to (1–3) $\times 10^{-4}$ (100-300 ppm) and a time resolution of $\sim15$ min \citep{Barnes2004} (well within Kepler and TESS capabilities for Sun-like or brighter stars) no ring signal has been detected so far (\citealt*{Heising2015}; \citealt{Aizawa2018}).  It could be possible that the photometric signatures of exorings and exomoons, at least for objects in the size range observed in the Solar System, are probably too faint to be distinguished from the noise in transiting exoplanets \citep{Heller2018}.

Still, a couple of candidates for exomoons and exorings have been proposed. However, in all cases they correspond to systems with oversized features. Planetary systems that defy current models of planetary and satellite formation are the still-debated Kepler-1625b I \citep*{Teachey2018} and Kepler-1708b I \citep{Kipping2022}: Neptune-sized exomoon candidates orbiting a Jupiter-like planet. Another example is the peculiar light curve of the sub-stellar object J1407b \citep{kenworthy2015}, which has been interpreted as a `super Saturn' with a gigantic ring system. Other anomalous signals include the case of very low-density planets \citep[among others]{Piro2020, Zuluaga2015}, the fairly atypical light curve of the so-called Tabby's star \citep{Boyajian2016}, and the unexplained behaviour of Fomalhaut b \citep[see, e.g.][]{Kalas2013}.  However, instead of closing the gap as the first key discoveries of exomoons and exorings, those findings are giving rise to unseen planetary phenomena, some of them being catastrophic in nature \citep{Currie2012,Janson2020}.

Transit photometry and radial velocity (RV) have been the most successful techniques for identifying and characterizing exoplanets (see e.g. \citealt{Perryman2018} and references therein). It is widely known, for instance, that planetary transits offer considerable physical and orbital information about specific exoplanets and planetary systems in general. Properties such as planetary radius and shape \citep{Seager2003,Barnes2003}, orbital elements \citep[]{Seager2003}, and even planetary surface albedo \citep{Serrano2018} or atmospheric composition \citep[and references therein]{Tinetti2018} can be inferred from light curves of transiting exoplanets. Moreover, if adequate sensitivity and cadence are ensured, exomoons and exorings could also be detected via transit photometry \citep[see e.g.][and others]{Brown2001,Barnes2004,Szabo2006,Kipping2009a,Kipping2009b,Zuluaga2015,Sucerquia2017,Heller2018,Sucerquia2019,Sucerquia2020b,Sucerquia2020a}. Additionally, combining transit photometry and RV measurements, the physical characterization of exoplanets and their host stars can be achieved to levels that were unthinkable a few decades ago.

Nevertheless, despite the great achievements of the aforementioned methods, they also suffer from two serious and well-recognized limitations: 1) Transits require the planetary orbital plane to be close to the observer's line of view (i.e. edge-on or high inclination orbits) and the planet's mass can be accurately constrained using RVs only if such configurations are also ensured; 2) despite the opportunities that transit photometry offers to measure atmospheric composition using wavelength-dependent transit depths \citep{Santos2015}, transits and RVs are only well-suited to provide orbital elements and bulk planetary properties.

The lack of key discoveries in the area of exomoons and exorings, as well as the limitations of the most prolific observational techniques in exoplanetary sciences should make us see in other directions. In the fringe of a revolution in abundance and quality of optical and infrared spectrophotometric observations, driven by the arrival of new and large ground- and space-based instruments, we should start thinking about the {\it bright side} of the light curve. Instead of desperately searching the signals of transiting exomoons and exorings or the absorption lines of wobbling stars (i.e. the dark side of the light curve), we should start shifting our attention to the starlight scattered from unresolved (ringed) exoplanets -- the vast majority of them.


There has been a growing interest in using scattered starlight from planetary systems to gather information about atmospheric dynamics \citep{Zugger2010}, magnetic fields \citep{Oklopcic2019}, polarised light reflected by atmospheres \citep*{Lietzow2021}, effects of shadows of planetary rings and planets on ingress/egress \citep{Arkhypov2021} or, in general, the presence of planetary rings \citep{Arnold2004,Santos2015,Sucerquia2020a}. Others have shown how advanced numerical techniques \citep*{Damiano2020a,Damiano2020b} can be applied to analyse reflected light and retrieve information on cloud properties and atmospheric composition of giant planets \citep{Hu2019, Damiano2021}.  Even the case of terrestrial exoplanets has been a matter of research in this area \citep{Damiano2022}.

In \citet{Sucerquia2020a} we used a simple model to show that the light curve of non-transiting ringed exoplanets can be used to easily retrieve its bulk physical properties and orbital parameters. However, this particular model was mainly restricted to systems with a face-on geometric orientation and a very simplistic description of light scattering on the rings. But planetary rings can be very complex systems and have intricate interactions with stellar and planetary light: they are granular in nature (and hence semitransparent) and may polarise the incoming starlight \citep{Johnson1980,Dollfus1984,Geake1990}. Also, they are not continuous, may have large divisions and/or ringlets, and depending upon their inclination and orientation they may cast large shadows on planetary surfaces. Conversely, a significant fraction of the rings' surface can be permanently shadowed by the planet. 

Disregarding some or all of these effects will impact the realism of any modelled reflected light curve. On the contrary, taking most of these effects into account and including other complex physical effects would definitively render reflected\footnote{Occasionally, we informally refer to backward scattering as `reflection' without necessarily implying that we assume that light is coherently reflected on any surface.} light curves of exorings into an incredible source of information.

The above is precisely the aim of this paper. We present and test here a general geometrical model to calculate the reflected light curve produced by a ringed exoplanet with arbitrary physical properties and geometrical configuration. In \autoref{sec:model}, we introduce the two basic ideas behind the model: 1) using an object-centred reference frame (\autoref{subsec:ref}) instead of a stellar-centred one as it is usual in dynamical applications and other photometric models; and 2) discretizing the surface of the objects involved using circular surface area elements or {\it spangles}, as explained in \autoref{subsec:spangles}.  In \autoref{sec:optics}, we describe the different optical effects (i.e. diffuse reflection, shadows, transits, and planet- and ring-shine) included into the present version of the model. The results of several numerical experiments are presented in \autoref{sec:numerical}, and these will be useful to illustrate the qualitative impact of different effects and geometrical configurations on the light curve of ringed exoplanets, while also providing us with first-order approximations of the magnitude of such effects. \comp{Pryngles}\footnote{\url{https://pypi.org/project/pryngles/}} is the package that implements the model described in this paper, and in \autoref{sec:validation} we compare it against different software packages intended to calculate the light curve of transiting objects that have been developed in recent years (at least regarding the dark side of the light curve). Finally, \autoref{sec:discussion} is devoted to a discussion about the limitations and potential of the model.

\section{The geometry of the model}
\label{sec:model}


The model we are about to describe is designed to solve the following general problem. Let us consider a planetary system with one or several primary sources of light (e.g. one or several stars) 
and various diffusive objects such as planets, moons, and rings that scatter the incoming light (diffuse reflection). Depending on the direction of observation, such objects could also block part of that light producing transits and occultations.

Light in the system not only comes from primary sources (i.e. stars) but also from close-by diffusive objects. Thus, for instance, planets partially illuminate their rings and moons (planet-shine) and,  conversely, light reflected from them can be a source of illumination for planets (ring- and moon-shine). Diffusive objects might be completely opaque (planets and moons) or semitransparent (rings), and totally or partially block the light from primary sources (transits) or from other diffusive objects (occultations). When blocking the light of primary sources in the direction of other objects, opaque or semitransparent objects cast a shadow on them and reduce the light reflected from their surface. 

What is the total amount of light coming from the whole system in a given direction (observer), spectral band (wavelength), and polarization as a function of time (light curve)?

Solving the aforementioned general problem in a single piece of work is not feasible.  Here, we restrict the problem to a system composed of a single star and one (or several) ringed planets. Including other objects such as stellar companions, circumstellar discs, moons, and even ringed-moons (cronomoons; \citealt{Sucerquia2022}) will be the goal of future improvements of this model. Still, since the model is modular (as demonstrated below), other objects, effects, and physical phenomena can easily be incorporated (see \autoref{sec:discussion}).

\subsection{Reference systems}
\label{subsec:ref}

\begin{figure*}
  \includegraphics[width=\textwidth]{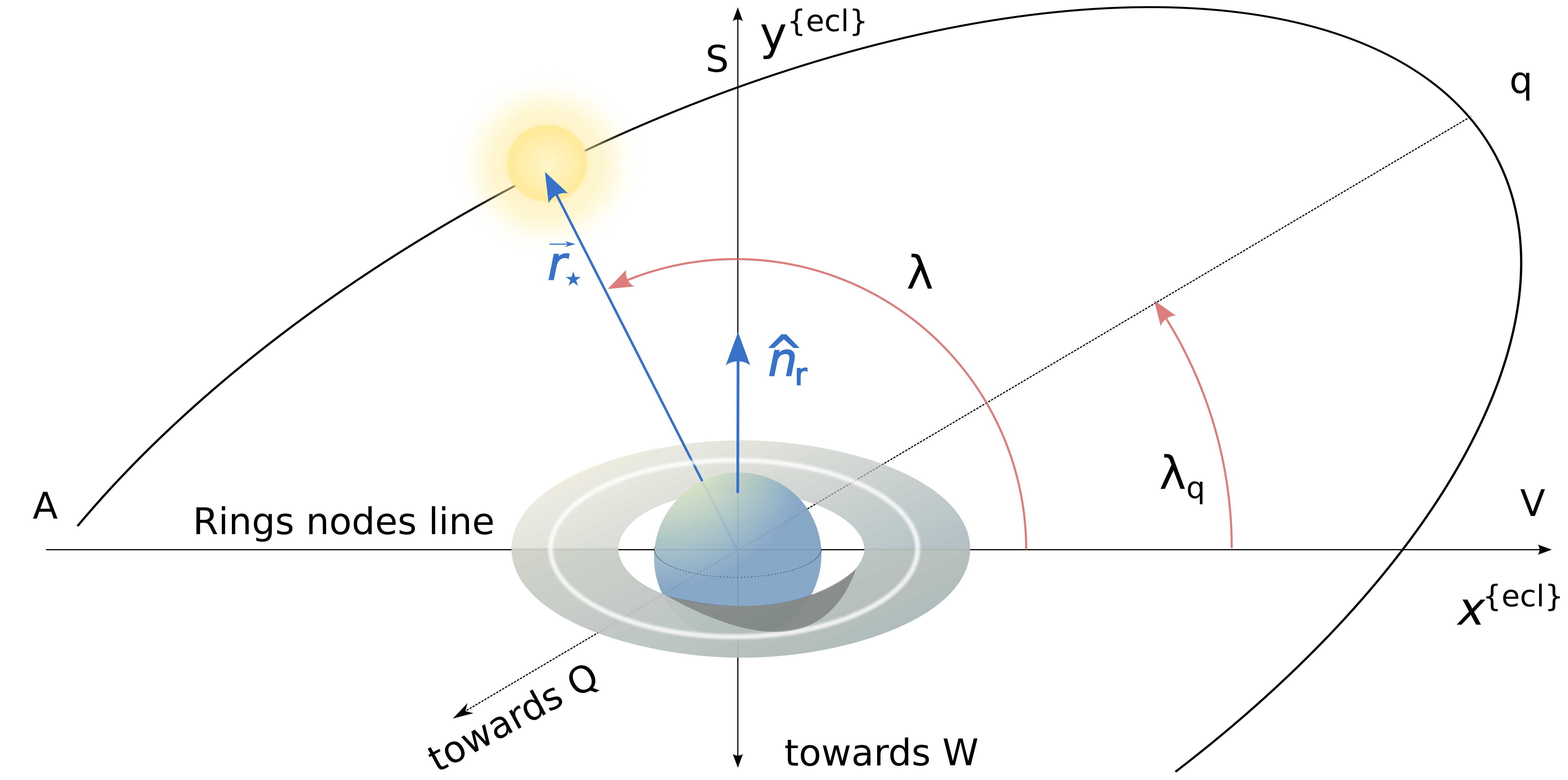}
    \caption{Orientation of the ring system and the orbit of the star around the planet as seen by an observer seeing towards the $-z\sup{\ecl}$ axis. The plane of the stellar orbit (ecliptic) is at the plane of the paper.  Since the normal vector to the ring is inclined with respect to the $+z\sup{\ecl}$ axis, the rings look elliptical. From this point of view, the upper side of the rings is inside the plane of the paper. The rotational axis of the planet points in the $\hat n_r$ direction, and although from this view it seems to coincide with the $+y\sup{\ecl}$ axis, the vector actually points towards the reader and is inclined with respect to the $+z\sup{\ecl}$ axis. There are several special points along the stellar orbit: V (vernal equinox, the star is in the node's line), q (periastron), S (summer solstice, the star is at the highest northern declination and the rings are illuminated as seen from the observer's point of view), A (autumn equinox), Q (apoastron), and W (winter solstice, the star is at its maximum southern declination and the rings will appear `dark' to the observer). The position of any point on the stellar orbit is measured using the ecliptic longitude $\lambda$.}
    \label{fig:orbit_ring_orientation}
\end{figure*}

When modelling the dynamics and even the photometry of planetary systems, it is customary to choose reference frames centred either at the barycenter of the system or its most massive object (i.e. the star). We find that the best way to model reflected light and even transits, is using an object-centric reference frame (irrespective of the mass or dynamical state of the object). In \autoref{fig:orbit_ring_orientation}, we show the orientation of a reference frame centred on a ringed planet. Accordingly, the star moves around the planet following an elliptical orbit with the same orbital parameters of the former. The star's orbit lies on a plane that we will call hereafter the `ecliptic' plane (in analogy to the plane of motion of the Sun with respect to Earth).

Following the analogy, the `equatorial' plane of the ringed planet, which is perpendicular to its rotational axis, is also the plane on which we assume the ring lies. This plane intersects the ecliptic plane at the `rings nodes line'. We arbitrarily assume that when seen from the $+z\sup{\ecl}$ axis, the star moves in the counterclockwise direction. The direction of the $+x\sup{\ecl}$ axis will be chosen so that the planet's `vernal equinox' (letter V in \autoref{fig:orbit_ring_orientation}) occurs when the star is on that point and moves towards points of its orbit where the `north' side of the rings is fully illuminated.  For notation such as $z\sup{\ecl}$, please refer to \autoref{tab:symbols} where a complete list of the symbols and quantities used in this work is presented.

\begin{table}
  \renewcommand{\arraystretch}{2.3} 
  \linespread{0.5}\selectfont\centering  
  \begin{tabularx}{\columnwidth}{lX}
    \hline\hline
    Symbol & Definition\\\hline
    \equ & A quantity referred to the equatorial system.\\
    \ecl & A quantity referred to the ecliptic system.\\
    \obs & A quantity referred to the observer system.\\
    \hor & A quantity referred to the horizontal system.\\
    $\lambda,\beta$ & Ecliptic longitude and latitude [rad].\\
    $\alpha,\delta$ & Equatorial right ascension and declination [rad].\\
    $A,h$ & Horizontal azimuth and elevation [rad].\\
    $l,b$ & Planetocentric latitude and longitude [rad].\\
    $\hat{n}\sub{r}$ & Normal vector to the ring surface.\\
    $\hat{n}\sub{p}$ & Normal vector to the planetary surface.\\
    $\hat{n}\sub{f}$ & Normal vector to a spangle.\\
    $\hat{n}\sub{o}$ & Unitary vector in the direction of the observer.\\
    $\hat{n}\sub{s}$ & Unitary vector in the direction of a light source as seen from a spangle.\\
    $I$ & Inclination of the planetary orbit with respect to the plane of the sky ($I=90\o$ edge-on orbit) [rad].\\
    $i$ & Inclination of the ring with respect to the plane of the orbit ($i=0\o$ rings are on the plane of the orbit) [rad].\\
    $i\sub{o}$ & Effective inclination of the ring with respect to the plane of the sky ($i=0\o$ rings are on the plane of the sky) [rad].\\
    $\Lambda$ & Angle between the direction of a light source on the sky of a spangle and the normal vector to the spangle [rad].\\
    $Z$ & Angle between the direction of the observer as seen on the sky of a spangle and the normal vector to the spangle [rad].\\
    $\theta$ & Angle between the direction of a light source and that of the observer as seen on the sky of a spangle [rad].\\
    $A'$ & Difference in azimuths of the direction of a light source and that of the observer as seen on the sky of a spangle [rad].\\
    $\gamma_0$ & Colour independent single scattering albedo.\\
    $\zeta$ & $\cos \Lambda$, cosine of incident polar angle.\\
    $\eta$ & $\cos \phi$, cosine of diffusely reflected polar angle.\\
    $\af$ & spangle area [m$^2$].\\
    $\gamma_\nu$ & Single scattering albedo for continuous medium.\\
    $\gamma'_\nu$ & Single scattering albedo for a surface.\\
    $\AS$ & Spherical albedo of the planet.\\
    $\AL$ & Lambertian albedo of spangles.\\
    $a$ & Planetary orbit semi major axis.\\
    $e$ & Planetary orbit eccentricity.\\
    \hline\hline
  \end{tabularx}
  \caption{Symbols, conventions, and units used in this work.\label{tab:symbols}}
\end{table}

\begin{figure}
  \includegraphics[width=\columnwidth]{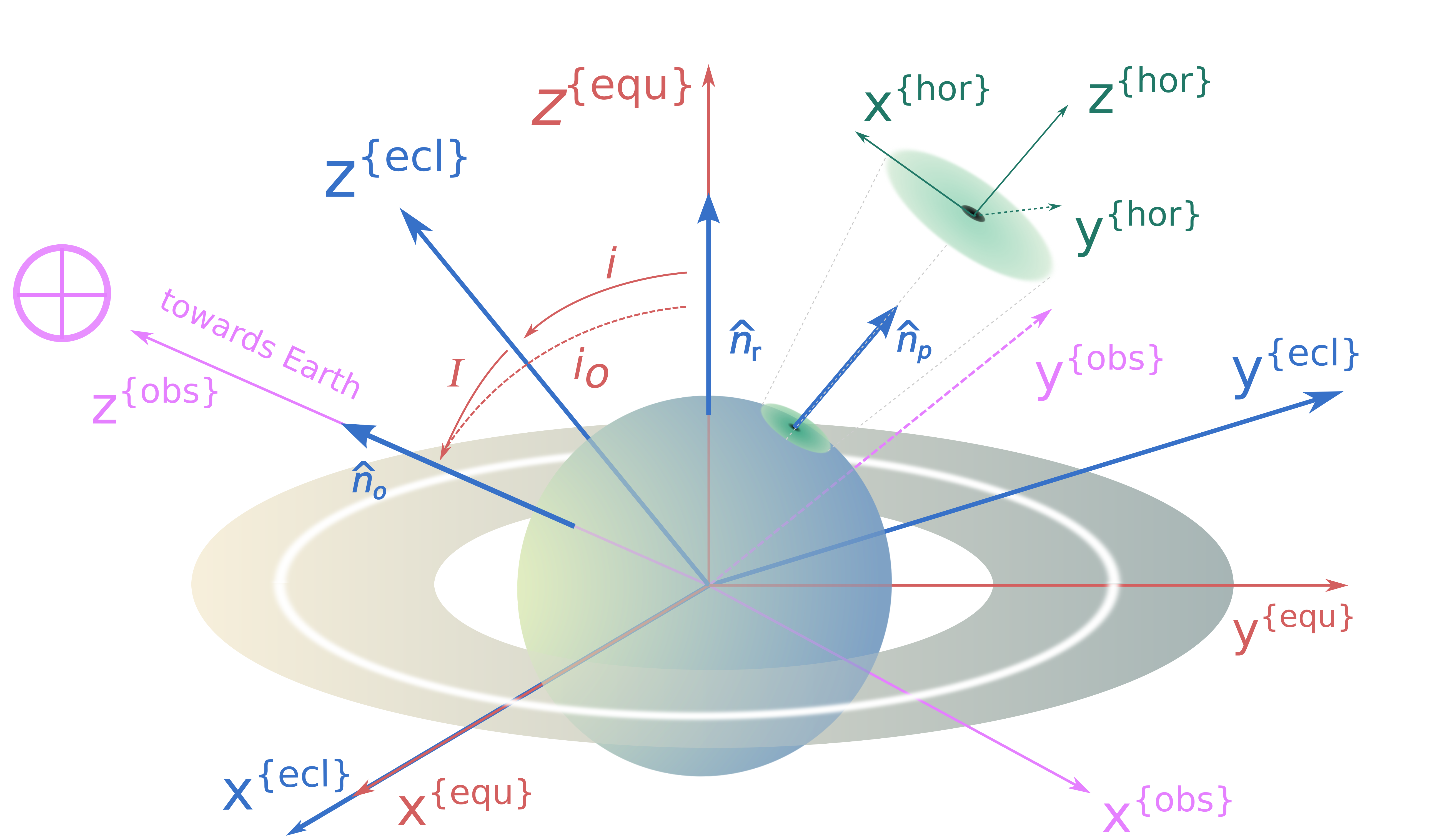}
    \caption{Illustration of the three coordinate systems used in our model: the ecliptic \ecl\ attached to the stellar orbit, the equatorial \equ\ attached to the planet's equator or the plane of the rings, and the horizontal \hor\ attached to a tangent plane at any point of the surface of the planet or ring. The inclination angles $I$, $i$, and $i\sub{o}$ are used in the model for different purposes: $I$ is the inclination of the planetary orbit with respect to the line-of-sight ($I=90^\circ$ corresponds to an edge-on orbit), $i$ is the rings' physical inclination with respect to the planetary orbit ($i=0^\circ$ corresponds to rings lying on the orbital plane), and $i\sub{o}$ is the apparent inclination of the rings' plane with respect to the observer ($i=0^\circ$ corresponds to face-on rings).}
    \label{fig:reference_frames}
\end{figure}

All the relevant geometrical and physical quantities defined hereafter, and used throughout our model, are referred to four different coordinate systems (see \autoref{fig:reference_frames}):

\begin{enumerate}
\item {\bf Ecliptic system, \ecl}.  A system attached to the orbit of the star. The spherical coordinates with respect to this system are $(r,\lambda,\beta)$, whe $r$ is the planetocentric distance, $\lambda$ and $\beta$ the ecliptic longitude and latitude, respectively.  The zero meridian in this system passes through the vernal equinox (point V in \autoref{fig:orbit_ring_orientation}).

\item {\bf Equatorial system, \equ}.  A system attached to a plane perpendicular to the planet's rotational axis (i.e. its equator). The $+z\sup{\equ}$ axis points in the direction of the planetary angular momentum: the rotation of the planet goes from the $+x\sup{\equ}$ axis to the $+y\sup{\equ}$ axis, that is in the counterclockwise direction as seen from $+z\sup{\equ}$. All points for which $z\sup{\equ}>0$ are to the `north' of the system, irrespective if they are on the planet, the rings, or other bodies thereof.  

For a ringed planet (in the current version of the model), all ring particles are assumed to be on the \xy\equ\  plane. The spherical coordinates with respect to this system are $(r,\alpha,\delta)$, where $\alpha$ is analogous to right ascension and $\delta$ to declination in the Earth's equatorial system. The zero meridian in this system also passes through the vernal equinox.

\item {\bf Observer system, \obs}.  A system attached to the plane of the sky of an observer on Earth. For convention, the observer will be located towards the $+z\sup{\obs}$ axis ($\oplus$ symbol in \autoref{fig:reference_frames}) and the $+x\sup{\obs}$ axis will always be on the plane of the ecliptic. The position of the observer in the model is described by its spherical coordinates with respect to the \ecl\ system $(\lambda\sub{obs},\beta\sub{obs})$, and we assume that $\beta\sub{obs}> 0$ (i.e. the observer is always to the north of the ecliptic).

The orbit's inclination  with respect to the plane of the sky, $I$, and the effective inclination of the rings with respect to the observer, $i\sub{o}$, are given by

\beq{eq:inclinations}
\begin{split}
&I = \frac{\pi}{2}-\beta\sub{obs}\\
&\cos i\sub{o} = \hat{n}\sub{o}\cdot \hat{n}\sub{r},
\end{split}
\eeq
where the unitary vector $\hat{n}\sub{o}$ points from the centre of the planet to the observer and $\hat{n}\sub{r}$ is normal to the rings' surface.
    
\item {\bf Horizontal (local) system, \hor}.  A system attached to the tangent plane of any point on the surface of the planet or ring (see the disc around the black point in \autoref{fig:reference_frames}). For convention, the $+z\sup{\hor}$ axis is normal to the surface of the object and, in the case of the planet, the $+x\sup{\hor}$ axis points towards the north pole. To make this system comparable to Earth's horizontal system, \hor\ coordinates are left-handed (i.e. $\hat{e}_y\sup{\hor}\times\hat{e}_x\sup{\hor}=\hat{e}_z\sup{\hor}$).

The spherical coordinates with respect to this system are $(r,A,h)$, where $A$ is analogous to azimuth and $h$ to elevation. For points on the ring, we use the equatorial system for the same purposes of \hor.  Therefore, in those cases we assume $A=\alpha$ and $h=\delta$.

\end{enumerate}  

\subsection{Sampling the surface of diffuse objects}
\label{subsec:spangles}

To simulate and integrate all possible optical effects affecting the light curve of the system, we discretize the surface of these bodies into small, plane circular area elements. These elements have a geometrical and optical similarity to sequins or spangles, those small reflective objects on the surface of elegant clothes. These particular discretization is precisely the reason why we coin our model {\bf P}laneta{\bf ry} spa{\bf ngles} or  \comp{\model}.\footnote{\url{https://github.com/seap-udea/pryngles-public}. All calculations and figures in this paper were performed using {\tt Pryngles} and are available for reproducibility purposes in \url{https://bit.ly/pryngles-paper-figures}.}

A single spangle on the surface of a planet or ring has a position specified by its coordinates on the \equ\ reference frame. If placed on the surface of a planet, a spangle's position and orientation is described by its spherical coordinates (planet-centric latitude $b$ and longitude $l$), a normal vector to the spangle ($\nf$), and its area ($\af$).

\begin{figure}
  \includegraphics[width=\columnwidth]{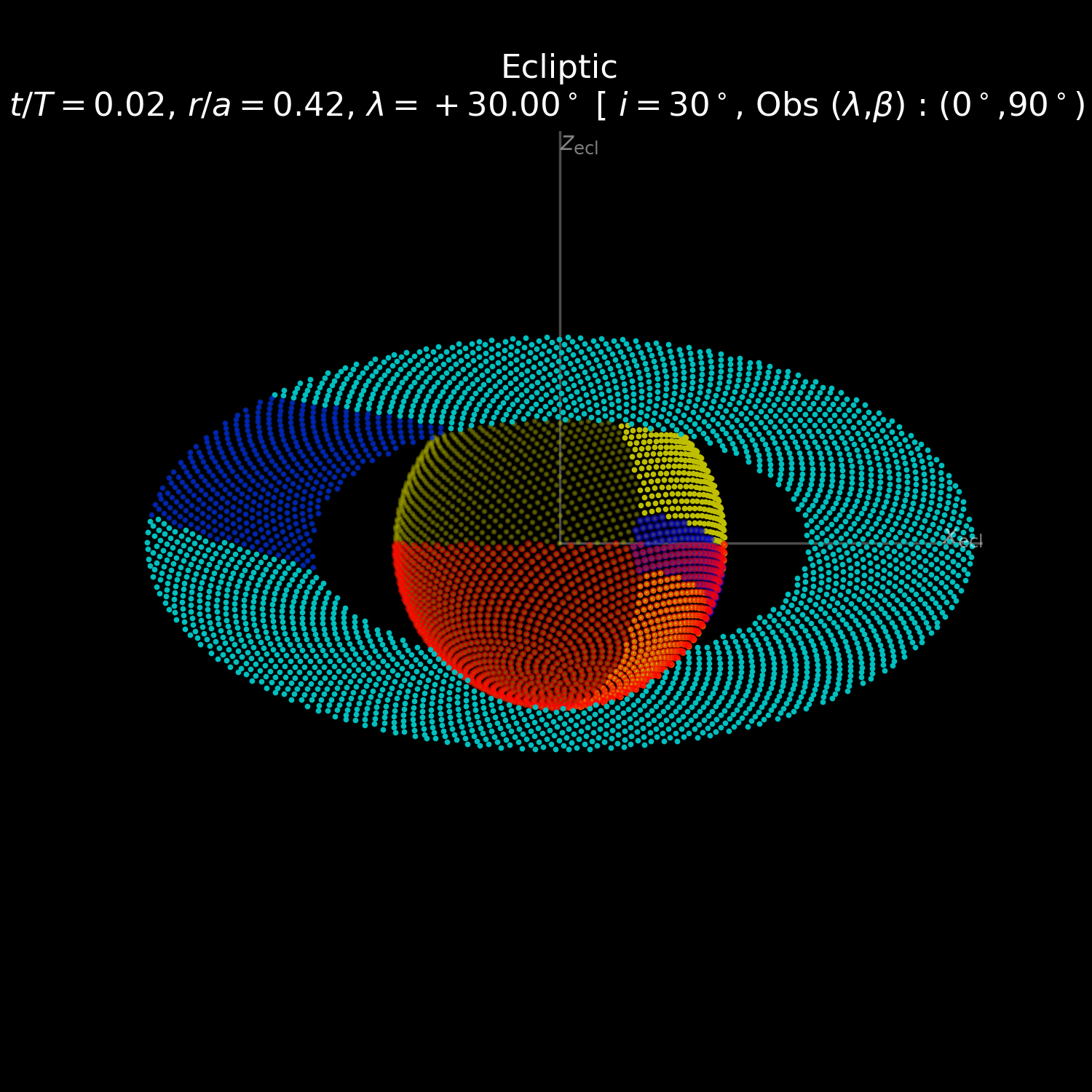}
    \caption{Example of planet and ring surface sampling using a Fibonacci spiral distribution (we use 5000 points for the planet's surface and 5000 points for the ring). The plot only shows the position of the spangles and not its extent.  Colours indicate the spangle state (see \autoref{subsec:spstate}): cyan points correspond to ring spangles illuminated by the star; yellow points correspond to planetary spangles illuminated by the star and visible from the vantage point of the observer (which is located in the direction of the $+z\sup{\ecl}$ axis, see legend); blue points mark the position of spangles inside the planetary or ring shadow; and red points correspond to planetary spangles which are invisible to the observer. For a colour (and animated) version of this figure see the electronic manuscript.}
    \label{fig:planet_ring_sampling}
\end{figure}

In order to cover the surface of planets and rings with spangles as uniformly as possibly, we use a Fibonacci spiral sampling\footnote{See \url{https://bit.ly/fibonacci-sampling} and references there in (visited \today).}.
In \autoref{fig:planet_ring_sampling}, we show an example of the positions of ring and planetary spangles in the \ecl\ system of coordinates.

\begin{figure}
  \includegraphics[width=\columnwidth]{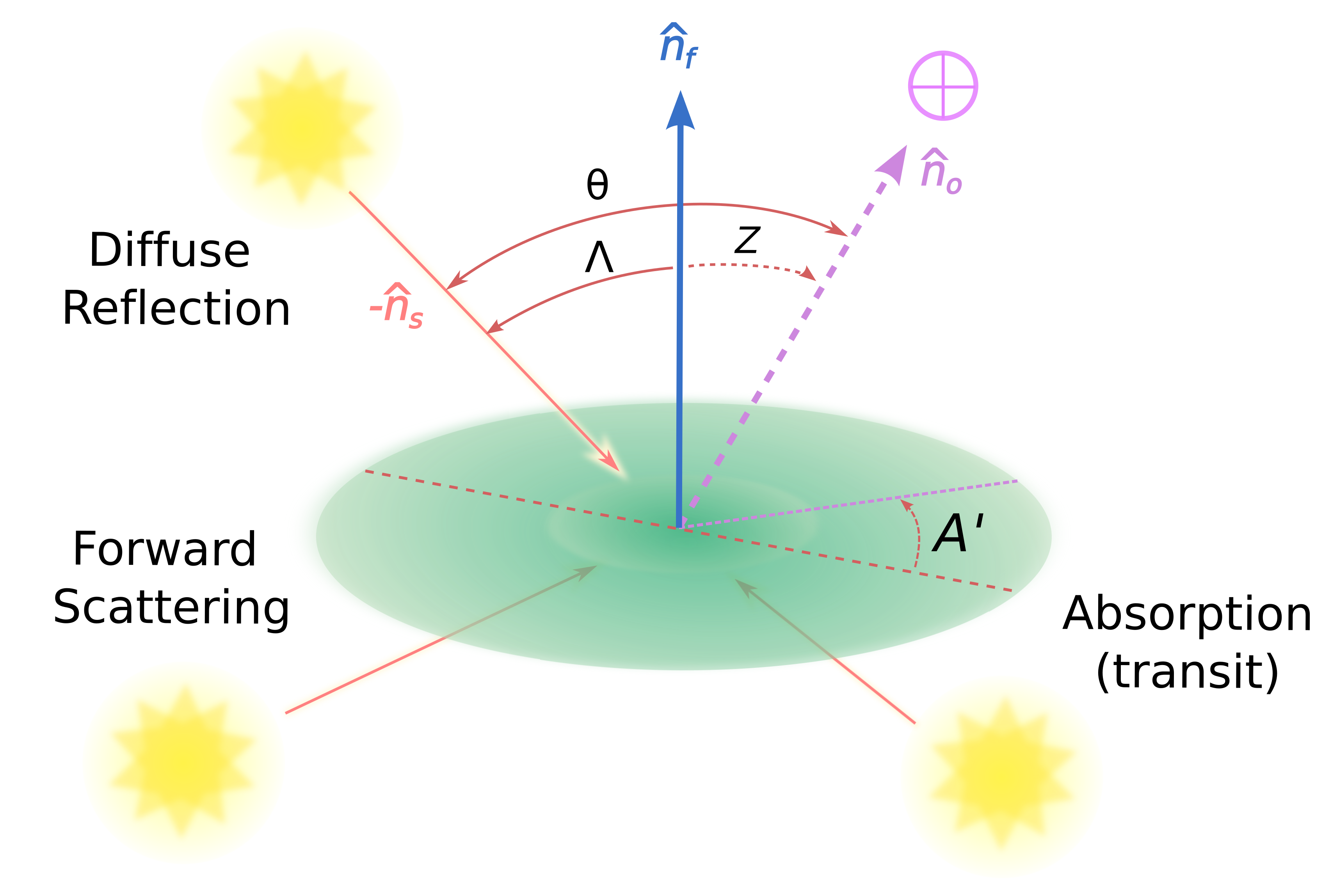}
    \caption{Geometry of light interacting with a planetary surface area element ({\em spangle})  of the planet or ring. Incoming light from a source (star or diffuse light from other spangles) forms an angle $\Lambda$ (for Lambertian, see text) with respect to the normal vector ($\nf$) of the spangle. In the more general case, the observer is in a direction with a zenith angle $Z$ and an azimuth difference $A'$ with respect to the direction of the incoming light. Diffuse reflection and absorption are possible on spangles of both the planet and ring, but forward scattering is only possible for semitransparent spangles (e.g. ring particles).}
    \label{fig:spangle_geometry}
\end{figure}

Light coming from a given source (primary or diffuse) on the sky of a spangle, may interact through different processes (see \autoref{fig:spangle_geometry}): diffusive reflection, absorption, and forward scattering.  As a result, the intensity of the light arriving to the observer will be modified in a complicated way, depending on the angles $\Lambda$ (incident angle), $Z$ (observer polar angle), $A'$ (relative azimuth), and $\theta$ (polar angle between the incoming and outgoing rays). In the next section, we describe the details of the adopted models for these processes.

\section{Modelling optical effects}
\label{sec:optics}

The frequency-dependent flux, $F_\nu(t)$ [W m$^{-2}$ Hz$^{-1}$], received on Earth from the extrasolar system at a given time $t$ and frequency $\nu$ is (in general) given by

\beq{eq:fluxnu}
F_\nu(t)=F\sub{\nu}\sup{\star}+F\sub{\nu}\sup{p}(t),
\eeq
where $F\sub{\nu}\sup{\star}$ is the natural stellar flux (i.e. with no planet) and $F\sub{\nu}\sup{p}(t)$ is the flux coming from all the diffusive objects in the system (planets, rings, and moons). If we have several planets, then $F\sub{\nu}\sup{p}(t)=\sum_i F\sub{\nu}^{\mathrm{p},i}(t)$, where $F\sub{\nu}^{\mathrm{p},i}(t)$ is the flux coming from the $i$-th diffusive object (in case of a ringed object, this flux would include both the flux of the planet and ring, as explained below). Note that $F\sub{\nu}\sup{p}(t)<0$ when the planet is transiting in front of the star.

In practical cases, we assume that any instrumental effect (e.g. trending), variability (e.g. stellar oscillations and variability), and blending have been removed from the observed stellar flux to produce the constant $F\sub{\nu}\sup{\star}$.

Here, we define the (dimensionless) frequency-dependent flux anomaly, or simply the light curve $L_\nu(t)$ as

\beq{eq:light_curve}
L_\nu(t)\equiv\frac{F\sub{\nu}(t)-F\sub{\nu}\sup{\star}}{F\sub{\nu}\sup{\star}}=\frac{F\sub{\nu}\sup{p}(t)}{F\sub{\nu}\sup{\star}}.
\eeq

According to \autoref{eq:light_curve}, the light curve will be $L_\nu(t)=0$ if no planetary system is present around the star; $L_\nu(t)>0$ when the diffusive objects in the system increase the stellar flux (i.e. the bright-side of the light curve); and $L_\nu(t)<0$ during planetary transits  (i.e. the dark side of the light curve).

For the case of a single ringed planet, the flux-anomaly contribution to the light curve of the $i$-th spangle can be written as

\beq{eq:light_curve_terms}
\delta L_\nu^i=R_\nu^{\star,i}+\sum_j R_\nu^{ij}+f_\nu^i-T_\nu^i,
\eeq
where $R_\nu^{\star,i}$ is the contribution of the diffusely reflected starlight on the $i$-th spangle; $R_\nu^{ij}$ is the diffusely reflected light coming from the $j$-th spangle (planet- or ring-shine); $f_\nu^i$ is the forward-scattered light of the $i$-th spangle; and $T_\nu^i$ is the fraction of light subtracted from the stellar flux when the $i$-th spangle transits the star. 

Note that some terms are zero at a given configuration; for instance, $T_\nu^i=0$ when the spangle does not transit and $f_\nu^i$ is always zero for opaque spangles like those on planets or moons.

In the following sections, we describe the physics of each contribution to $\delta L_\nu^i$. For this purpose, we use the assumptions and notation of the classical works by \cite{Russell1916} and \cite{Sobolev1975}.

\subsection{Diffuse reflection}
\label{subsubsec:scattering}

When light interacts with a system (e.g. planetary atmosphere or rings), it is partially or completely scattered by particles in the medium. The fraction of specific intensity at frequency $\nu$, $I_\nu$ [W m$^2$ sr$^{-1}$ Hz$^{-1}$], scattered by a volume element at an angle $\theta$ with respect to the incident direction is

\beq{eq:light_intensity_nu}
I_{\nu}=\gamma_\nu I_{0\nu} \Phi_\nu(\theta),
\eeq
where $0\leq\gamma_\nu\leq 1$ is a dimensionless quantity called the {\em single scattering albedo} and $\Phi_\nu(\theta)$ is called the phase function.

The value of $\gamma_\nu$ and the specific functional form of $\Phi_\nu(\theta)$ will depend on many complex factors (for a detailed discussion and specific expressions of both quantities, see section 1.1 of \citealt{Sobolev1975}). For the current version of our model, we use simplified expressions for both quantities that depend on the type of diffusive surfaces involved (see next sections). However, given that our method is general and the software package is modular, any complex generalisation of the physics of diffuse reflection can easily be implemented and incorporated into the model.

\subsubsection{Planetary spangles}
\label{subsubsec:reflection_planet}

In order to calculate the flux of light diffusely scattered by a planetary atmosphere, we must use a specific form of $\Phi_\nu(\theta)$, assume a value of $\gamma_\nu$, and then solve the equation of radiative transfer. However, for the purposes of this work, we consider the atmosphere as a semi-infinite layer of matter (optical opacity $\tau\rightarrow\infty$). In this case, the flux $B\sub{\nu}$ [W m$^{-2}$ Hz$^{-1}$] diffusely reflected by an area element of the atmosphere's surface is given by Eq. (9.4) in \citet{Sobolev1975}:

\beq{eq:diffuse_reflection}
B\sub{\nu}(\zeta,\eta,A')=\rho_\nu(\zeta,\eta,A')  B_{\nu0}\zeta\eta,
\eeq
with $B_{\nu0}$ the incoming flux; $\eta=\cos \Lambda$ (incident polar angle, see \autoref{fig:spangle_geometry}) and $\zeta=\cos Z$ (scattered polar angle) are geometrical factors. $\rho_\nu(\zeta,\eta,A')$ is called the {\em reflection coefficient}, and in general this function results from solving the radiative transfer equation.

We focus here on the most simple case of an {\em isotropically-gray} phase function (i.e. $\Phi_\nu=1$) and a colour-independent single scattering albedo $\gamma_0$. In this case, an analytical expression for $\rho$ is given by Eq. (2.43) in \citet{Sobolev1975}:

\beq{eq:reflection_coefficient}
\rho_0(\zeta,\eta)=\frac{\gamma_0}{4}\frac{f(\gamma_0,\eta)f(\gamma_0,\zeta)}{\eta+\zeta},
\eeq
where $f(\gamma_0,\mu)$ obeys equation (2.44) in \cite{Sobolev1975}. To calculate $\rho_0(\zeta,\eta)$ and its related physical quantities, we use in our model the values provided in table 2.3 of the same work 

By averaging $\rho_0(\zeta,\eta)$ over all the possible observing directions $\zeta$, and assuming the surface reflectance to be independent of the observing direction ({\it Lambertian reflectance}), the {\em bulk albedo} of a planetary spangle in the direction $\eta$ can be written as

\beq{eq:lambp}
\ALp(\eta)=2\int_0^1 \rho_0(\zeta,\eta)\zeta\;\mathrm{d}\zeta,
\eeq
and its contribution to the light curve from diffuse reflection could be finally written from \autoref{eq:diffuse_reflection} as

\beq{eq:diffuse_spangle_planet}
R\sub{p}^{\star,i,\planet}=\frac{1}{4\pi}\frac{\af}{r_{\star,i}^2}\ALp(\cos \Lambda_{\star,i}) \cos\Lambda_{\star,i} 
\cos Z_{\obsr,i}.
\eeq

In \autoref{eq:diffuse_spangle_planet}, $r_{\star,i}$ is the distance from the star to the $i$-th spangle, $\Lambda_{\star,i}$ the angle between its normal vector and the direction towards the star, while $Z_{\obsr,i}$ is the angle between such normal vector and the direction of the observer. That being said, if we provide a value for the single scattering albedo ($\gamma_0$), the bulk Lambertian albedo of the spangle can be computed with \autoref{eq:lambp} and then its contribution to the diffuse light flux via \autoref{eq:diffuse_spangle_planet}.

However, if instead of providing $\gamma_0$ (which is a microscopic uncertain property) we assume a value for the {\em spherical Albedo} of the planet (equation 1.87 in \citealt{Sobolev1975}),

\beq{eq:spherical_albedo}
\AS=\frac{2}{\pi}\int\int\int \rho_\nu(\zeta,\eta,A)\eta \zeta\;\mathrm{d}\eta\;\mathrm{d}\zeta\;\mathrm{d}A,
\eeq
which can be estimated for the Solar System as well as for extrasolar planets \citep{Dyudina2016}, we can obtain a value of $\gamma_0$ by combining both \autoref{eq:reflection_coefficient} and \autoref{eq:spherical_albedo}. This will be the parametrization used in this work for the current version of \comp{\model}.

\subsubsection{Ring spangles}
\label{subsubsubsec:reflection_rings}

The contribution of each ring spangle to the diffusely reflected light will be computed via an analogous formula to \autoref{eq:diffuse_spangle_planet},

\beq{eq:diffuse_spangle_ring}
R\sub{\nu}^{\star,{\ring}i}=\frac{1}{4\pi}\frac{\af}{r_{\star,i}^2}\ALr(\Lambda_{\star,i}) \cos\Lambda_{\star,i} 
\cos Z_{\obsr,i}.
\eeq

Since a ring system is a large collection of particles with sizes much larger than light wavelength ($s/\lambda\gg 1$), we calculate albedo using Eq. (1) in \citet{Russell1916}:

\beq{eq:albedo}
\ALr(\Lambda)=2\pi\gamma'_{\nu}\int_0^{\pi/2} \frac{f(\Lambda,Z)}{\cos\Lambda}\sin Z\;\mathrm{d}Z,
\eeq
where $\gamma'_\nu$ is analogous to the single scattering albedo of the surface and $f(\Lambda,Z)$ is a phase-like function known as the `law of diffuse reflection'. For $f(\Lambda,Z)$ there are two well-motivated and tested possibilities  (see \citealt{Hameen-Anttila1972}):

\begin{enumerate}

\item The Lambertian law, which assumes that a spangle's brightness is the same regardless of the angle from which it is illuminated or observed. According to this, the reflection law can be written as
  
\beq{eq:reflection_law_lambertian}
f\sub{L}(\Lambda,Z)=\cos\Lambda,
\eeq
and 

\beq{eq:lambertian_albedo}
\ALr=2\pi\gamma'_\nu
\eeq
irrespective of the incoming angle.

\item The Lommel-Seeliger law:

\beq{eq:reflection_law_LS}
f\sub{LS}(\Lambda,Z)=\frac{\cos\Lambda\cos Z}{\cos\Lambda+\cos Z},
\eeq
where the normal Lambertian albedo is given by

\beq{eq:normal_lambertian_albedo}
\mathrm{Lommel-Seeliger:}\;
\ALr(0)=2\pi\gamma'_\nu(1-\ln 2).
\eeq
\end{enumerate}

As we did in the case of planetary spangles, in the current version of \comp{\model} we set the value of the normal Lambertian albedo $\ALo$ for the rings. Then, using \autoref{eq:lambertian_albedo} we find the value of the corresponding $\gamma'_\nu$ for the selected reflection law. Once set, we use this quantity to compute the incident-angle-dependent albedo needed in \autoref{eq:diffuse_spangle_ring}.

\bigskip

In summary, in the current version of the model, the spherical (wavelength-independent) albedo $A\sup{(sph)}$ of the planet; and the (normal) Lambert's albedo $\ALr$ and law of diffuse reflection $f(\Lambda,Z)$ for the ring, should be provided. From these quantities; and using \autoref{eq:spherical_albedo} and \autoref{eq:lambertian_albedo} (or \autoref{eq:normal_lambertian_albedo}) the single scattering albedo $\gamma_0$ (in the case of the planet) and $\gamma'_\nu$ (for the ring) can be estimated.  Using these two quantities, we can finally compute the direction-dependent albedo of the spangle (\autoref{eq:lambp} and \autoref{eq:lambertian_albedo}) and finally its contribution $R^{\star,i}$ to the light curve.

\subsection{Spangle transit}
\label{subsubsec:spangle_transit}

If a spangle (either opaque or semitransparent) transits in front of the star, we should remove the corresponding blocked flux from the light curve. For completely opaque spangles (e.g. those of a planet), their contribution to the light curve (equation \ref{eq:light_curve_terms}) will be given by

\beq{eq:spangle_transit}
T\sub{\nu}^{\mathrm{p},i}=-\af \frac{I\sub{\nu}(\mu)}{I\sub{\nu}(0)} \cos \Lambda_{i},
\eeq

Here $I\sub{\nu}(\mu)/I\sub{\nu}(0)$ is the intensity of the stellar disc in the direction of the spangle's centre and $\mu\equiv\sqrt{1-R^2/\Rs^2}$ is a geometrical factor used to fit empirically the limb-darkening of the stellar disc. $R$ the projected distance to the stellar centre in the \obs\ coordinate system (see  \autoref{fig:reference_frames}) and $\Rs$ the stellar radius.

In its current version, \comp{\model} uses the non-linear limb darkening law \citep{Claret2000}:

\beq{eq:non_linear_limb_darkening}
\frac{I\sub{\nu}(\mu)}{I\sub{\nu}(0)}=1-\sum_{n=1}^{4} c_{\nu,n}\left(1-\mu^{n / 2}\right),
\eeq
and the band-dependent coefficients ($c_{\nu,n}$) published by \citet{Sing2010}\footnote{Available at \url{https://bit.ly/limb-darkening-coefficients} (visited \today).}

On the contrary, if the spangle is semitransparent (e.g. ring spangles), the contribution to the light curve will be calculated as 

\beq{eq:ring_spangle_transit}
T_{i}^{\mathrm{r},i}=-\beta\af \frac{I(\mu)}{I(1)} \cos \Lambda_{i},
\eeq
where $\beta$ is an attenuation factor given by \citep{Barnes2004}:

\beq{eq:attenuation_factor}
\beta\equiv\left(1-\frac{\tau\sub{\nu}\sec\Lambda}{2}e^{-\tau_\nu\sec\Lambda}\right).
\eeq

Here, $\tau_\nu$ is the rings' total optical depth, which includes the effect of light absorbed (the so-called geometrical optical depth $\tau\sub{\nu,g}$) and diffracted by ring particles. 

For large particle size ($s\gg\lambda$), the Babinet's principle states that light diffracted by large objects is the same as the light absorbed by its cross-section (geometrical absorption), so the total optical depth is $\tau_\nu=2\tau\sub{\nu,g}$ \citep{French2000}.

On the other hand, if rings are composed of small particles, a non-negligible contribution to the light curve will be due to forward-scattered light. This will especially affect the ingress/egress phases of the transit \citep{Barnes2004}. However, for most of the time when rings are transiting, the effect of such particles can be modelled simply by changing the geometrical optical depth by a factor $Q\sub{sc}$ which accounts for the forward-scattering efficiency.

As the focus of the current version of our model is the bright side of the light curve, we have not attempted a more realistic description of forward-scattering. Still, since the package is modular, it would be easy to include these interesting effects in future versions.

\subsection{Planet- and ring-shine}
\label{subsubsec:shine}

\begin{figure}
  \includegraphics[width=\columnwidth]{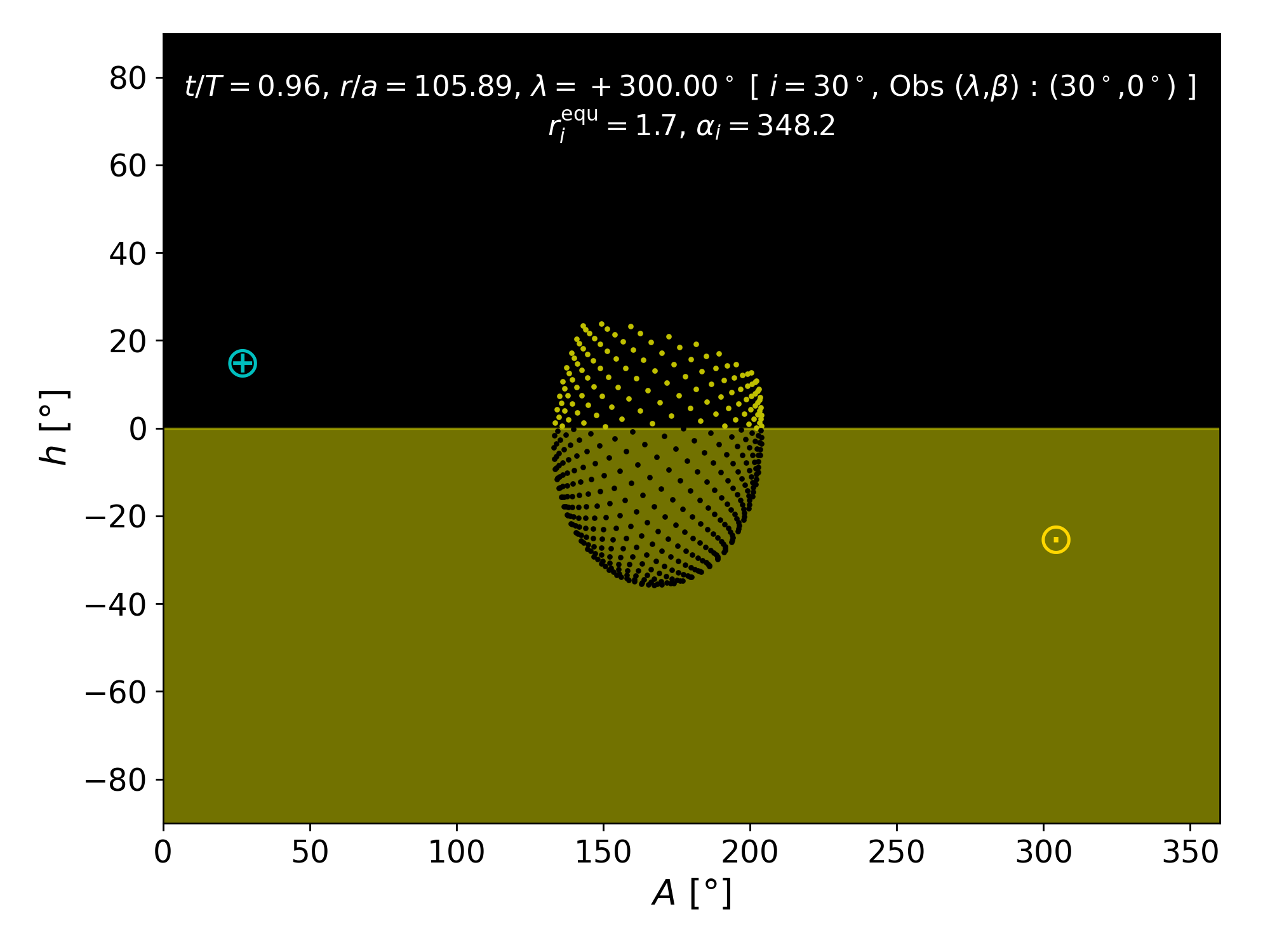}
  \includegraphics[width=\columnwidth]{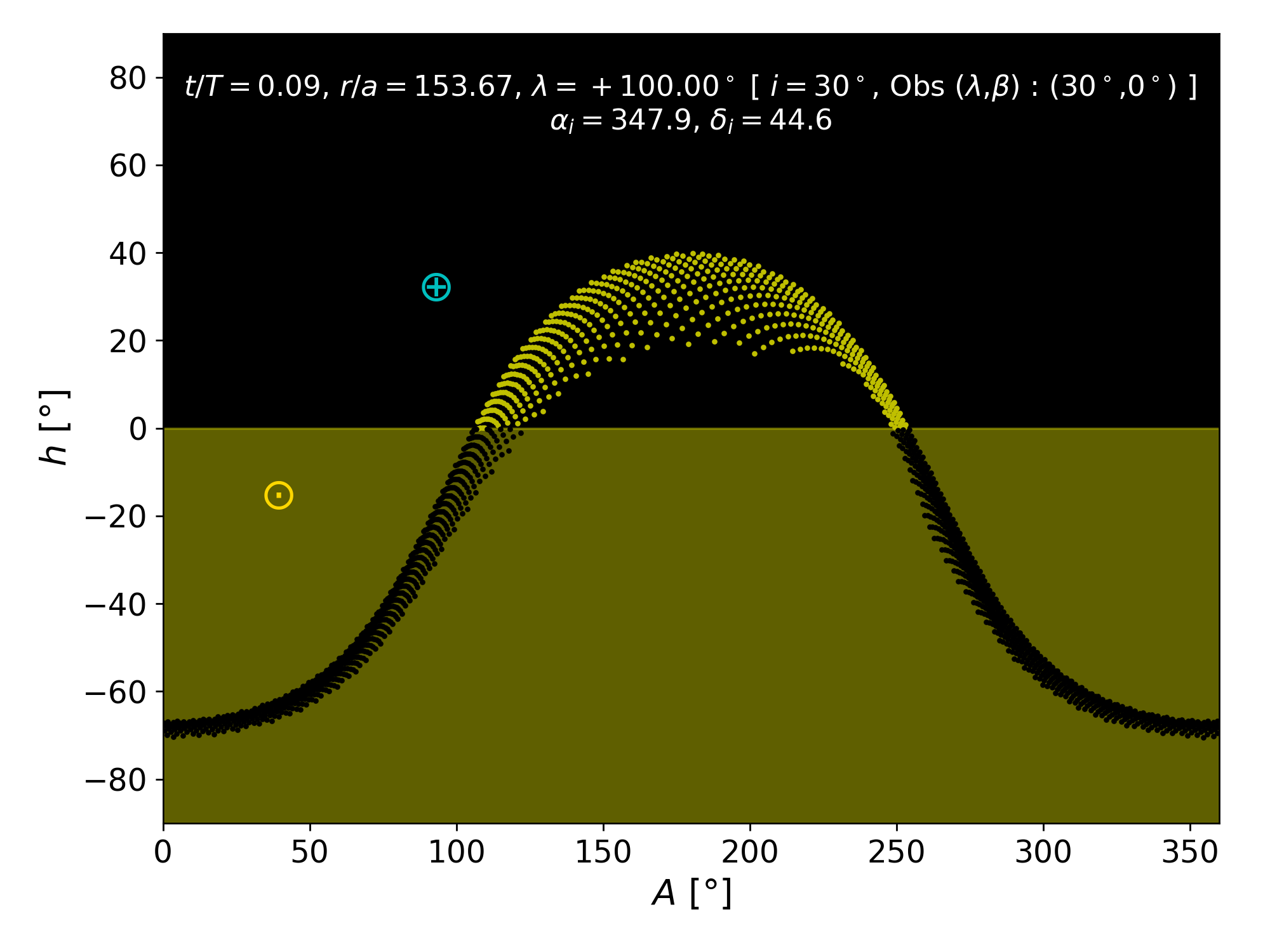}
    \caption{Schematic representation of the position of all the spangles on a simulated ringed planet, over and under (shaded area) the horizon of a ring spangle (upper panel) and a planetary spangle (lower panel). The $\odot$ and $\oplus$ symbols indicate the direction on the sky of the star and the observer, respectively. In this configuration the star is under the horizon, but the spangle receives light from illuminated spangles (yellow dots) of the planet and ring.}
    \label{fig:planet_ring_shine}
\end{figure}

Spangles are not only illuminated by the star. They also receive light from the planet (planet-shine) and/or from the ring (ring-shine). To determine which spangles on the planet or the ring illuminate a given spangle, we compute (in the local reference frame of the illuminated spangle) the horizontal coordinates $(x\sup{\hor},y\sup{\hor},z\sup{\hor})$ of all spangles in the simulation (see \autoref{fig:planet_ring_shine}).

Spangles with $z\sup{\hor}\geq 0$ potentially shed light on the local spangle, so their indirect flux should be added to the contribution to the light curve of the latter. Thus, the ring-shine on a planetary spangle increases its brightness by

\beq{eq:ring_shine}
R_\nu^{{\planet}i,{\ring}j}=\frac{R_\nu^{\star,{\ring}j}}{4\pi} 
\frac{\af}{r_{ij}^2} \ALp(\cos \Lambda_{ij}) \cos \Lambda_{ij}\cos Z_{ij},
\eeq
where $R_\nu^{\star,{\ring}j}\cos Z_{ij}$ is the normalised light intensity coming from the $j$-th ring spangle (calculated with \autoref{eq:diffuse_spangle_ring}), $\Lambda_{ij}$ is the angle between the normal of the $i$-th planetary spangle (the illuminated one) and the position vector of the $j$-th ring spangle ($\vec r_{ij}$), and $Z_{ij}$ is the angle between the normal of the $j$-th ring spangle and $\vec r_{ij}$.

Similarly, planet-shine will increase the brightness of the ring spangles following

\beq{eq:planet}
R_\nu^{{\ring}j,{\planet}i}=\frac{R_\nu^{\star,{\planet}i}
}{4\pi} 
\frac{\af}{r_{ji}^2} \ALr(\Lambda_{ji}) \cos \Lambda_{ji}
\cos Z_{ji},
\eeq
and now $\Lambda_{ji}$ is the angle of the position vector with respect to the normal vector of the $j$-th ring spangle. The role of $\Lambda_{ji}$ and $Z_{ji}$ is inverted ($\Lambda_{ji}$ is the angle of the position vector with respect to the normal vector of the $j$-th ring spangle).

\subsection{Spangles states}
\label{subsec:spstate}

In a given configuration for the star and the observer, the spangles in the simulation can be in different conditions: they might be illuminated or shadowed by the star, be visible or invisible from the vantage-point of the observer, etc. We call `state' to the different illumination and visibility conditions of a spangle. A state determines the way a spangle will interact with light or contribute to the light curve.

In order to characterise spangle states, we define six Boolean state variables:

\begin{enumerate}

  \item {\bf Visibility, v}. True, if the spangle is visible from Earth. That is, there is no blocking object (planet or ring) in the line-of-sight directed to the centre of the spangle. 

  \item {\bf Illumination, i}. True, if the spangle is potentially illuminated by the star.  In the case of planetary spangles, {\bf i} is defined by the condition $\hat{n}\sub{f}\cdot\hat{n}\sub{s}\geq 0$. Ring spangles, while semitransparent, are always potentially illuminated by the star except when $\hat{n}\sub{f}\cdot\hat{n}\sub{s}=0$ (i.e. the rings are edge-on as seen from the star).

  \item {\bf Shadow, s}. True, if the spangle is inside the shadow produced by the planet or rings. The value of {\bf s} is determined by a non-trivial geometrical condition (see \autoref{app:shadow}).

  \item {\bf Indirect illumination, n}. True, if the spangle can be indirectly illuminated by other spangles (planet- or ring-shine).

  \item {\bf Transit, t}. True, if the spangle is transiting in front of the star. The value of ${\bf t}$ is determined by the condition:

    \beq{eq:spangle_transit_condition}
    R\equiv\sqrt{(x\sub{f}\sup{obs}-x\sub{\star}\sup{obs})^2+(y\sup{obs}-y\sub{\star}\sup{obs})^2}\leq \Rs\;\mathrm{and}\;z\sub{\star}\sup{obs}< 0,
    \eeq
    where $\Rs$ is the radius of the star.
    
  \item {\bf Occultation, o}. True, if the spangle is obscured by the star. The value of ${\bf o}$ is determined by a condition analogous to that in  \autoref{eq:spangle_transit_condition}, but with $z\sub{\star}\sup{obs}>0$.

\end{enumerate}

Once the state of each spangle is evaluated, the light curve is computed by summing their contribution to the optical output from the system as described in  \autoref{sec:optics}. For efficiency purposes, we only include in the summation those spangles which, at a given observer and stellar configuration, contribute actively to the light curve. We call these the {\em active spangles}. 

In order to determine if a spangle is active or not at a given time in the simulation, we combine all the aforementioned state variables; thus, for instance, the necessary condition for a spangle to be active is that it is visible ({\bf i}). But, even if it is visible, a ring spangle will not be active if: 1) it is obscured ({\bf o}); 2) it is not illuminated ($\neg${\bf i}); or 3) it is not transiting ($\neg${\bf t}).

Symbolically, the non-active condition for ring spangles can be obtained with the logical expression:

\beq{eq:non_active_spangles}
\neg\mathbf{v} \vee [ \mathbf{o} \vee (\neg\mathbf{i} \wedge \neg\mathbf{t})].
\eeq

Similarly, the non-active condition for planet spangles is:

\beq{eq:active_spangles}
\neg\mathbf{v} \vee [ \mathbf{o} \vee (\neg\mathbf{i} \wedge \neg\mathbf{t} \wedge \neg\mathbf{n})].
\eeq

Although assessing the state and activity of the spangles may seem to be an irrelevant aspect of the physical and astronomical problem, it is key to the general method described here.

The model described so far has been implemented into a PYTHON package that we have called \comp{\model}. This package is highly modular, easy to use, and it is fully documented for both users and developers. Many of the physical algorithms can be changed to include modifications in the law of diffuse reflection and to implement forward scattering or other complex physical effects (e.g. emission, polarization, wavelength dependent albedos, etc.).

In the following sections, we show some examples of the outputs produced by the the package for several astronomically cases of interest.  We also test the validity of these results against already existing tools, at least for the dark side of the light curve.

\section{Numerical experiments}
\label{sec:numerical}

\begin{figure*}%
\centering
\subfigure{%
\label{fig:lc_scatter}%
\includegraphics[width=\fullfig\textwidth]{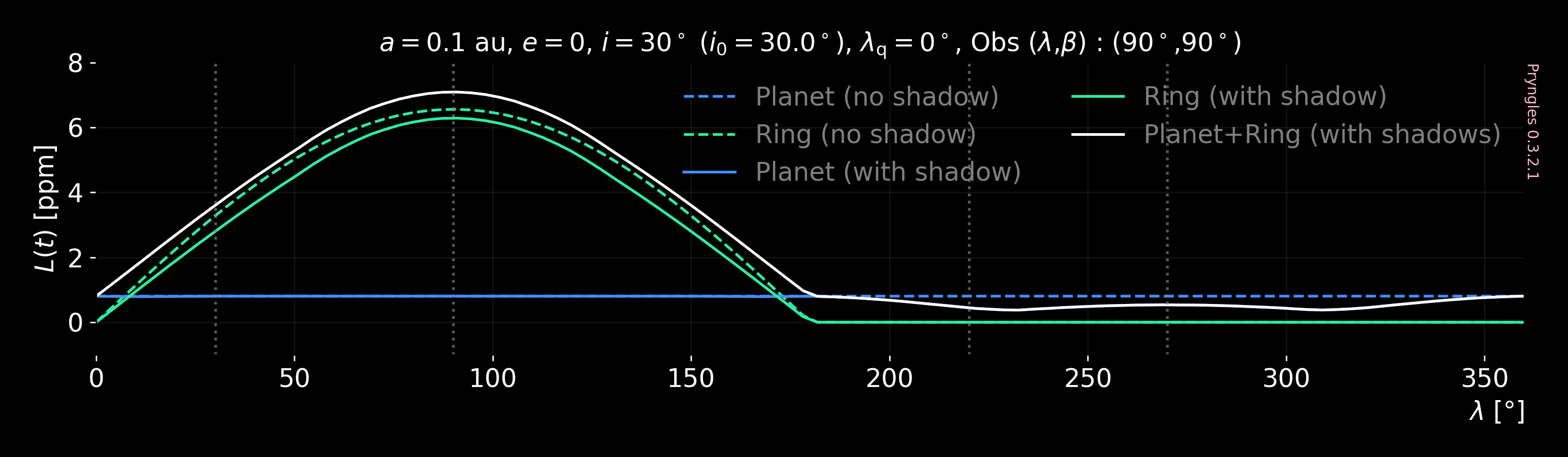}}\\%
\vspace{-0.39cm}
\subfigure{%
\label{fig:lc_scatter_30}%
\includegraphics[width=\halffig\columnwidth]{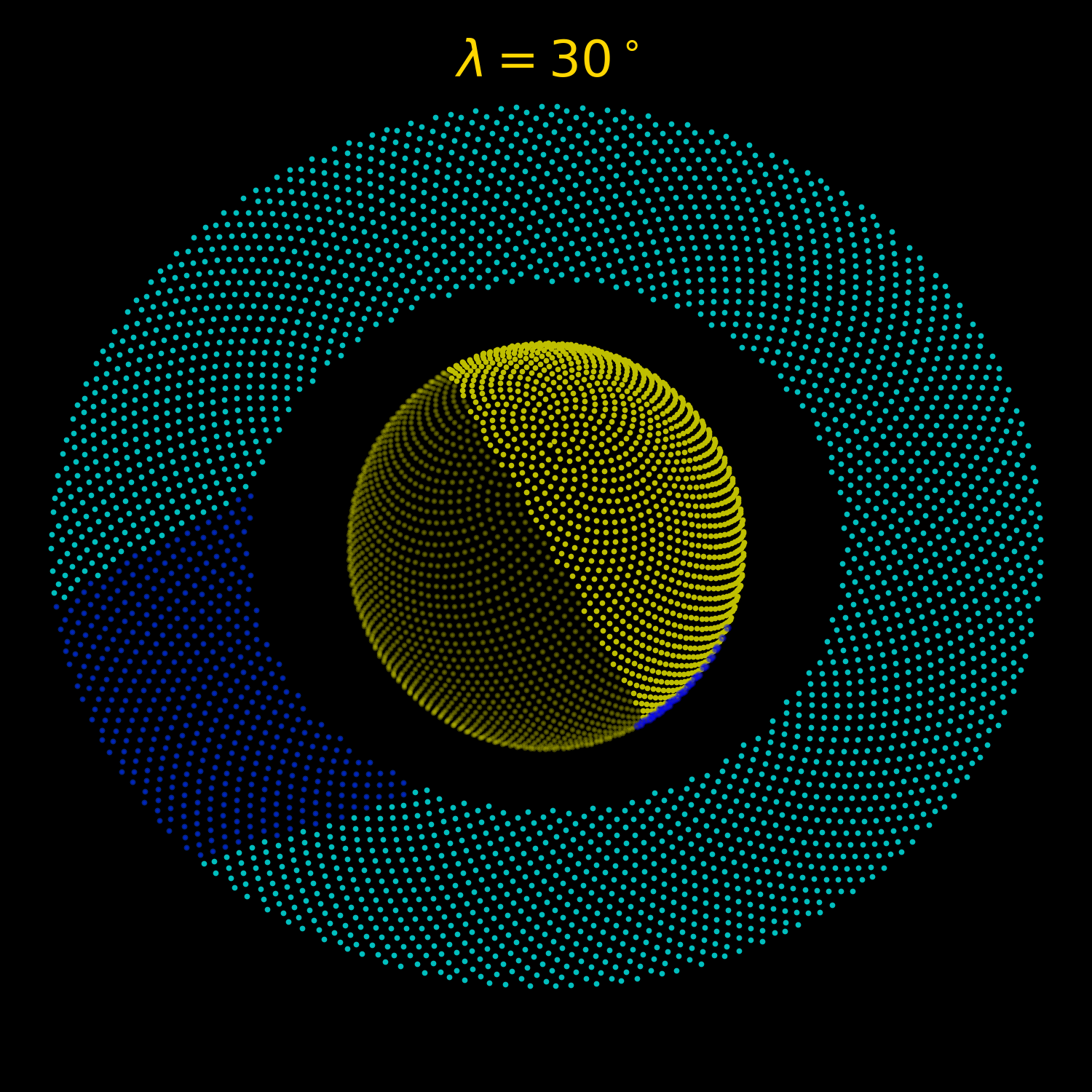}}%
\subfigure{%
\label{fig:lc_scatter_90}%
\includegraphics[width=\halffig\columnwidth]{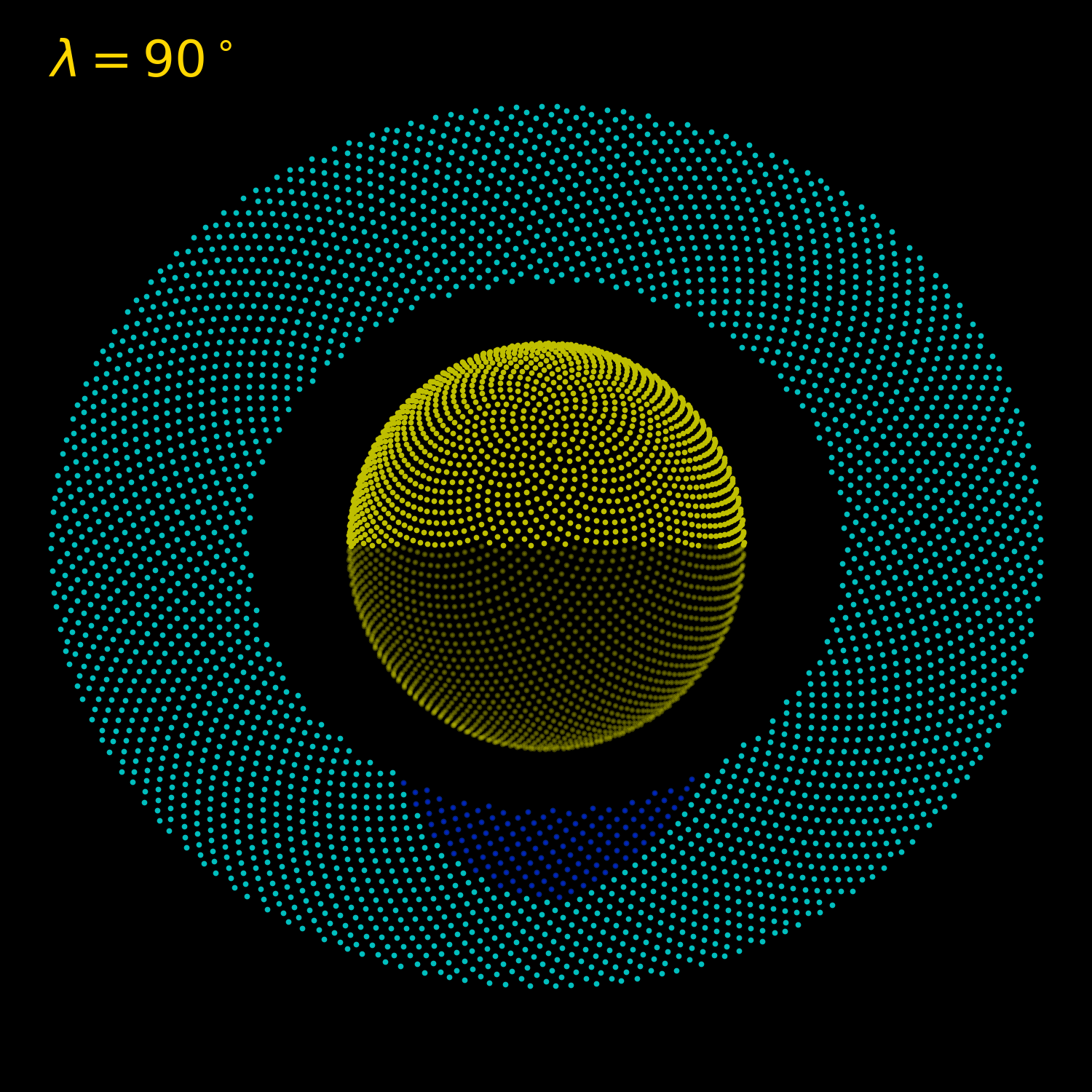}}%
\subfigure{%
\label{fig:lc_scatter_220}%
\includegraphics[width=\halffig\columnwidth]{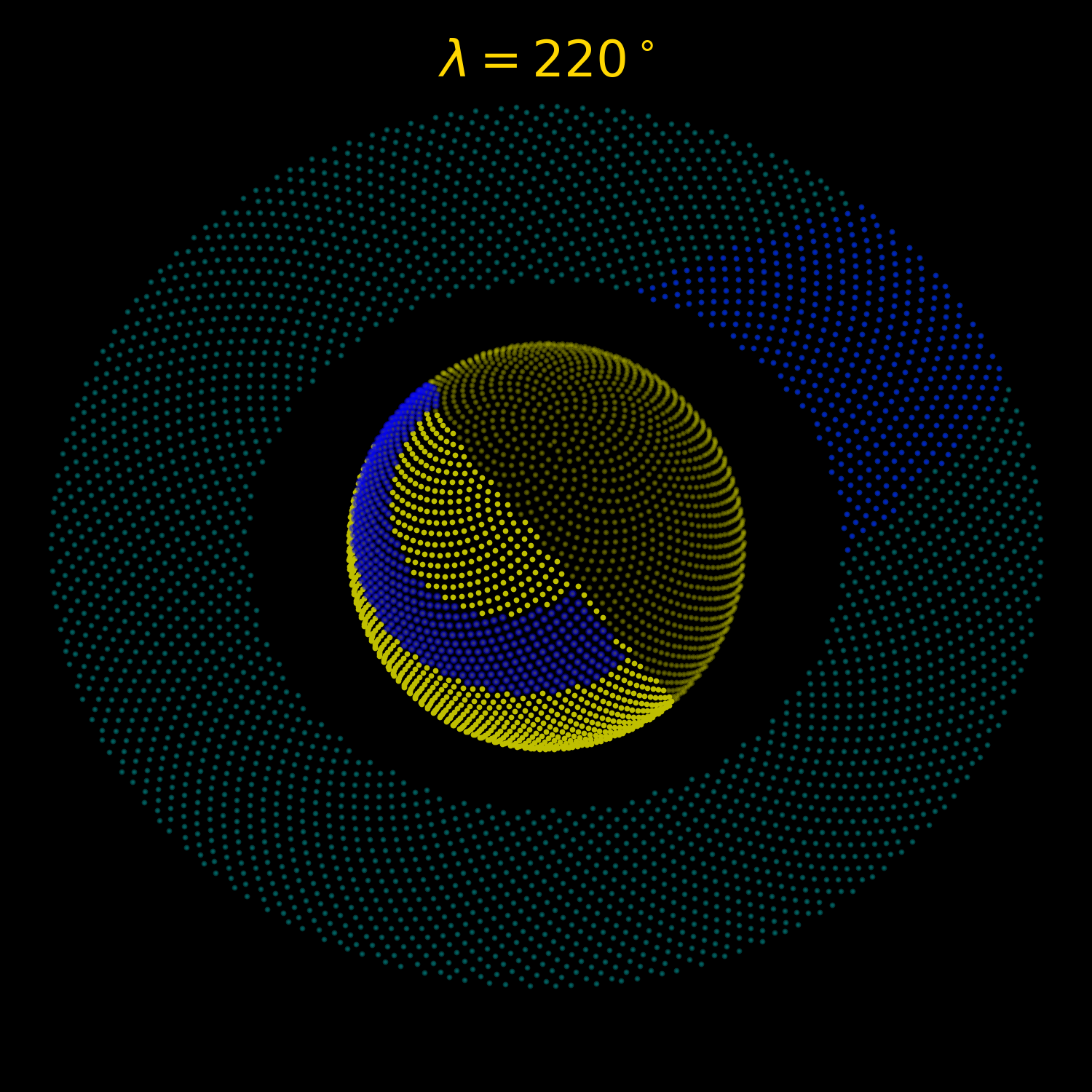}}%
\subfigure{%
\label{fig:lc_scatter_270}%
\includegraphics[width=\halffig\columnwidth]{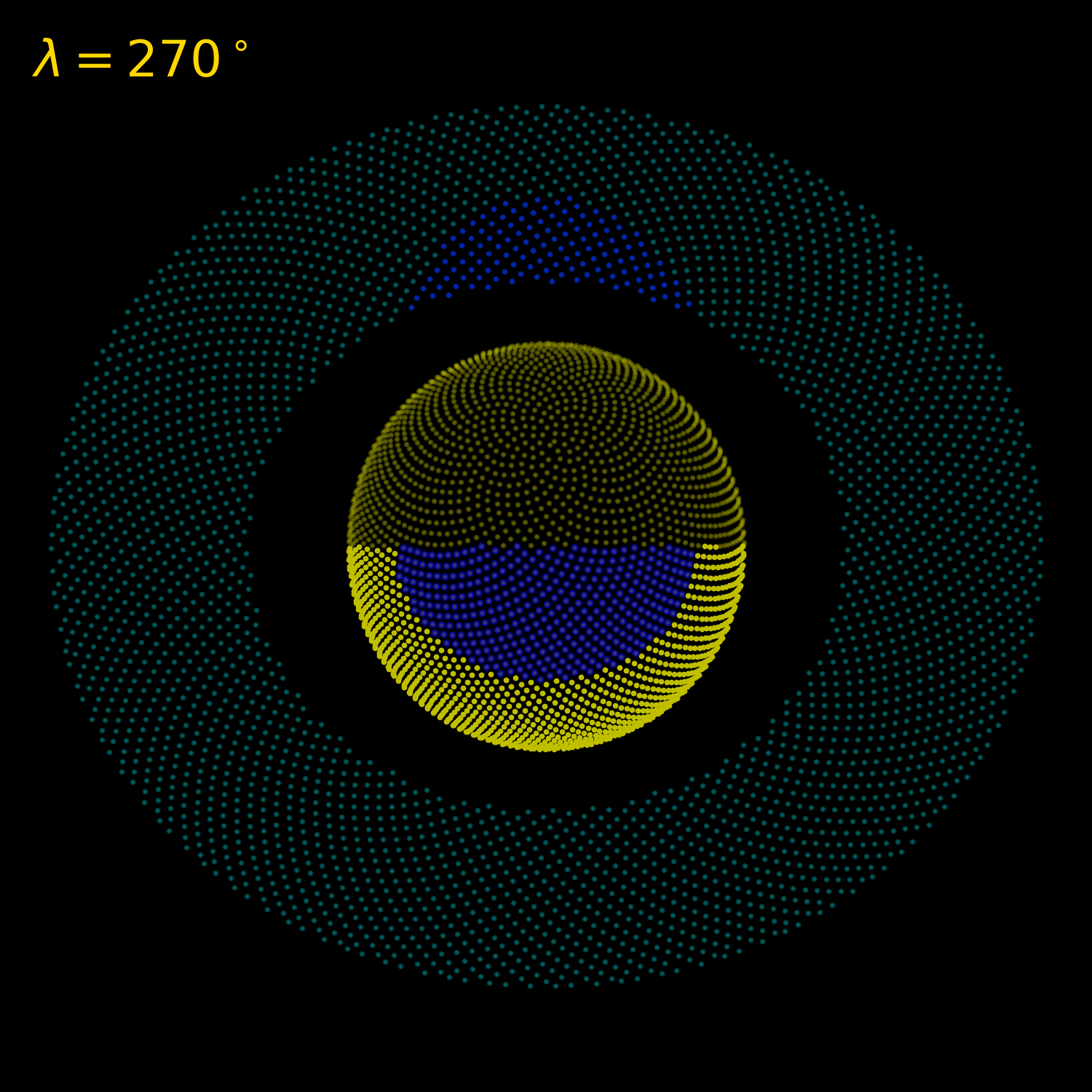}}%

\caption{The reflected light curve of a Saturn-like planet in a circular orbit around a Sun-like star. Dotted grey lines show especially selected phases of the light curve where shadows are enhanced. In the lower panel, we show the illumination conditions of the planet and ring spangles at those selected phases.
\label{fig:samplelc}
}
\end{figure*}

To illustrate the typical results that can be obtained with the model, we show in \autoref{fig:samplelc} the reflected light curve of a Saturn-like planet in a circular orbit around a Sun-like star. The planetary semi-major axis has been set to $a=0.1$ to enhance the effect produced by the rings. Whether these can survive high isolation conditions and the adverse dynamical environment around close-in planets is beyond the scope of this paper. However, numerical simulations have shown that warm planetary rings could survive thermal and gravitational stresses as long as they are composed of refractory material instead of ices \citep{Schlichting2011}.

In our basic example, rings have dimensions similar to those of Saturn and are inclined with respect to the orbit in an angle $i=30^\circ$.  The observer is looking from above (towards the $-z\sup{\ecl}$ or equivalently located at $\beta\sub{obs}=90^\circ$), and in the direction of the $+y\sup{\ecl}$ axis ($\lambda\sub{obs}=90^\circ$). 

In the upper inset panels of \autoref{fig:samplelc}, we show schematically the illumination and shadows of the planet and the rings at key positions of the orbit (dotted vertical lines). When $\lambda>180^\circ$ (two rightmost snapshots in the upper panel), the side of the rings visible from the vantage point of the observer does not receive light from the star (rings' winter) and they are barely visible. This characteristic is what \citet{Sucerquia2020a} proposed to exploit the most basic characterisation of non-transiting exorings.

The photometric signature produced by a ringed planet of this size and orbit is barely detectable with current photometric sensitivities ($\sim 10$ ppm). Still, flux anomalies as low as a few ppm have been measured in secondary transits (see e.g. \citealt{Eftekhar2022}) and, therefore, the detection of the characteristic `bump' of an extended ring around a close-in planet is somewhat feasible.

The effect of shadows on the light curve is non-negligible (see the lower panel of \autoref{fig:samplelc}). At the orbital/observational configuration of the example considered here, shadows are especially noticeable if the planet is the only body illuminated by the star ($\lambda>180^\circ$).

\begin{figure*}
\begin{center}
  \includegraphics[width=\overfig\columnwidth]{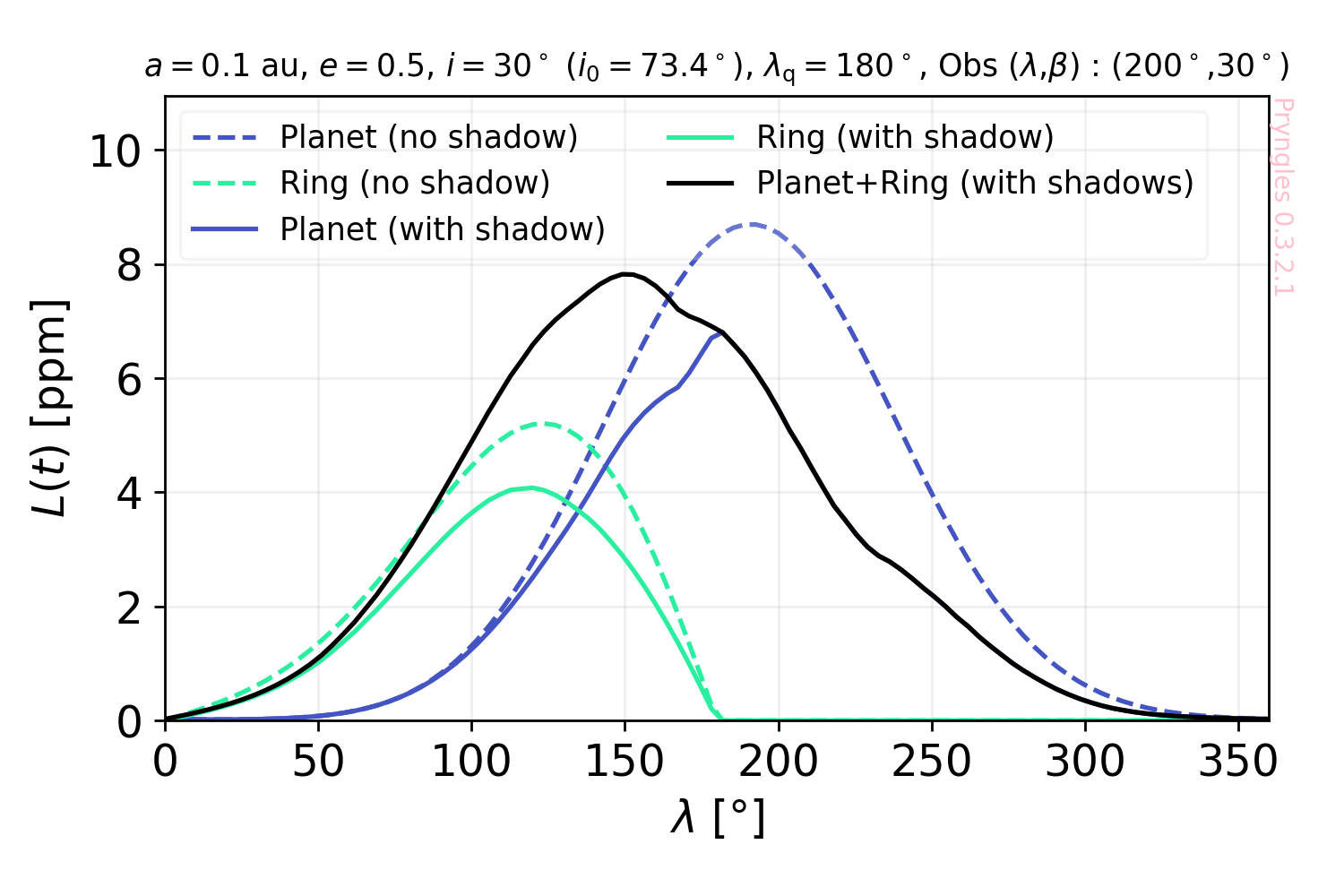}
    \caption{The reflected light curve of an eccentric Saturn-like planet around a Sun-like star. The planet is at the perihelion when $\lambda=180^\circ$ and the light coming from the star reaches a maximum. The position of the observer, $30^\circ$ above the ecliptic, allows them to perceive the light reflected by $>50\%$ of the planet's surface at certain phases of its orbit.}
    \label{fig:samplelc-eccentric}
\end{center}
\end{figure*}

More interesting results are obtained when planets are located on eccentric orbits and observed from oblique line-of-sights (the most probable configuration). In \autoref{fig:samplelc-eccentric}, we show the light curve of a planet with the same size and physical properties as before, but with $e=0.5$. Also, an increased stellar illumination at a critical transition phase of the light curve (e.g. right when the planet becomes the only visible object) has an effect on the system's photometric signature. To study such an effect, we have set the periapsis position at $\lambda_q=180^\circ$.

At this orbital inclination, the observer may perceive a larger portion of the planetary surface at certain phases of the orbit.  As a result, the maximum reflected light (after correcting for shadows) becomes comparable and even larger to that coming from the ring. This is particularly important at the periapsis when the illuminated face of the rings is invisible for the observer. In this case, the effect of shadows is more pronounced due mainly to the fact that the light reflected from the planet is more affected when the latter is closer to the star.

\begin{figure*}%
\centering
\subfigure[Eccentricity effect]{%
\label{fig:samplelc-parameters-eccentricity}%
\includegraphics[width=\widfig\columnwidth]{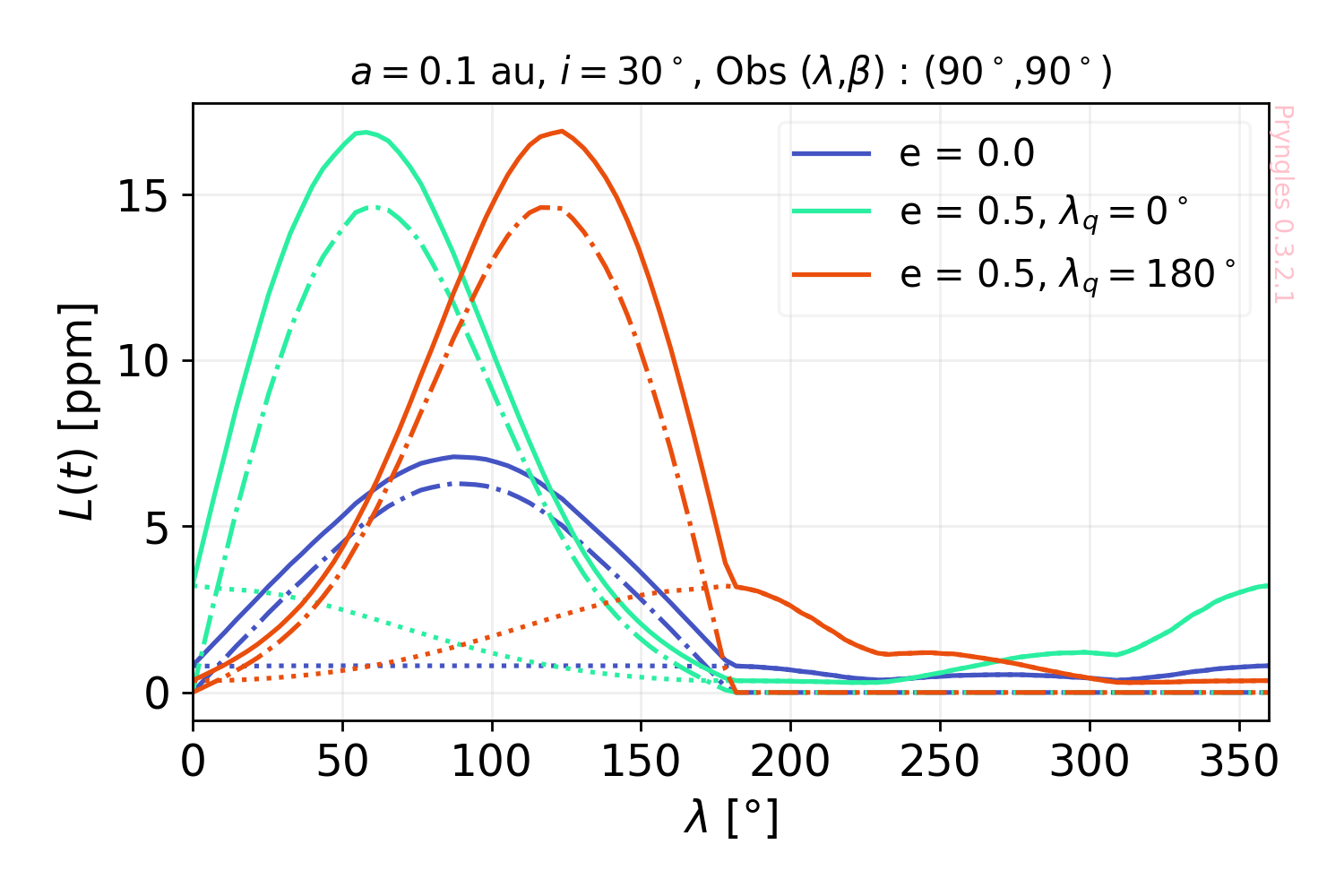}}%
\qquad
\subfigure[Inclination effect]{%
\label{fig:samplelc-parameters-inclination}%
\includegraphics[width=\widfig\columnwidth]{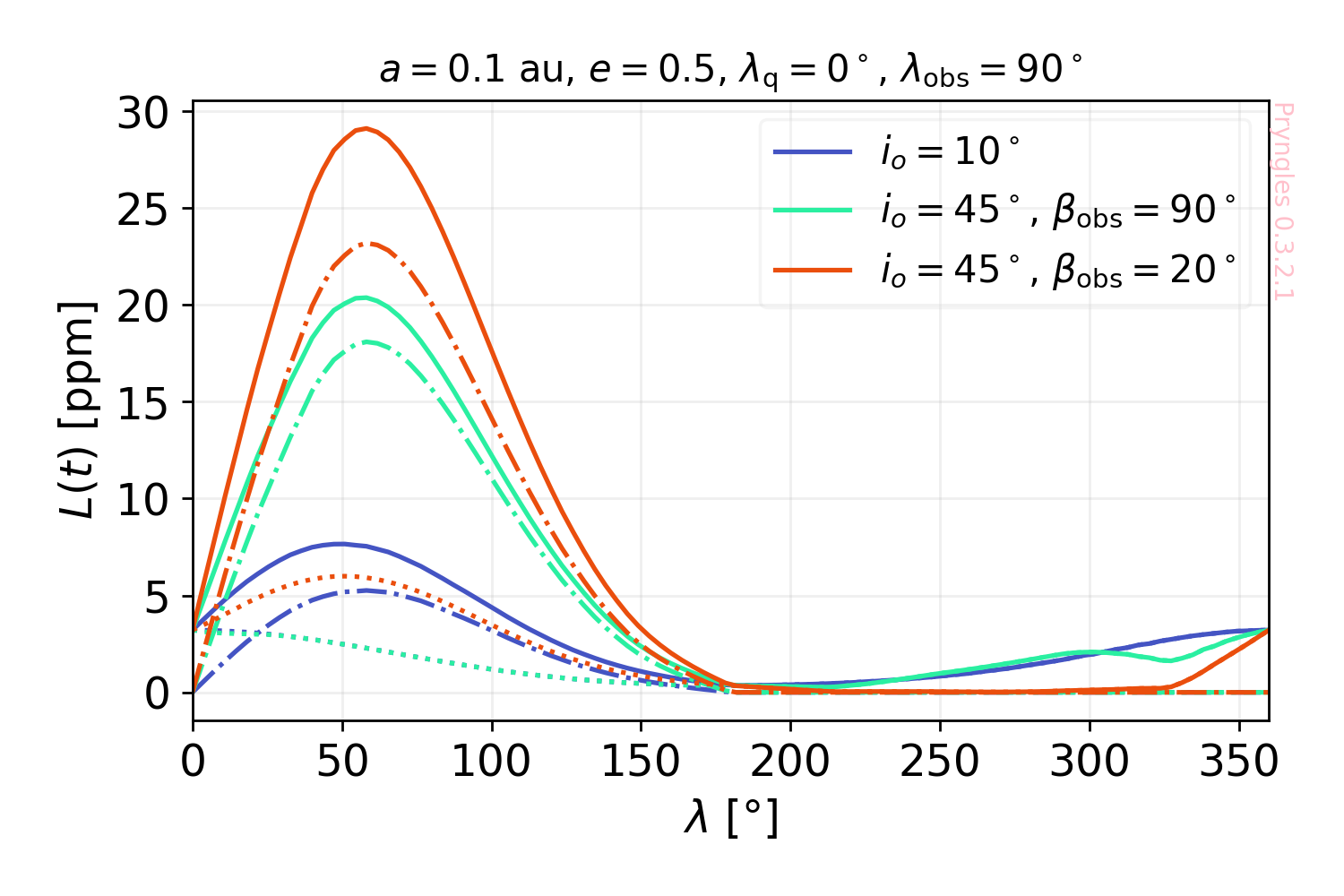}}%
\qquad

\subfigure[Observer effect]{%
\label{fig:samplelc-parameters-observer}%
\includegraphics[width=\widfig\columnwidth]{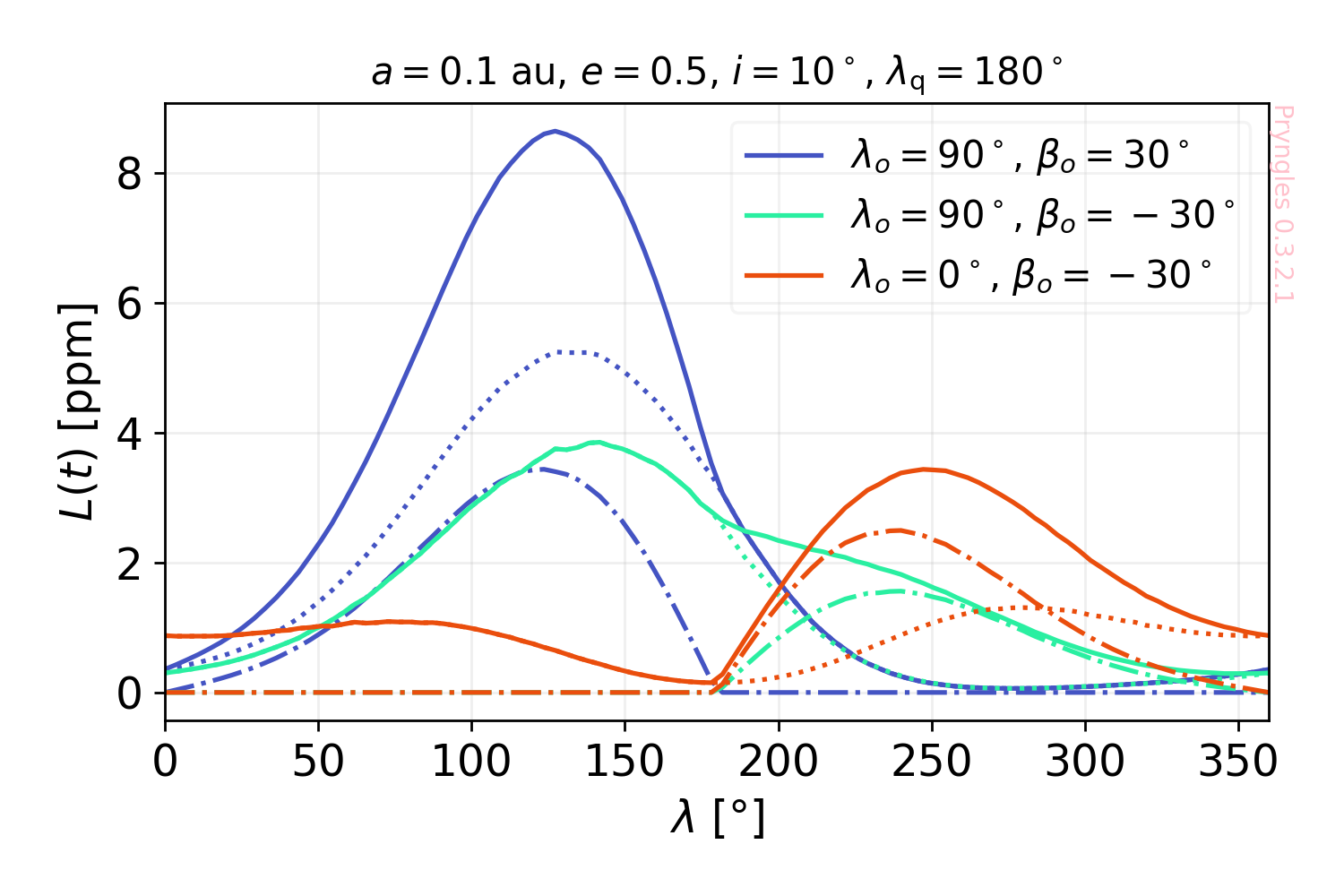}}%
\qquad
\subfigure[Distance effect]{%
\label{fig:samplelc-parameters-distance}%
\includegraphics[width=\widfig\columnwidth]{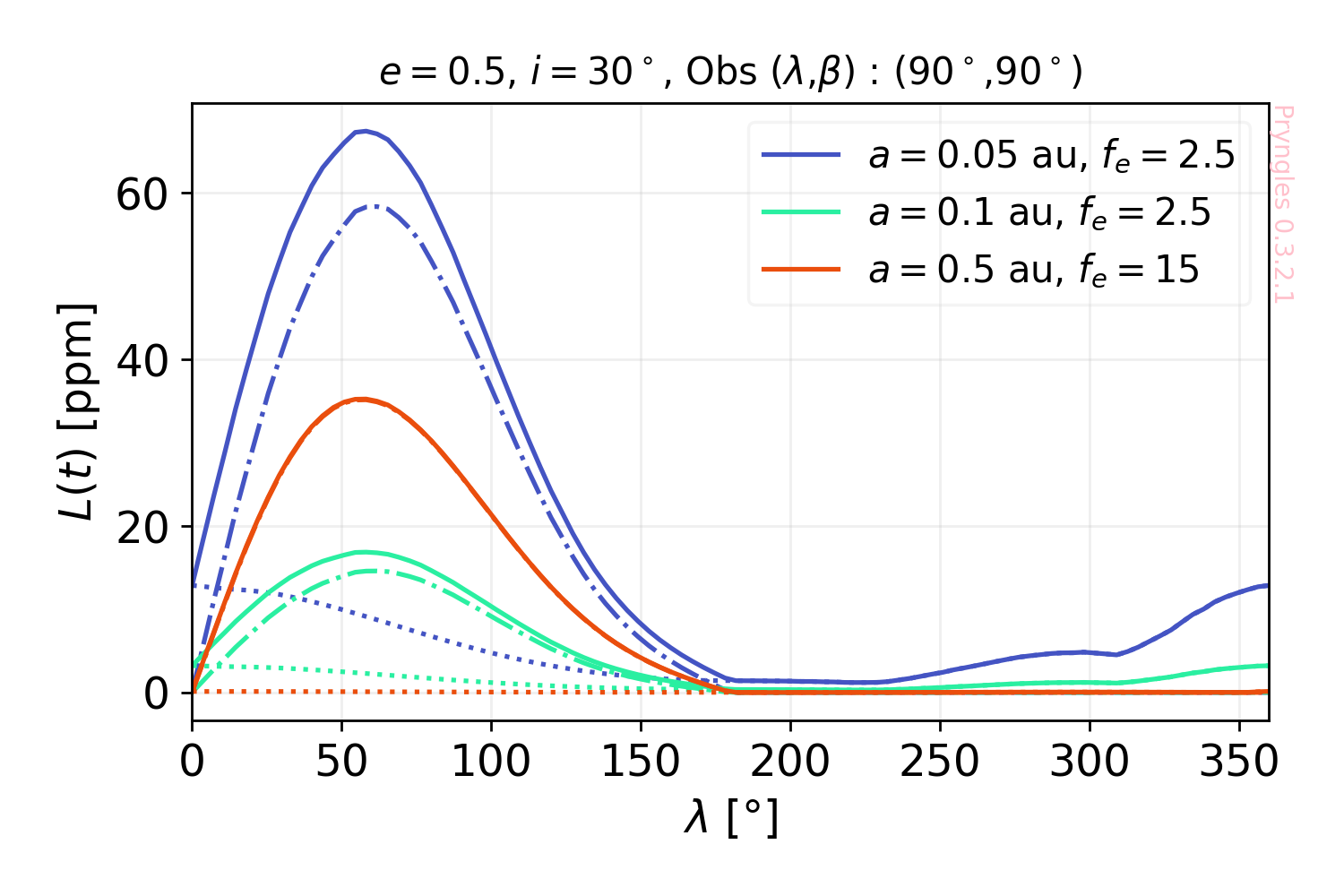}}%
\caption{Basic exploration of the parameter space of the reflected light of a ringed exoplanet.}
\label{fig:samplelc-parameters}
\end{figure*}

The shape of the light curve depends on many intertwined parameters: the orbital eccentricity, the argument of the periapsis, the albedo of the planet, the position of the observer; as well as the size, inclination, and albedo of the ring, etc. \comp{\model} is especially well-suited to study qualitatively this complex parameter space. In \autoref{fig:samplelc-parameters}, we explore the effects of changing some of the key parameters in the model, and analyse the resulting shape of the light curve.

e. To study such an effect, we have set the periapsis position at $\lambda_q=180^\circ$.

At this orbital inclination, the observer may perceive a larger portion of the planetary surface at certain phases of the orbit.  As a result, the maximum reflected light (after correcting for shadows) becomes comparable and even larger to that coming from the ring. This is particularly important at the periapsis when the illuminated face of the rings is invisible for the observer. In this case, the effect of shadows is more pronounced due mainly to the fact that the light reflected from the planet is more affected when the latter is closer to the star.

The shape of the light curve depends on many intertwined parameters: the orbital eccentricity, the argument of the periapsis, the albedo of the planet, the position of the observer; as well as the size, inclination, and albedo of the ring, etc. \comp{\model} is especially well-suited to study qualitatively this complex parameter space. In \autoref{fig:samplelc-parameters}, we explore the effects of changing some of the key parameters in the model, and analyse the resulting shape of the light curve.

Three parameters are key to determining the magnitude of the photometric signature of reflected light from a ringed planet: distance to the star ($a$), rings' size ($f_e$), and inclination ($i$). In Figure \autoref{fig:samplelc-parameters-distance}, we see how at very close distances the characteristic photometric bump might easily reach the 100-ppm level. However, the survival of normal exorings might be hampered at such close distances. At large distances, the reflected light is much smaller, but if the rings are large enough they may still be detectable.

The parameter having the largest effect on the shape of the light curve is the inclination of the planetary orbit with respect to the sky plane (see Figure \autoref{fig:samplelc-parameters-observer}). In our parametrization, this property is determined by the observer's ecliptic latitude $\beta_\mathrm{obs}$. The combination of observer's latitude and rings' inclination make the light curve to be dominated by the ring or by the planet.

\begin{figure*}
  \includegraphics[width=\textwidth]{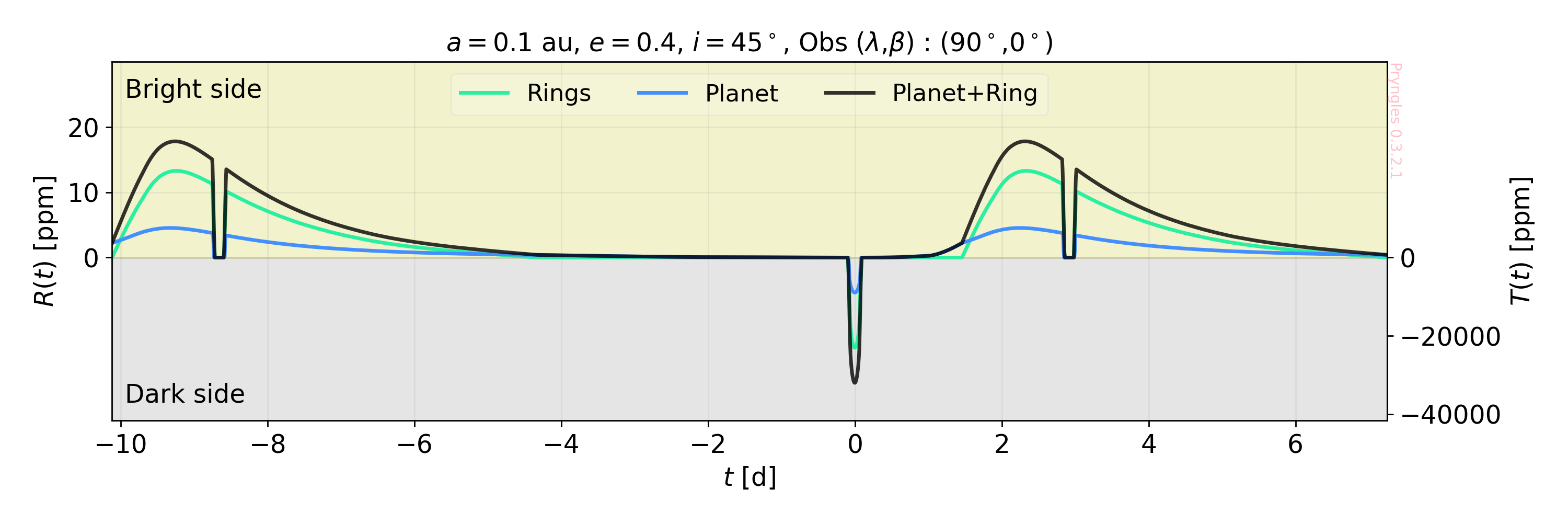}
    \caption{Complete light curve for one of the planets studied here. The vertical scale for the diffuse reflection component $R(t)$ (positive values of the light curve) are in the left vertical axis. The scale of the transit at the centre is on the scale to the right.}
    \label{fig:samplelc-transit}
\end{figure*}

The most valuable feature of \comp{\model} is the possibility of calculating the whole light curve, including diffusely reflected light, transits, and occultations. In \autoref{fig:samplelc-transit} we show the light curve of one of the planets studied before. In contrast to what we have done so far, the system is edge-on and thus primary and secondary transits are observed in the simulated light curve.

\begin{figure*}%
\centering
\subfigure[Eccentricity effect on primary transit]{%
\label{fig:lc_transit_1e}%
\includegraphics[width=\widfig\columnwidth]{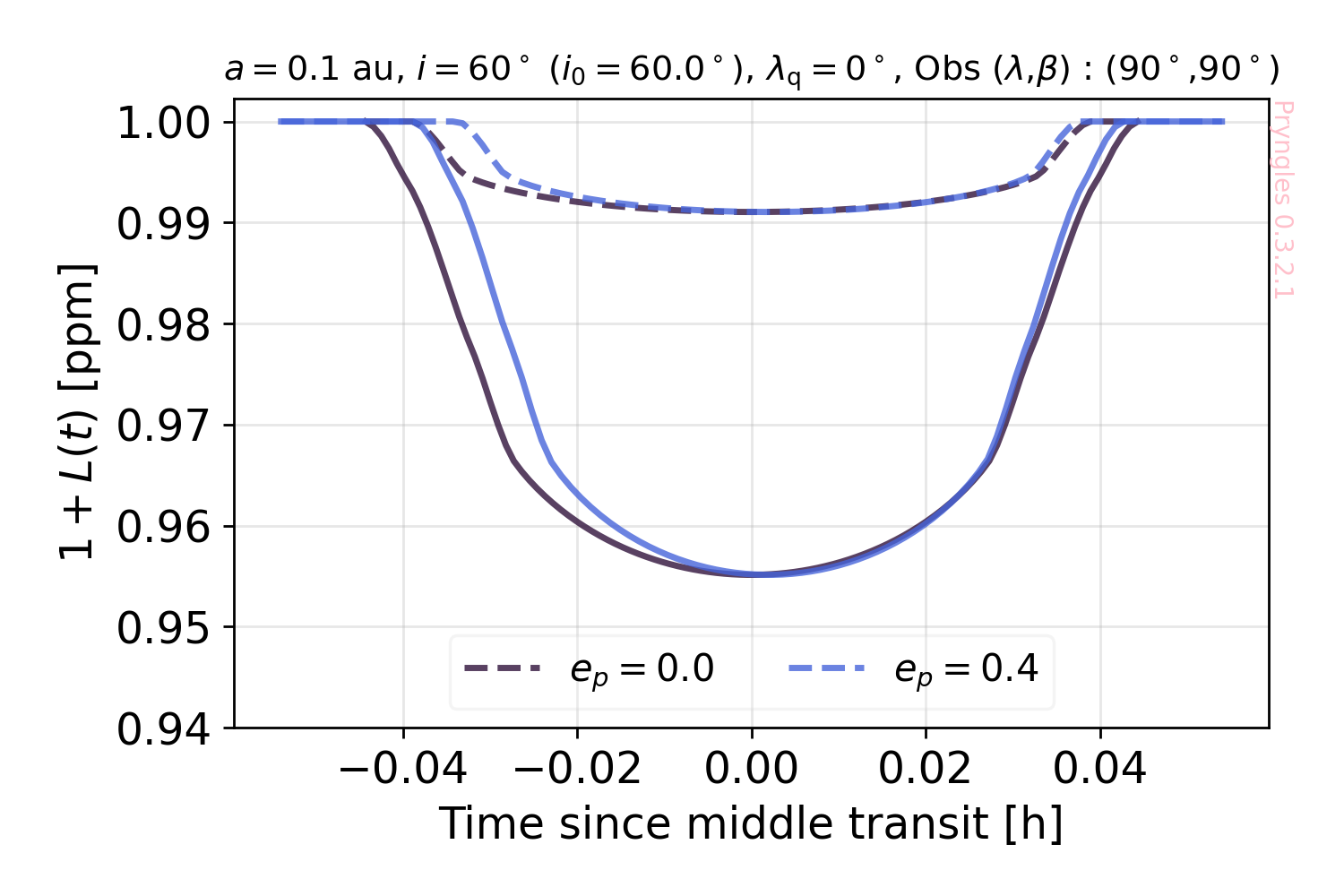}}%
\qquad
\subfigure[Inclination effect on primary transit]{%
\label{fig:lc_transit_1i}%
\includegraphics[width=\widfig\columnwidth]{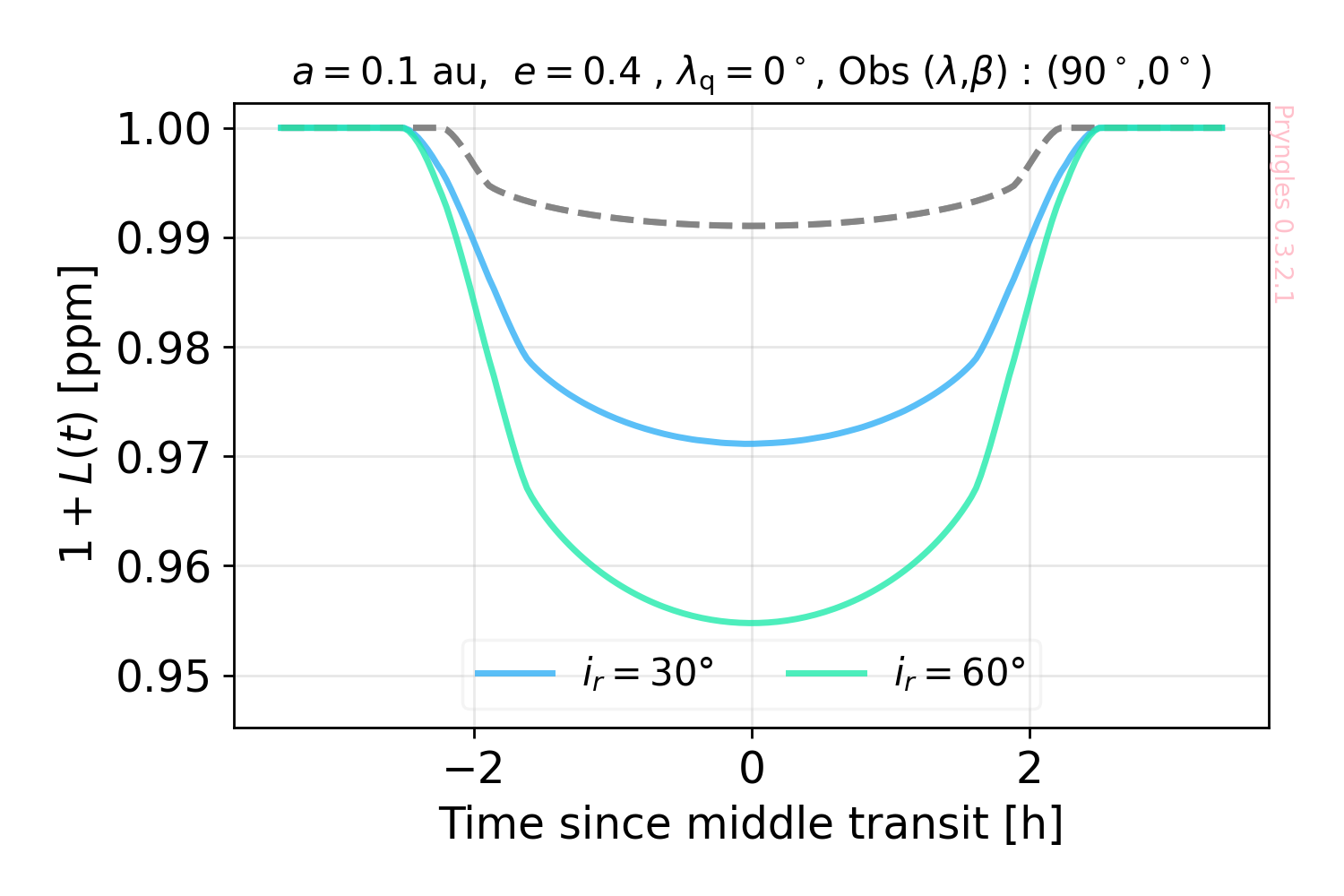}}
\qquad
\subfigure[Eccentricity effect on secondary transit]{%
\label{fig:lc_occultation_1e}%
\includegraphics[width=\widfig\columnwidth]{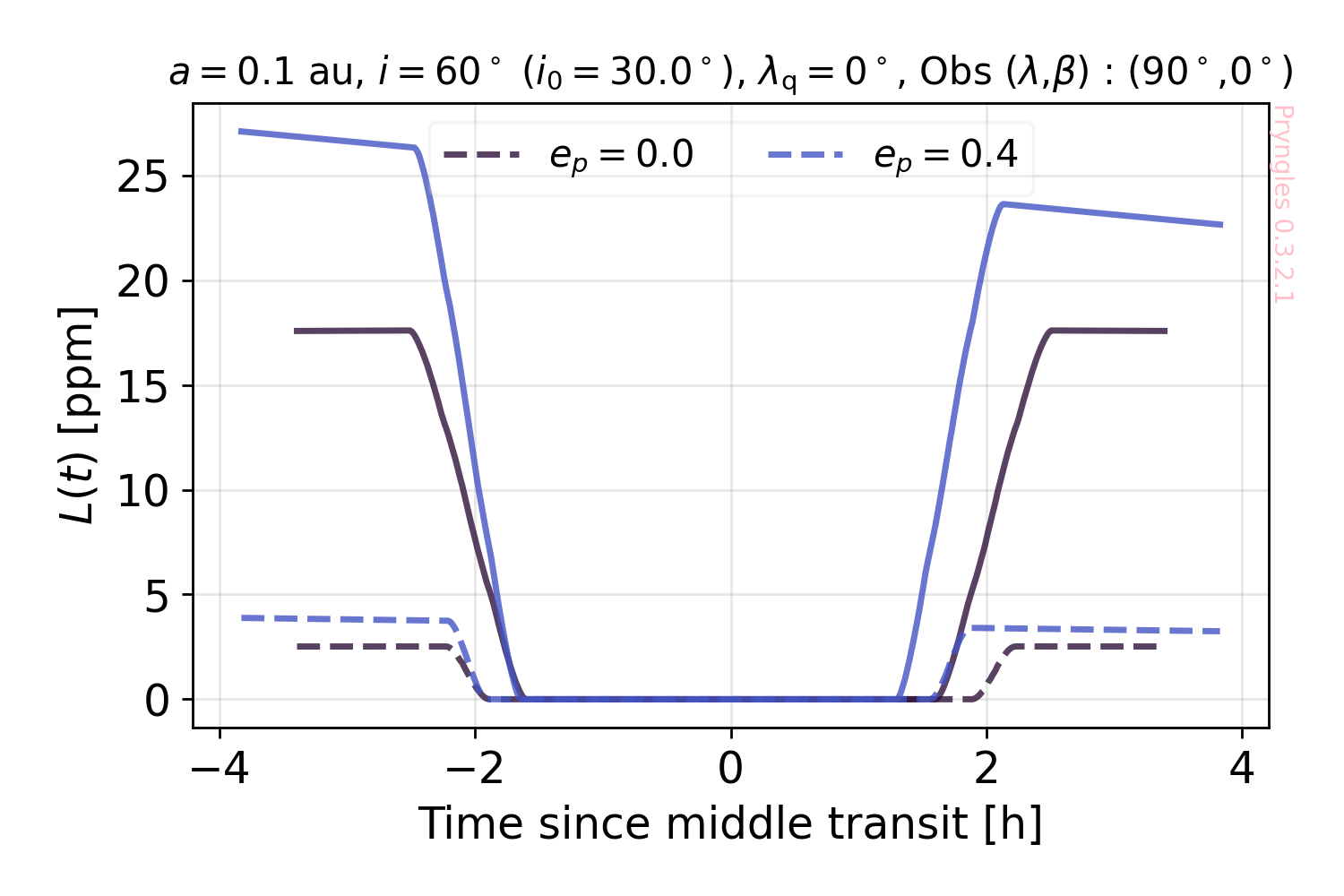}}%
\qquad
\subfigure[Inclination effect on secondary transit]{%
\label{fig:lc_occultation_1i}%
\includegraphics[width=\widfig\columnwidth]{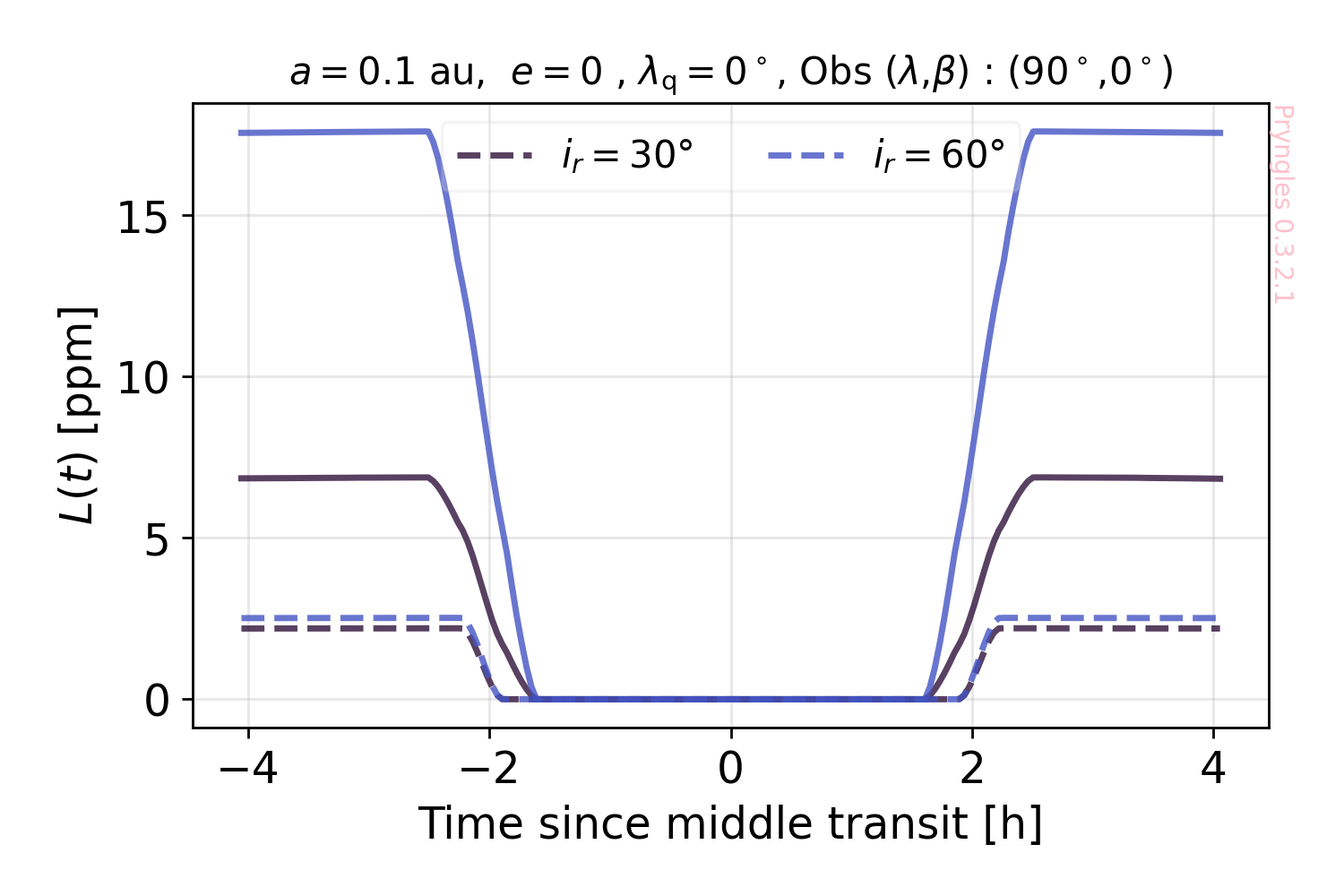}}%
\caption{The primary transit (upper row) and secondary transit (lower row) components of the light curve of a Saturn-like planet around a Sun-like star, computed with \comp{\model}. For consistency with literature, we have plotted in the two upper panels the normalised stellar flux $1+L(t)$ while for the secondary transit we use the convention used in this work, and put in the y-axis the light curve $L(t)$.}
\label{fig:samplelc-primary_secondary}
\end{figure*}

We have repeated the exercise we did in the case of the pure diffuse reflection component of the light curve, and studied the effect that different parameters of the system have on the primary and secondary transits. We show the result in \autoref{fig:samplelc-primary_secondary}.

\comp{\model} allows to recover the orbital effects on the light curve, such as the eccentricity and, of course, the inclination of the ring and the orbit as a function of the observer's position. On one hand, in Figure~\ref{fig:lc_transit_1e}, we show the transit profile of a planet (dashed lines) with rings (solid lines), for a circular (dark solid-lines) and eccentric orbit (blue solid-line). The asymmetries between the shapes at the ingress and egress stages of the light curve are a consequence of the eccentric shape of the orbit. This effect is known as the \textit{Photoeccentric effect} \citep{Burke2007, Kipping2008}, and it also affects the timing, duration, and occurrence of secondary transits \citep*{Kane2009,Dong2013}.

In summary, we have shown in this section the versatility of \comp{\model} to calculate not only the diffuse reflection light curve of a ringed exoplanet (the main purpose of its design), but also the primary and secondary transits. We have used the model to explore the qualitative properties of light curves and the dependence of their magnitude and shapes on some of the key parameters previously mentioned. In the following section, we will compare the predictions of our model with those obtained by simple analytical approximations and photometry packages extensively tested.

\section{Model comparison and validation}
\label{sec:validation}

The novel discretization of planetary surfaces using circular spangles, the physical approximations used to describe the diffuse reflection on planetary surfaces, and the transmission of light through the rings in \comp{\model} make us wonder if the results computed with the model are: 1) actually correct, and 2) have the precision required for exoplanetary research.

Fortunately, \comp{\model} arises at a time when several publicly available packages have been designed and tested to study, at least, the dark side of light curves. The case of the bright side is more complicated. To date, no single package has been developed to compute systematically the diffuse reflection from exoplanetary surfaces (including rings), or not at least with the general aim of \comp{\model}. Still, a significant amount of literature has been written about the physics of how light is reflected from planetary bodies, and many analytical and semi-analytical approximations have been developed. To start with, we perform a comparison of the predictions of \comp{\model} with those analytical models.

Although these validations are simply not enough as the ultimate test of a real package is its use for studying real observations (not yet available for most configurations), the tests described here may increase the confidence in the package and the mathematical and physical approximations on which it relies.

\subsection{Validation against other transit packages}
\label{subsec:validation_transits}

Transit photometry is probably one the most developed observational techniques in exoplanetary research. Several well-known packages have been developed in recent years (see e.g. \citealt{Kreidberg2015, LightkurveCollaboration2018, Akinsanmi2018, Rein2019}), and most of them implement the experience accumulated in 20 years of transit observations and research.  Those packages focus on two special features: precision and performance. From a synthetic light curve generator, we expect it to compute its products as efficiently as possible so we can use them, for instance, to fit a light curve. Although \comp{\model} still needs more work to achieve such computational efficiency, the least we expect is that the package will predict the same results of more mature tools.

\begin{figure*}%
\centering
\subfigure[Planet transit]{%
\label{fig:lc_occultation_i90}%
\includegraphics[width=\widfig\columnwidth]{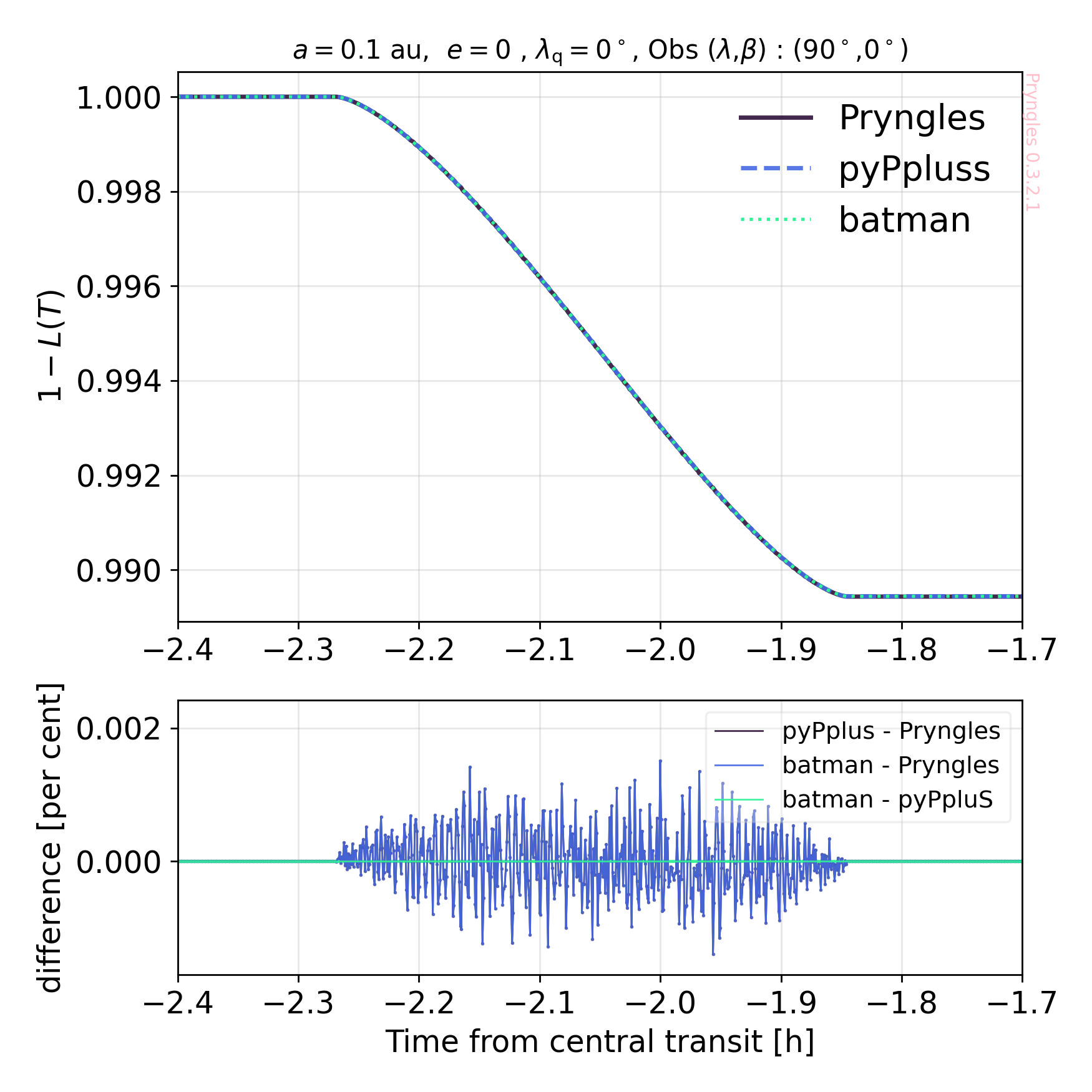}}%
\qquad
\subfigure[Ringed planet transit]{%
\label{fig:lc_occultation_i45}%
\includegraphics[width=\widfig\columnwidth]{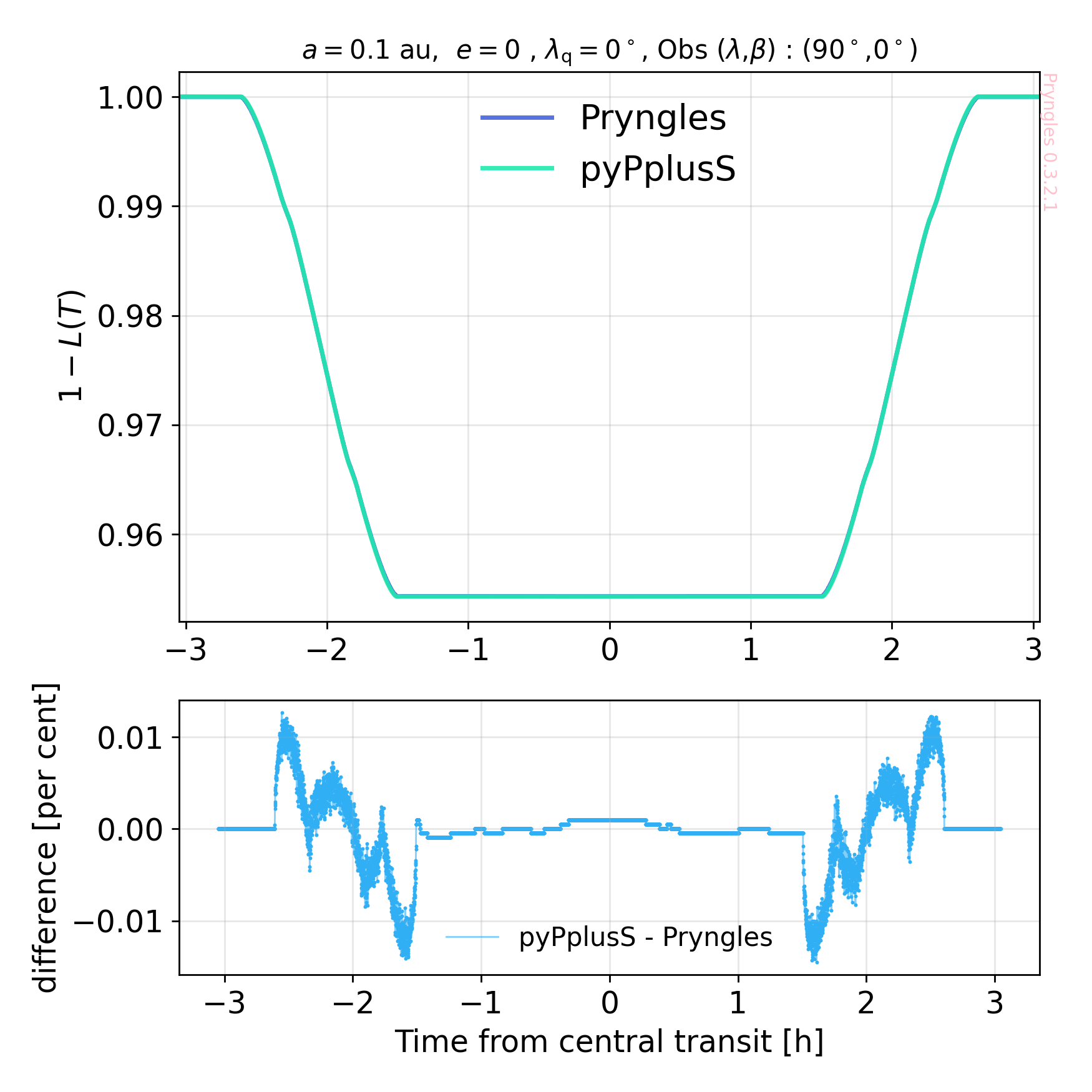}}%
\caption{light curve comparison between \comp{\model} and two well-known packages: \comp{batman} \citep{Kreidberg2015} and \comp{pyPplusS} \citep{Rein2019}}.
\label{fig:validation}
\end{figure*}

In \autoref{fig:validation}, we show the results of comparing the predictions of \comp{\model} with the well-known and widely used package \comp{batman} \citep{Kreidberg2015}. This package still does not calculate the transit signal of a ringed planet. For that reason, we have compared the transit light curve predicted by \comp{\model} for a spherical planet with the predictions of \comp{batman}.

Recently, \citet{Rein2019} released a new transit package able to efficiently compute the transit light curve of ringed and non-ringed exoplanets. We have also included comparisons with this package in \autoref{fig:validation}. For this case, we compare the transit light curve calculated by \comp{\model} of the ringed planet we studied in \autoref{sec:numerical} to that calculated by \comp{pyPplusS}.

In the case of the transit of a non-ringed planet, the predictions of \comp{batman} and \comp{pyPplusS} essentially coincide. When compared to \comp{\model}, the difference between the predictions are still below 0.001\%. This implies that the discretization of the planetary surface via spangles (which mostly explains the noisy differences observed with the other packages) is good enough to describe the transit of the planet in front of the star.

The case of a ringed planet deserves more attention. Although the relative difference between \comp{pyPplusS} and \comp{\model} is still very small, $\sim 10^{-4}$, the discrepancy is systematically larger at transit ingress and egress. The oscillating structure of the difference between the predictions, suggest that it arises from the fact that, in contrast to \comp{\model}, \comp{pyPplusS} actually models the effect of forward-scattering on ring particles. However, this is not a true defect of our model, but simply a feature that needs to be implemented in future version of the package. 

\subsection{Analytical models of diffusely reflected light}
\label{subsec:validation_analytical}

Even more importantly than validating the transit predictions of \comp{\model}, is to check that the predicted amount of light diffusely reflected by the planet and the ring (under different geometrical circumstances) is correct.  It is out of the scope of this paper to validate the model for a wide range of different assumptions concerned to the complex physics of atmospheric, surface and ring scattering.  Still, we can test it in simplified and well-known cases.

Let us assume, for instance, that both the surface of the planet and the rings reflect light like Lambertian surfaces (see \autoref{sec:optics}).
We can consider two geometrical configurations where relatively simple analytical and semi-analytical calculations can be used to predict the diffusely reflected light. 

In the first one, the ringed planet is observed from above (face-on observations). This is precisely the case that was studied in \citet{Sucerquia2020a}. Using the convention of our model, the observer is located at $\beta\sub{obs}=90^\circ$. Under this configuration, the observer always sees one fourth of the planet illuminated by the star. The amount of light $\mathrm{d}B\sup{p}$ diffusely reflected by a differential surface area element located at planetocentric spherical coordinates $\phi,\theta$ (longitude and co-latitude with respect to the ecliptic plane) and with an ideal spherical albedo $A\sup{sph}=1$ (see \autoref{eq:spherical_albedo}) will be

\beq{eq:differential_area}
\begin{aligned}
dB\sup{p}&= B\sup{\star} (\hat n\sub{r}\cdot\hat n\sub{s})(\hat n\sub{r}\cdot\hat n\sub{o}) R_p^2 d\Omega\\
&=B\sup{\star}R_p^2(\sin\theta\cos\phi)(\cos \theta)\sin\theta\;\mathrm{d}\theta\mathrm{d}\phi
\end{aligned}
\eeq
where $B\sup{*}=L\sup{*}/(4\pi r^2)$ is the total flux of light coming from the star at the instantaneous distance $r$, and $\hat{n_r}$, $\hat{n_o}=\hat{e_z}$ and $\hat{n_s}=\hat{e_x}$ are the unitary vectors normal to the surface element, pointing to the observer and to the star respectively (see \autoref{fig:reference_frames}).  Integrating over the illuminated surface, $\theta\in[0,\pi/2], \phi\in[-\pi/2,\pi/2]$, the total diffused reflected light by the planet under this configuration will be:

\beq{eq:lambertian_planet}
\frac{B\sup{p}}{B\sup{\star}}=\frac{2}{3} R_p^2.
\eeq

This result is in contrast with what is naively expected (and nor rarely assumed) $B\sup{p}/B\sup{\star}=\pi R_p^2/2$.

On the other hand, the amount of light diffusely reflected by a Lambertian ring having inner ($R_i$) and outer ($R_e$) radii, will simply be \citep{Sucerquia2020a}:

\beq{eq:lambertian_ring}
\frac{B\sup{r}}{B\sup{\star}}=A_L(\hat n\sub{r}\cdot\hat n\sub{s})(\hat n\sub{r}\cdot\hat n\sub{o})\pi(R_e^2-R_i^2),
\eeq
where $A_L=\ALo$ (see \autoref{eq:lambertian_albedo}) is the Lambertian albedo of the ring.

\begin{figure*}%
\centering
\subfigure[Face-on configuration]{%
\label{fig:validation_analyic_faceon}%
\includegraphics[width=\widfig\columnwidth]{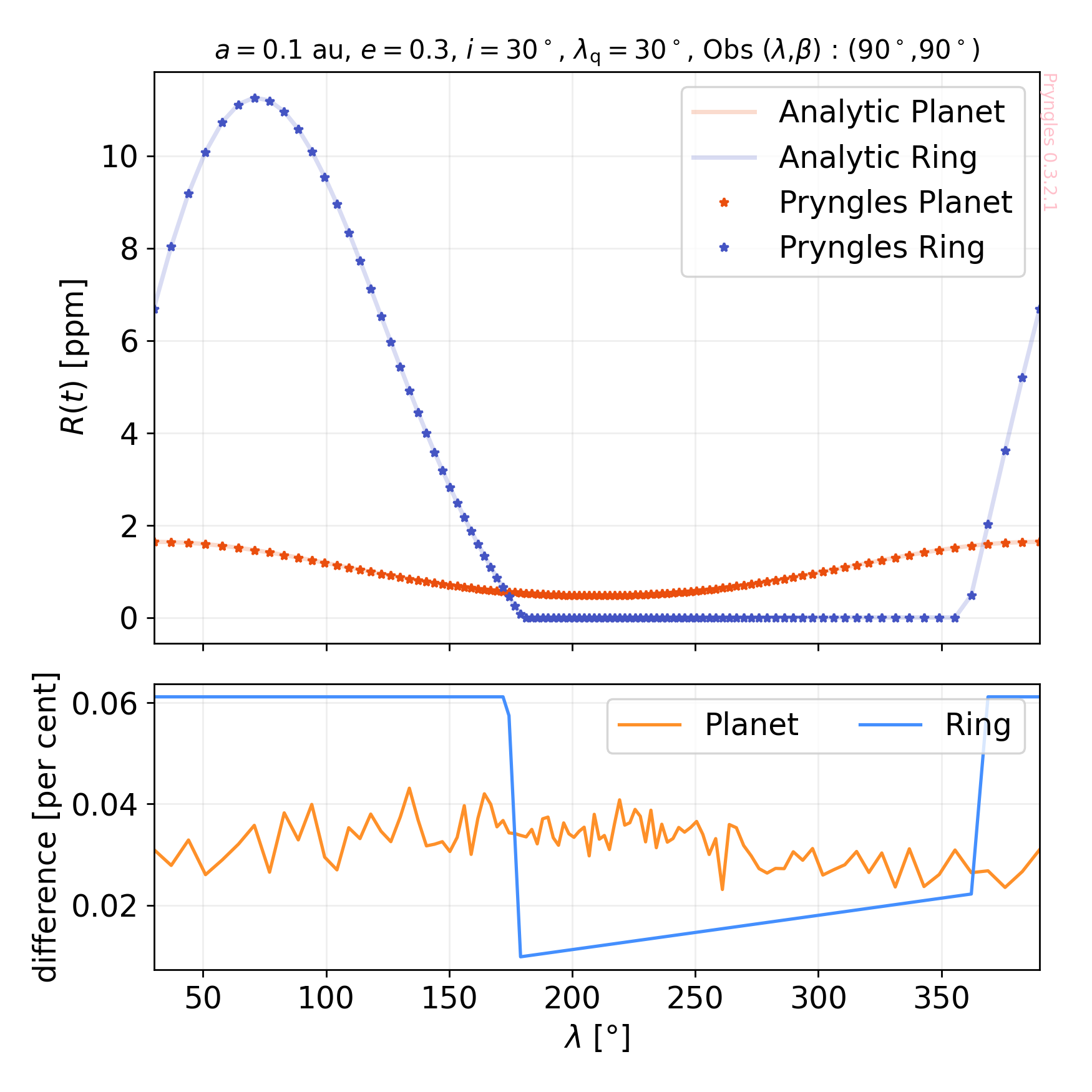}}%
\qquad
\subfigure[Edge-on configuration]{%
\label{fig:validation_analyic_edgeon}%
\includegraphics[width=\widfig\columnwidth]{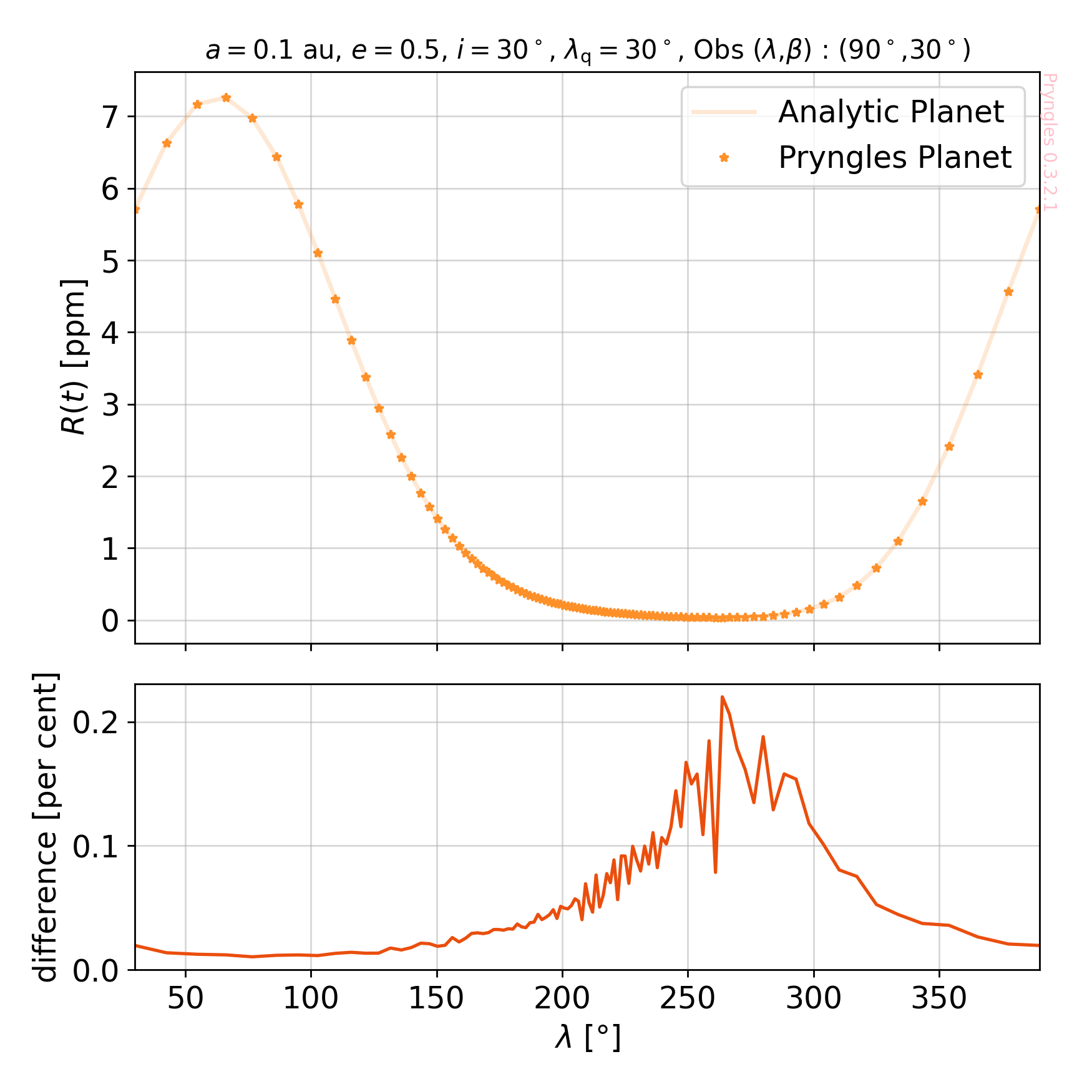}}%
\qquad
\caption{Reflected light curve calculated by \comp{\model} for two simple geometrical configurations: a face-on (a) and edge-on (b) planetary system. The predictions are compared against the corresponding analytical and semi-analytical predictions (see text).}
\label{fig:validation_analytic}
\end{figure*}

In \autoref{fig:validation_analytic} we compare the predictions of \comp{\model} (without the effect of shadows) and that obtained using analytical formulae (\autoref{eq:lambertian_planet} and \autoref{eq:lambertian_ring}), for different orbital and ring configurations. As we expect, the predicted and analytical values of the diffusely reflected light for both the planet and the ring fairly coincide. The noisy nature of the errors, at least in the case of the planet, is a product of the discretization of the surface in spangles.

The last but not less important validation test of the model is to compute the diffusely reflected light from the planet under different {\it phase angles} $\alpha$, defined as the angle between the direction of the primary light source (the star) and the direction of the observer.

In the previous test, the planet was illuminated under a constant phase angle $\alpha=90^\circ$. Using \comp{\model}, we now simulate a planetary system from an almost oblique face-on view (i.e. $\beta\sub{obs}<90^\circ$).

For a perfect Lambertian planetary surface (spherical albedo equal to 1) illuminated at phase angle $\alpha$, the diffusely reflected light will be given by \citep*{Seager2000}:

\beq{eq:flux_phase_angle}
\frac{B\sup{p}(\alpha)}{B\sup{\star}}=\frac{\pi\phi(\alpha)}{q}R_p^2,
\eeq
where $\phi(\alpha)$ is the so-called phase function and 

\beq{eq:integral_q}
q=\int_0^\pi \phi(\alpha)\sin\alpha\;\mathrm{d}\alpha.
\eeq

For the Lambertian-law of reflection the phase function is \citep{Russell1916}:

\beq{eq:phase_function_lambertian}
\phi(\alpha)=\frac{1}{\pi}[\sin \alpha+(\pi-\alpha) \cos \alpha],
\eeq
and $q=3/2$. In the case studied before, $\alpha=\pi/2$ and $B\sup{p}(\pi/2)/B\sup{\star}=2R_p^2/3$ as independently obtained from \autoref{eq:lambertian_planet}.

In \autoref{fig:validation_analytic}, we show the comparison of the light curve for a single planet in an almost edge-on configuration, calculated using \autoref{eq:flux_phase_angle} and \comp{\model}. As we expect both results fairly coincide. The largest non-significant differences arise at phase angles close to $180^\circ$. At these angles, the errors due to the discrete nature of \comp{\model} are more noticeable (a few spangles produce the light reflected by the planet when $\phi\approx 180^\circ$ and the statistical error is larger).

In summary, the validation tests presented in this section show, that the model implemented by \comp{\model} is able to reproduce the characteristics, both in reflected (the bright-side) as in transit and occultations (the dark-side) of the light curve of ringed and non-ringed exoplanets.  Despite the discrete approximation in spangles and the novel geometrical description implemented in \comp{\model}, the model is accurate to $\lesssim 0.01\%$ levels as compared with widely-used and tested models and analytical formulae, at least for the resolution and for the cases discussed in this section.

\section{Discussion and prospects}
\label{sec:discussion}

Modelling reflected and transit light curves using a discretization of planetary and ring surfaces is certainly not new (for recent examples see e.g. \citealt{Dyudina2016,Sandford2019}). However, using a general and modular computational architecture to calculate not only reflected-light (with the potential of easily including complex physical effects like wavelength dependent scattering, polarization, a non uniform surface, etc.), but also transits and occultations, as well as the capacity of capturing all the subtle effects of rings, gaps, shadows, planet- and ring-shine is unprecedented. The goal for the future is that most effects in this area can be modelled with \comp{\model}. As the validation tests presented in \autoref{fig:validation} show, there are phenomena which are fairly well described even with analytical formulae.  Moreover, transits can be efficiently and reliably modelled with modern widely-tested specialised packages. Still, for a qualitative understanding of the complex light curves of ringed exoplanets and/or the quantitative assessment of the reflected light curve of those complex systems and all their subtleties, \comp{\model} (and the general model that it implements) is very well-suited.

Although some of the most subtle effects modelled by \comp{\model} are far below the photometric sensitivity of ground- and space-based observations, it is not unreasonable to expect that in the future ppm and even sub-ppm sensitivities can be achieved. We just must recall that only 50 years ago we did not have the photometric or spectral sensitivity to detect planets. If we permanently limit the scope of our models to current instrumental capabilities, we will not be prepared to perform new discoveries or to extract most information from them.

The model implemented in \comp{\model} is general enough to include many other objects and effects than those discussed in this first paper.  We may add multiple ringed- and non-ringed planets to a single system and simply add-up their contribution to the light curve. Exomoons could also be included \citep{Moskovitz2009} with the corresponding algorithms to calculate their shadows on the host planet and the ring, and their mutual shine.  Even ringed moons or cronomoons \citep{Sucerquia2022} could be simulated using the spangles approximation. More interestingly, and probably more general in scope, is the possibility to include circumsecondary discs and sub-stellar companions \citep{vanDam2020,Matthews2021}. However, to study those scenarios a major redesign of the package must be tackled.

The kind of surface discretization used in the current version of the model, namely using small circular area elements or spangles, was chosen due to their geometrical simplicity.  However, more complex discretizations using, for instance, triangles, hexagons or other tessellation patterns is also possible without compromising significantly the core of the model\footnote{In those cases the name of the package is still the same. For a different tessellation pattern, instead of calling it {\bf P}laneta{\bf ry} spa{\bf ngles} we can call it {\bf P}laneta{\bf ry} tria{\bf ngles}, {\bf P}laneta{\bf ry} hexa{\bf ngles}, etc.}

In the upcoming years of JWST the interest in planetary infrared emission will be certainly increased. As expected, including emission effects into \comp{\model} is rather `simple'. Besides the potential complexity of the corresponding physics, a simple way to do it is to assign to each spangle in the simulation a black body temperature depending on the amount of light received from primary sources and the time spent inside the shadow. With a proper chosen wavelength-dependent emissivity, the light curve of actual planets such as HAT-P-1b \citep{Wakeford2013}, HD 209458b \citep{Zellem2014}, HAT-P-7b \citep{Armstrong2016}, HD 80606b, and other similar planets could also be modelled with \comp{\model}.

Adding colour and albedo variations across the surface of planets (a surface map) is also straightforward using \comp{\model}. In future versions of the package, we will develop a map-to-albedo transformation that will allow us to create planetary light curves with arbitrary complex surfaces. This is another example of our model's versatility.

Improving the performance of \comp{\model} is a priority, so the package can be used efficiently to fit observed light curves. In its current version, the package is entirely written in \comp{Python} while other similar packages are combinations of \comp{Python}, \comp{Cython}, and \comp{C} code. However, none of them were originally as fast as packages such as \comp{batman} or \comp{PyPpluS}. Only with time, efficient solutions arise. We hope this will also be the case with \comp{\model}.


\section*{Acknowledgements}
JIZ is funded by Vicerrector\'ia de Docencia, UdeA. MS acknowledges support by Agencia Nacional de Investigaci\'on y Desarrollo (ANID) through FONDECYT postdoctoral 3210605, and from ANID – Millennium Science Initiative Program – NCN19\_171. JAAM acknowledges funding support by Macquarie University through the International Macquarie University Research Excellence Scholarship (`iMQRES'). JIZ thanks the {\it Núcleo Milenio de Formación Planetaria} (NPF) and the {\it Instituto de Física y Astronomía} (IFA) of the University of Valparaiso for an invitation to stay as a visitor scholar at the IFA/UV where some parts of this work were developed.

\appendix

\section{Calculation of rings and planetary shadow}
\label{app:shadow}

In order to compute the contribution of shadows to our model, we define the spangle state variable {\bf s} (see \autoref{subsec:spstate}).  This variable is true if the spangle is inside a shadow (planetary of ring shadow).

If we call $(\alpha_s,\delta_s)$ the object-centric equatorial coordinates of the primary light source, the conditions to be within a shadow are given by the formulae below:

\begin{enumerate}
    
\item {\bf Planetary shadow}.  A spangle on the ring having spherical coordinates $(r_i,\alpha_i,\delta_i)$ in \equ, is inside the planet shadow, if the following condition is met:
    
  \beq{eq:r_i}
  r_i<r_s\equiv\frac{\Rp}{\sqrt{\cos^2\Delta\alpha\sin^2\delta_s+\sin^2\Delta\alpha}},
  \eeq
  where $\Delta\alpha=|\alpha_s-\alpha_i|$ and $\Rp$ is the radius of the planet.
  
\item {\bf Ring shadow}.  A spangle on the planet having spherical coordinates $(r_i,\alpha_i,\delta_i)$ in \equ, is inside the ring shadow if the following condition is met:

  \beq{eq:elta_min_max}
  \delta\sub{min}\le|\delta_i|\le\delta\sub{max},
  \eeq
  where

  \beq{eq:deltasubminmax}
  \delta\sub{min,max}=\tan^{-1}\left(\frac{\sin\delta}{D\sub{e,i}-\cos\Delta\alpha\cos\delta}\right),
  \eeq
  and

  \beq{eq:D_ei}
  D\sub{e,i}=\sqrt{R\sub{e,i}^2/\Rp^2-\sin^2\Delta\alpha\cos^2\delta},
  \eeq
  and the e and i indexes stand for the external and internal borders of the ring.
\end{enumerate} 

\printcredits



\begin{thebibliography}{73}
\expandafter\ifx\csname natexlab\endcsname\relax\def\natexlab#1{#1}\fi
\providecommand{\url}[1]{\texttt{#1}}
\providecommand{\href}[2]{#2}
\providecommand{\path}[1]{#1}
\providecommand{\DOIprefix}{doi:}
\providecommand{\ArXivprefix}{arXiv:}
\providecommand{\URLprefix}{URL: }
\providecommand{\Pubmedprefix}{pmid:}
\providecommand{\doi}[1]{\href{http://dx.doi.org/#1}{\path{#1}}}
\providecommand{\Pubmed}[1]{\href{pmid:#1}{\path{#1}}}
\providecommand{\bibinfo}[2]{#2}
\ifx\xfnm\relax \def\xfnm[#1]{\unskip,\space#1}\fi
\bibitem[{{Aizawa} et~al.(2018){Aizawa}, {Masuda}, {Kawahara} and
  {Suto}}]{Aizawa2018}
\bibinfo{author}{{Aizawa}, M.}, \bibinfo{author}{{Masuda}, K.},
  \bibinfo{author}{{Kawahara}, H.}, \bibinfo{author}{{Suto}, Y.},
  \bibinfo{year}{2018}.
\newblock \bibinfo{title}{{Systematic Search for Rings around Kepler Planet
  Candidates: Constraints on Ring Size and Occurrence Rate}}.
\newblock \bibinfo{journal}{\aj} \bibinfo{volume}{155}, \bibinfo{pages}{206}.
\newblock \DOIprefix\doi{10.3847/1538-3881/aab9a1},
  \href{http://arxiv.org/abs/1803.09114}{\tt arXiv:1803.09114}.
\bibitem[{{Akinsanmi} et~al.(2018){Akinsanmi}, {Oshagh}, {Santos} and
  {Barros}}]{Akinsanmi2018}
\bibinfo{author}{{Akinsanmi}, B.}, \bibinfo{author}{{Oshagh}, M.},
  \bibinfo{author}{{Santos}, N.C.}, \bibinfo{author}{{Barros}, S.C.C.},
  \bibinfo{year}{2018}.
\newblock \bibinfo{title}{{Detecting transit signatures of exoplanetary rings
  using SOAP3.0}}.
\newblock \bibinfo{journal}{\aap} \bibinfo{volume}{609}, \bibinfo{pages}{A21}.
\newblock \DOIprefix\doi{10.1051/0004-6361/201731215},
  \href{http://arxiv.org/abs/1709.06443}{\tt arXiv:1709.06443}.
\bibitem[{{Alvarado-Montes} et~al.(2017){Alvarado-Montes}, {Zuluaga} and
  {Sucerquia}}]{Alvarado2017}
\bibinfo{author}{{Alvarado-Montes}, J.A.}, \bibinfo{author}{{Zuluaga}, J.I.},
  \bibinfo{author}{{Sucerquia}, M.}, \bibinfo{year}{2017}.
\newblock \bibinfo{title}{{The effect of close-in giant planets' evolution on
  tidal-induced migration of exomoons}}.
\newblock \bibinfo{journal}{\mnras} \bibinfo{volume}{471},
  \bibinfo{pages}{3019--3027}.
\newblock \DOIprefix\doi{10.1093/mnras/stx1745},
  \href{http://arxiv.org/abs/1707.02906}{\tt arXiv:1707.02906}.
\bibitem[{{Arkhypov} et~al.(2021){Arkhypov}, {Khodachenko} and
  {Hanslmeier}}]{Arkhypov2021}
\bibinfo{author}{{Arkhypov}, O.V.}, \bibinfo{author}{{Khodachenko}, M.L.},
  \bibinfo{author}{{Hanslmeier}, A.}, \bibinfo{year}{2021}.
\newblock \bibinfo{title}{{Revealing peculiar exoplanetary shadows from transit
  light curves}}.
\newblock \bibinfo{journal}{\aap} \bibinfo{volume}{646}, \bibinfo{pages}{A136}.
\newblock \DOIprefix\doi{10.1051/0004-6361/202039050}.
\bibitem[{{Armstrong} et~al.(2016){Armstrong}, {de Mooij}, {Barstow}, {Osborn},
  {Blake} and {Saniee}}]{Armstrong2016}
\bibinfo{author}{{Armstrong}, D.J.}, \bibinfo{author}{{de Mooij}, E.},
  \bibinfo{author}{{Barstow}, J.}, \bibinfo{author}{{Osborn}, H.P.},
  \bibinfo{author}{{Blake}, J.}, \bibinfo{author}{{Saniee}, N.F.},
  \bibinfo{year}{2016}.
\newblock \bibinfo{title}{{Variability in the atmosphere of the hot giant
  planet HAT-P-7 b}}.
\newblock \bibinfo{journal}{Nature Astronomy} \bibinfo{volume}{1},
  \bibinfo{pages}{0004}.
\newblock \DOIprefix\doi{10.1038/s41550-016-0004},
  \href{http://arxiv.org/abs/1612.04225}{\tt arXiv:1612.04225}.
\bibitem[{{Arnold} and {Schneider}(2004)}]{Arnold2004}
\bibinfo{author}{{Arnold}, L.}, \bibinfo{author}{{Schneider}, J.},
  \bibinfo{year}{2004}.
\newblock \bibinfo{title}{{The detectability of extrasolar planet surroundings.
  I. Reflected-light photometry of unresolved rings}}.
\newblock \bibinfo{journal}{\aap} \bibinfo{volume}{420},
  \bibinfo{pages}{1153--1162}.
\newblock \DOIprefix\doi{10.1051/0004-6361:20035720},
  \href{http://arxiv.org/abs/astro-ph/0403330}{\tt arXiv:astro-ph/0403330}.
\bibitem[{{Barnes} and {Fortney}(2003)}]{Barnes2003}
\bibinfo{author}{{Barnes}, J.W.}, \bibinfo{author}{{Fortney}, J.J.},
  \bibinfo{year}{2003}.
\newblock \bibinfo{title}{{Measuring the Oblateness and Rotation of Transiting
  Extrasolar Giant Planets}}.
\newblock \bibinfo{journal}{\apj} \bibinfo{volume}{588},
  \bibinfo{pages}{545--556}.
\newblock \DOIprefix\doi{10.1086/373893},
  \href{http://arxiv.org/abs/astro-ph/0301156}{\tt arXiv:astro-ph/0301156}.
\bibitem[{{Barnes} and {Fortney}(2004)}]{Barnes2004}
\bibinfo{author}{{Barnes}, J.W.}, \bibinfo{author}{{Fortney}, J.J.},
  \bibinfo{year}{2004}.
\newblock \bibinfo{title}{{Transit Detectability of Ring Systems around
  Extrasolar Giant Planets}}.
\newblock \bibinfo{journal}{\apj} \bibinfo{volume}{616},
  \bibinfo{pages}{1193--1203}.
\newblock \DOIprefix\doi{10.1086/425067},
  \href{http://arxiv.org/abs/astro-ph/0409506}{\tt arXiv:astro-ph/0409506}.
\bibitem[{{Boyajian} et~al.(2016){Boyajian}, {LaCourse}, {Rappaport},
  {Fabrycky}, {Fischer}, {Gandolfi}, {Kennedy}, {Korhonen}, {Liu}, {Moor},
  {Olah}, {Vida}, {Wyatt}, {Best}, {Brewer}, {Ciesla}, {Cs{\'a}k}, {Deeg},
  {Dupuy}, {Handler}, {Heng}, {Howell}, {Ishikawa}, {Kov{\'a}cs}, {Kozakis},
  {Kriskovics}, {Lehtinen}, {Lintott}, {Lynn}, {Nespral}, {Nikbakhsh},
  {Schawinski}, {Schmitt}, {Smith}, {Szabo}, {Szabo}, {Viuho}, {Wang},
  {Weiksnar}, {Bosch}, {Connors}, {Goodman}, {Green}, {Hoekstra}, {Jebson},
  {Jek}, {Omohundro}, {Schwengeler} and {Szewczyk}}]{Boyajian2016}
\bibinfo{author}{{Boyajian}, T.S.}, \bibinfo{author}{{LaCourse}, D.M.},
  \bibinfo{author}{{Rappaport}, S.A.}, \bibinfo{author}{{Fabrycky}, D.},
  \bibinfo{author}{{Fischer}, D.A.}, \bibinfo{author}{{Gandolfi}, D.},
  \bibinfo{author}{{Kennedy}, G.M.}, \bibinfo{author}{{Korhonen}, H.},
  \bibinfo{author}{{Liu}, M.C.}, \bibinfo{author}{{Moor}, A.},
  \bibinfo{author}{{Olah}, K.}, \bibinfo{author}{{Vida}, K.},
  \bibinfo{author}{{Wyatt}, M.C.}, \bibinfo{author}{{Best}, W.M.J.},
  \bibinfo{author}{{Brewer}, J.}, \bibinfo{author}{{Ciesla}, F.},
  \bibinfo{author}{{Cs{\'a}k}, B.}, \bibinfo{author}{{Deeg}, H.J.},
  \bibinfo{author}{{Dupuy}, T.J.}, \bibinfo{author}{{Handler}, G.},
  \bibinfo{author}{{Heng}, K.}, \bibinfo{author}{{Howell}, S.B.},
  \bibinfo{author}{{Ishikawa}, S.T.}, \bibinfo{author}{{Kov{\'a}cs}, J.},
  \bibinfo{author}{{Kozakis}, T.}, \bibinfo{author}{{Kriskovics}, L.},
  \bibinfo{author}{{Lehtinen}, J.}, \bibinfo{author}{{Lintott}, C.},
  \bibinfo{author}{{Lynn}, S.}, \bibinfo{author}{{Nespral}, D.},
  \bibinfo{author}{{Nikbakhsh}, S.}, \bibinfo{author}{{Schawinski}, K.},
  \bibinfo{author}{{Schmitt}, J.R.}, \bibinfo{author}{{Smith}, A.M.},
  \bibinfo{author}{{Szabo}, G.}, \bibinfo{author}{{Szabo}, R.},
  \bibinfo{author}{{Viuho}, J.}, \bibinfo{author}{{Wang}, J.},
  \bibinfo{author}{{Weiksnar}, A.}, \bibinfo{author}{{Bosch}, M.},
  \bibinfo{author}{{Connors}, J.L.}, \bibinfo{author}{{Goodman}, S.},
  \bibinfo{author}{{Green}, G.}, \bibinfo{author}{{Hoekstra}, A.J.},
  \bibinfo{author}{{Jebson}, T.}, \bibinfo{author}{{Jek}, K.J.},
  \bibinfo{author}{{Omohundro}, M.R.}, \bibinfo{author}{{Schwengeler}, H.M.},
  \bibinfo{author}{{Szewczyk}, A.}, \bibinfo{year}{2016}.
\newblock \bibinfo{title}{{Planet Hunters IX. KIC 8462852 - where's the flux?}}
\newblock \bibinfo{journal}{\mnras} \bibinfo{volume}{457},
  \bibinfo{pages}{3988--4004}.
\newblock \DOIprefix\doi{10.1093/mnras/stw218},
  \href{http://arxiv.org/abs/1509.03622}{\tt arXiv:1509.03622}.
\bibitem[{{Brown} et~al.(2001){Brown}, {Charbonneau}, {Gilliland}, {Noyes} and
  {Burrows}}]{Brown2001}
\bibinfo{author}{{Brown}, T.M.}, \bibinfo{author}{{Charbonneau}, D.},
  \bibinfo{author}{{Gilliland}, R.L.}, \bibinfo{author}{{Noyes}, R.W.},
  \bibinfo{author}{{Burrows}, A.}, \bibinfo{year}{2001}.
\newblock \bibinfo{title}{{Hubble Space Telescope Time-Series Photometry of the
  Transiting Planet of HD 209458}}.
\newblock \bibinfo{journal}{\apj} \bibinfo{volume}{552},
  \bibinfo{pages}{699--709}.
\newblock \DOIprefix\doi{10.1086/320580},
  \href{http://arxiv.org/abs/astro-ph/0101336}{\tt arXiv:astro-ph/0101336}.
\bibitem[{{Burke} et~al.(2007){Burke}, {McCullough}, {Valenti}, {Johns-Krull},
  {Janes}, {Heasley}, {Summers}, {Stys}, {Bissinger}, {Fleenor}, {Foote},
  {Garc{\'\i}a-Melendo}, {Gary}, {Howell}, {Mallia}, {Masi}, {Taylor} and
  {Vanmunster}}]{Burke2007}
\bibinfo{author}{{Burke}, C.J.}, \bibinfo{author}{{McCullough}, P.R.},
  \bibinfo{author}{{Valenti}, J.A.}, \bibinfo{author}{{Johns-Krull}, C.M.},
  \bibinfo{author}{{Janes}, K.A.}, \bibinfo{author}{{Heasley}, J.N.},
  \bibinfo{author}{{Summers}, F.J.}, \bibinfo{author}{{Stys}, J.E.},
  \bibinfo{author}{{Bissinger}, R.}, \bibinfo{author}{{Fleenor}, M.L.},
  \bibinfo{author}{{Foote}, C.N.}, \bibinfo{author}{{Garc{\'\i}a-Melendo}, E.},
  \bibinfo{author}{{Gary}, B.L.}, \bibinfo{author}{{Howell}, P.J.},
  \bibinfo{author}{{Mallia}, F.}, \bibinfo{author}{{Masi}, G.},
  \bibinfo{author}{{Taylor}, B.}, \bibinfo{author}{{Vanmunster}, T.},
  \bibinfo{year}{2007}.
\newblock \bibinfo{title}{{XO-2b: Transiting Hot Jupiter in a Metal-rich Common
  Proper Motion Binary}}.
\newblock \bibinfo{journal}{\apj} \bibinfo{volume}{671},
  \bibinfo{pages}{2115--2128}.
\newblock \DOIprefix\doi{10.1086/523087},
  \href{http://arxiv.org/abs/0705.0003}{\tt arXiv:0705.0003}.
\bibitem[{{Claret}(2000)}]{Claret2000}
\bibinfo{author}{{Claret}, A.}, \bibinfo{year}{2000}.
\newblock \bibinfo{title}{{A new non-linear limb-darkening law for LTE stellar
  atmosphere models. Calculations for -5.0 <= log[M/H] <= +1, 2000 K <=
  T$_{eff}$ <= 50000 K at several surface gravities}}.
\newblock \bibinfo{journal}{\aap} \bibinfo{volume}{363},
  \bibinfo{pages}{1081--1190}.
\bibitem[{{Currie} et~al.(2012){Currie}, {Debes}, {Rodigas}, {Burrows}, {Itoh},
  {Fukagawa}, {Kenyon}, {Kuchner} and {Matsumura}}]{Currie2012}
\bibinfo{author}{{Currie}, T.}, \bibinfo{author}{{Debes}, J.},
  \bibinfo{author}{{Rodigas}, T.J.}, \bibinfo{author}{{Burrows}, A.},
  \bibinfo{author}{{Itoh}, Y.}, \bibinfo{author}{{Fukagawa}, M.},
  \bibinfo{author}{{Kenyon}, S.J.}, \bibinfo{author}{{Kuchner}, M.},
  \bibinfo{author}{{Matsumura}, S.}, \bibinfo{year}{2012}.
\newblock \bibinfo{title}{{Direct Imaging Confirmation and Characterization of
  a Dust-enshrouded Candidate Exoplanet Orbiting Fomalhaut}}.
\newblock \bibinfo{journal}{\apjl} \bibinfo{volume}{760}, \bibinfo{pages}{L32}.
\newblock \DOIprefix\doi{10.1088/2041-8205/760/2/L32},
  \href{http://arxiv.org/abs/1210.6620}{\tt arXiv:1210.6620}.
\bibitem[{{Damiano} and {Hu}(2020)}]{Damiano2020b}
\bibinfo{author}{{Damiano}, M.}, \bibinfo{author}{{Hu}, R.},
  \bibinfo{year}{2020}.
\newblock \bibinfo{title}{{EXOREL \$\{\}\^\{\{\textbackslashmathfrak\{R\}\}\} :
  A Bayesian Inverse Retrieval Framework for Exoplanetary Reflected Light
  Spectra}}.
\newblock \bibinfo{journal}{\aj} \bibinfo{volume}{159}, \bibinfo{pages}{175}.
\newblock \DOIprefix\doi{10.3847/1538-3881/ab79a5},
  \href{http://arxiv.org/abs/2003.01814}{\tt arXiv:2003.01814}.
\bibitem[{{Damiano} and {Hu}(2021)}]{Damiano2021}
\bibinfo{author}{{Damiano}, M.}, \bibinfo{author}{{Hu}, R.},
  \bibinfo{year}{2021}.
\newblock \bibinfo{title}{{Reflected Spectroscopy of Small Exoplanets I:
  Determining the Atmospheric Composition of Sub-Neptunes Planets}}.
\newblock \bibinfo{journal}{\aj} \bibinfo{volume}{162}, \bibinfo{pages}{200}.
\newblock \DOIprefix\doi{10.3847/1538-3881/ac224d},
  \href{http://arxiv.org/abs/2109.11659}{\tt arXiv:2109.11659}.
\bibitem[{{Damiano} and {Hu}(2022)}]{Damiano2022}
\bibinfo{author}{{Damiano}, M.}, \bibinfo{author}{{Hu}, R.},
  \bibinfo{year}{2022}.
\newblock \bibinfo{title}{{Reflected spectroscopy of small exoplanets II:
  characterization of terrestrial exoplanets}}.
\newblock \bibinfo{journal}{arXiv e-prints} ,
  \bibinfo{pages}{arXiv:2204.13816}\href{http://arxiv.org/abs/2204.13816}{\tt
  arXiv:2204.13816}.
\bibitem[{{Damiano} et~al.(2020){Damiano}, {Hu} and
  {Hildebrandt}}]{Damiano2020a}
\bibinfo{author}{{Damiano}, M.}, \bibinfo{author}{{Hu}, R.},
  \bibinfo{author}{{Hildebrandt}, S.R.}, \bibinfo{year}{2020}.
\newblock \bibinfo{title}{{Multi-orbital-phase and Multiband Characterization
  of Exoplanetary Atmospheres with Reflected Light Spectra}}.
\newblock \bibinfo{journal}{\aj} \bibinfo{volume}{160}, \bibinfo{pages}{206}.
\newblock \DOIprefix\doi{10.3847/1538-3881/abb76a},
  \href{http://arxiv.org/abs/2009.08579}{\tt arXiv:2009.08579}.
\bibitem[{{Dollfus}(1984)}]{Dollfus1984}
\bibinfo{author}{{Dollfus}, A.}, \bibinfo{year}{1984}.
\newblock \bibinfo{title}{{The Saturn Ring Particles from Optical Reflectance
  Polarimetry}}, in: \bibinfo{editor}{{Brahic}, A.} (Ed.),
  \bibinfo{booktitle}{Planetary Rings}, p. \bibinfo{pages}{121}.
\bibitem[{{Dong} et~al.(2013){Dong}, {Katz} and {Socrates}}]{Dong2013}
\bibinfo{author}{{Dong}, S.}, \bibinfo{author}{{Katz}, B.},
  \bibinfo{author}{{Socrates}, A.}, \bibinfo{year}{2013}.
\newblock \bibinfo{title}{{Exploring a ``Flow'' of Highly Eccentric Binaries
  with Kepler}}.
\newblock \bibinfo{journal}{\apjl} \bibinfo{volume}{763}, \bibinfo{pages}{L2}.
\newblock \DOIprefix\doi{10.1088/2041-8205/763/1/L2},
  \href{http://arxiv.org/abs/1201.4399}{\tt arXiv:1201.4399}.
\bibitem[{{Dyudina} et~al.(2016){Dyudina}, {Zhang}, {Li}, {Kopparla},
  {Ingersoll}, {Dones}, {Verbiscer} and {Yung}}]{Dyudina2016}
\bibinfo{author}{{Dyudina}, U.}, \bibinfo{author}{{Zhang}, X.},
  \bibinfo{author}{{Li}, L.}, \bibinfo{author}{{Kopparla}, P.},
  \bibinfo{author}{{Ingersoll}, A.P.}, \bibinfo{author}{{Dones}, L.},
  \bibinfo{author}{{Verbiscer}, A.}, \bibinfo{author}{{Yung}, Y.L.},
  \bibinfo{year}{2016}.
\newblock \bibinfo{title}{{Reflected Light Curves, Spherical and Bond Albedos
  of Jupiter- and Saturn-like Exoplanets}}.
\newblock \bibinfo{journal}{\apj} \bibinfo{volume}{822}, \bibinfo{pages}{76}.
\newblock \DOIprefix\doi{10.3847/0004-637X/822/2/76},
  \href{http://arxiv.org/abs/1511.04415}{\tt arXiv:1511.04415}.
\bibitem[{{Eftekhar} and {Abedini}(2022)}]{Eftekhar2022}
\bibinfo{author}{{Eftekhar}, M.}, \bibinfo{author}{{Abedini}, Y.},
  \bibinfo{year}{2022}.
\newblock \bibinfo{title}{{Revisiting the secondary eclipses of KELT-1b using
  TESS observations}}.
\newblock \bibinfo{journal}{arXiv e-prints} ,
  \bibinfo{pages}{arXiv:2204.14190}\href{http://arxiv.org/abs/2204.14190}{\tt
  arXiv:2204.14190}.
\bibitem[{{French} and {Nicholson}(2000)}]{French2000}
\bibinfo{author}{{French}, R.G.}, \bibinfo{author}{{Nicholson}, P.D.},
  \bibinfo{year}{2000}.
\newblock \bibinfo{title}{{Saturn's Rings II. Particle Sizes Inferred from
  Stellar Occultation Data}}.
\newblock \bibinfo{journal}{\icarus} \bibinfo{volume}{145},
  \bibinfo{pages}{502--523}.
\newblock \DOIprefix\doi{10.1006/icar.2000.6357}.
\bibitem[{{Geake} and {Geake}(1990)}]{Geake1990}
\bibinfo{author}{{Geake}, J.E.}, \bibinfo{author}{{Geake}, M.},
  \bibinfo{year}{1990}.
\newblock \bibinfo{title}{{A Remote Sensing Method for Sub-Wavelength Grains on
  Planetary Surfaces by Optical Polarimetry}}.
\newblock \bibinfo{journal}{\mnras} \bibinfo{volume}{245}, \bibinfo{pages}{46}.
\newblock \DOIprefix\doi{10.1093/mnras/245.1.46}.
\bibitem[{{Hameen-Anttila} and {Pyykko}(1972)}]{Hameen-Anttila1972}
\bibinfo{author}{{Hameen-Anttila}, K.A.}, \bibinfo{author}{{Pyykko}, S.},
  \bibinfo{year}{1972}.
\newblock \bibinfo{title}{{Photometric Behaviour of Saturn's Rings as a
  Function of the Saturn ocentric Latitudes of the Earth and the Sun}}.
\newblock \bibinfo{journal}{\aap} \bibinfo{volume}{19}, \bibinfo{pages}{235}.
\bibitem[{{Heising} et~al.(2015){Heising}, {Marcy} and
  {Schlichting}}]{Heising2015}
\bibinfo{author}{{Heising}, M.Z.}, \bibinfo{author}{{Marcy}, G.W.},
  \bibinfo{author}{{Schlichting}, H.E.}, \bibinfo{year}{2015}.
\newblock \bibinfo{title}{{A Search for Ringed Exoplanets Using Kepler
  Photometry}}.
\newblock \bibinfo{journal}{\apj} \bibinfo{volume}{814}, \bibinfo{pages}{81}.
\newblock \DOIprefix\doi{10.1088/0004-637X/814/1/81},
  \href{http://arxiv.org/abs/1511.01083}{\tt arXiv:1511.01083}.
\bibitem[{{Heller}(2018)}]{Heller2018}
\bibinfo{author}{{Heller}, R.}, \bibinfo{year}{2018}.
\newblock \bibinfo{title}{{Detecting and Characterizing Exomoons and
  Exorings}}, in: \bibinfo{editor}{{Deeg}, H.J.}, \bibinfo{editor}{{Belmonte},
  J.A.} (Eds.), \bibinfo{booktitle}{Handbook of Exoplanets},
  p.~\bibinfo{pages}{35}.
\newblock \DOIprefix\doi{10.1007/978-3-319-55333-7\_35}.
\bibitem[{{Hu}(2019)}]{Hu2019}
\bibinfo{author}{{Hu}, R.}, \bibinfo{year}{2019}.
\newblock \bibinfo{title}{{Information in the Reflected-light Spectra of Widely
  Separated Giant Exoplanets}}.
\newblock \bibinfo{journal}{\apj} \bibinfo{volume}{887}, \bibinfo{pages}{166}.
\newblock \DOIprefix\doi{10.3847/1538-4357/ab58c7},
  \href{http://arxiv.org/abs/1911.06274}{\tt arXiv:1911.06274}.
\bibitem[{{Janson} et~al.(2020){Janson}, {Wu}, {Cataldi} and
  {Brandeker}}]{Janson2020}
\bibinfo{author}{{Janson}, M.}, \bibinfo{author}{{Wu}, Y.},
  \bibinfo{author}{{Cataldi}, G.}, \bibinfo{author}{{Brandeker}, A.},
  \bibinfo{year}{2020}.
\newblock \bibinfo{title}{{Tidal disruption versus planetesimal collisions as
  possible origins for the dispersing dust cloud around Fomalhaut}}.
\newblock \bibinfo{journal}{\aap} \bibinfo{volume}{640}, \bibinfo{pages}{A93}.
\newblock \DOIprefix\doi{10.1051/0004-6361/202038589},
  \href{http://arxiv.org/abs/2007.06912}{\tt arXiv:2007.06912}.
\bibitem[{{Johnson} et~al.(1980){Johnson}, {Kemp}, {King}, {Parker} and
  {Barbour}}]{Johnson1980}
\bibinfo{author}{{Johnson}, P.E.}, \bibinfo{author}{{Kemp}, J.C.},
  \bibinfo{author}{{King}, R.}, \bibinfo{author}{{Parker}, T.E.},
  \bibinfo{author}{{Barbour}, M.S.}, \bibinfo{year}{1980}.
\newblock \bibinfo{title}{{New results from optical polarimetry of Saturn's
  rings}}.
\newblock \bibinfo{journal}{\nat} \bibinfo{volume}{283},
  \bibinfo{pages}{146--149}.
\newblock \DOIprefix\doi{10.1038/283146a0}.
\bibitem[{{Kalas} et~al.(2013){Kalas}, {Graham}, {Fitzgerald} and
  {Clampin}}]{Kalas2013}
\bibinfo{author}{{Kalas}, P.}, \bibinfo{author}{{Graham}, J.R.},
  \bibinfo{author}{{Fitzgerald}, M.P.}, \bibinfo{author}{{Clampin}, M.},
  \bibinfo{year}{2013}.
\newblock \bibinfo{title}{{STIS Coronagraphic Imaging of Fomalhaut: Main Belt
  Structure and the Orbit of Fomalhaut b}}.
\newblock \bibinfo{journal}{\apj} \bibinfo{volume}{775}, \bibinfo{pages}{56}.
\newblock \DOIprefix\doi{10.1088/0004-637X/775/1/56},
  \href{http://arxiv.org/abs/1305.2222}{\tt arXiv:1305.2222}.
\bibitem[{{Kane} and {von Braun}(2009)}]{Kane2009}
\bibinfo{author}{{Kane}, S.R.}, \bibinfo{author}{{von Braun}, K.},
  \bibinfo{year}{2009}.
\newblock \bibinfo{title}{{Exoplanetary Transit Constraints Based upon
  Secondary Eclipse Observations}}.
\newblock \bibinfo{journal}{\pasp} \bibinfo{volume}{121},
  \bibinfo{pages}{1096}.
\newblock \DOIprefix\doi{10.1086/606062},
  \href{http://arxiv.org/abs/0908.0519}{\tt arXiv:0908.0519}.
\bibitem[{{Kanodia} et~al.(2021){Kanodia}, {Stefansson}, {Ca{\~n}as}, {Maney},
  {Lin}, {Ninan}, {Jones}, {Monson}, {Parker}, {Kobulnicky}, {Rothenberg},
  {Beard}, {Lubin}, {Robertson}, {Gupta}, {Mahadevan}, {Cochran}, {Bender},
  {Diddams}, {Fredrick}, {Halverson}, {Hawley}, {Hearty}, {Hebb}, {Kopparapu},
  {Metcalf}, {Ramsey}, {Roy}, {Schwab}, {Schutte}, {Terrien}, {Wisniewski} and
  {Wright}}]{Kanodia2021}
\bibinfo{author}{{Kanodia}, S.}, \bibinfo{author}{{Stefansson}, G.},
  \bibinfo{author}{{Ca{\~n}as}, C.I.}, \bibinfo{author}{{Maney}, M.},
  \bibinfo{author}{{Lin}, A.S.J.}, \bibinfo{author}{{Ninan}, J.P.},
  \bibinfo{author}{{Jones}, S.}, \bibinfo{author}{{Monson}, A.},
  \bibinfo{author}{{Parker}, B.A.}, \bibinfo{author}{{Kobulnicky}, H.A.},
  \bibinfo{author}{{Rothenberg}, J.}, \bibinfo{author}{{Beard}, C.},
  \bibinfo{author}{{Lubin}, J.}, \bibinfo{author}{{Robertson}, P.},
  \bibinfo{author}{{Gupta}, A.F.}, \bibinfo{author}{{Mahadevan}, S.},
  \bibinfo{author}{{Cochran}, W.D.}, \bibinfo{author}{{Bender}, C.F.},
  \bibinfo{author}{{Diddams}, S.A.}, \bibinfo{author}{{Fredrick}, C.},
  \bibinfo{author}{{Halverson}, S.}, \bibinfo{author}{{Hawley}, S.},
  \bibinfo{author}{{Hearty}, F.}, \bibinfo{author}{{Hebb}, L.},
  \bibinfo{author}{{Kopparapu}, R.}, \bibinfo{author}{{Metcalf}, A.J.},
  \bibinfo{author}{{Ramsey}, L.W.}, \bibinfo{author}{{Roy}, A.},
  \bibinfo{author}{{Schwab}, C.}, \bibinfo{author}{{Schutte}, M.},
  \bibinfo{author}{{Terrien}, R.C.}, \bibinfo{author}{{Wisniewski}, J.},
  \bibinfo{author}{{Wright}, J.T.}, \bibinfo{year}{2021}.
\newblock \bibinfo{title}{{TOI-532b: The Habitable-zone Planet Finder confirms
  a Large Super Neptune in the Neptune Desert orbiting a metal-rich M-dwarf
  host}}.
\newblock \bibinfo{journal}{\aj} \bibinfo{volume}{162}, \bibinfo{pages}{135}.
\newblock \DOIprefix\doi{10.3847/1538-3881/ac1940},
  \href{http://arxiv.org/abs/2107.13670}{\tt arXiv:2107.13670}.
\bibitem[{{Kenworthy} and {Mamajek}(2015)}]{kenworthy2015}
\bibinfo{author}{{Kenworthy}, M.A.}, \bibinfo{author}{{Mamajek}, E.E.},
  \bibinfo{year}{2015}.
\newblock \bibinfo{title}{{Modeling Giant Extrasolar Ring Systems in Eclipse
  and the Case of J1407b: Sculpting by Exomoons?}}
\newblock \bibinfo{journal}{\apj} \bibinfo{volume}{800}, \bibinfo{pages}{126}.
\newblock \DOIprefix\doi{10.1088/0004-637X/800/2/126},
  \href{http://arxiv.org/abs/1501.05652}{\tt arXiv:1501.05652}.
\bibitem[{{Kipping} et~al.(2022){Kipping}, {Bryson}, {Burke}, {Christiansen},
  {Hardegree-Ullman}, {Quarles}, {Hansen}, {Szul{\'a}gyi} and
  {Teachey}}]{Kipping2022}
\bibinfo{author}{{Kipping}, D.}, \bibinfo{author}{{Bryson}, S.},
  \bibinfo{author}{{Burke}, C.}, \bibinfo{author}{{Christiansen}, J.},
  \bibinfo{author}{{Hardegree-Ullman}, K.}, \bibinfo{author}{{Quarles}, B.},
  \bibinfo{author}{{Hansen}, B.}, \bibinfo{author}{{Szul{\'a}gyi}, J.},
  \bibinfo{author}{{Teachey}, A.}, \bibinfo{year}{2022}.
\newblock \bibinfo{title}{{An exomoon survey of 70 cool giant exoplanets and
  the new candidate Kepler-1708 b-i}}.
\newblock \bibinfo{journal}{Nature Astronomy} \bibinfo{volume}{6},
  \bibinfo{pages}{367--380}.
\newblock \DOIprefix\doi{10.1038/s41550-021-01539-1},
  \href{http://arxiv.org/abs/2201.04643}{\tt arXiv:2201.04643}.
\bibitem[{{Kipping}(2008)}]{Kipping2008}
\bibinfo{author}{{Kipping}, D.M.}, \bibinfo{year}{2008}.
\newblock \bibinfo{title}{{Transiting planets - light-curve analysis for
  eccentric orbits}}.
\newblock \bibinfo{journal}{\mnras} \bibinfo{volume}{389},
  \bibinfo{pages}{1383--1390}.
\newblock \DOIprefix\doi{10.1111/j.1365-2966.2008.13658.x},
  \href{http://arxiv.org/abs/0807.0096}{\tt arXiv:0807.0096}.
\bibitem[{{Kipping}(2009a)}]{Kipping2009b}
\bibinfo{author}{{Kipping}, D.M.}, \bibinfo{year}{2009}a.
\newblock \bibinfo{title}{{Transit timing effects due to an exomoon}}.
\newblock \bibinfo{journal}{\mnras} \bibinfo{volume}{392},
  \bibinfo{pages}{181--189}.
\newblock \DOIprefix\doi{10.1111/j.1365-2966.2008.13999.x},
  \href{http://arxiv.org/abs/0810.2243}{\tt arXiv:0810.2243}.
\bibitem[{{Kipping}(2009b)}]{Kipping2009a}
\bibinfo{author}{{Kipping}, D.M.}, \bibinfo{year}{2009}b.
\newblock \bibinfo{title}{{Transit timing effects due to an exomoon - II}}.
\newblock \bibinfo{journal}{\mnras} \bibinfo{volume}{396},
  \bibinfo{pages}{1797--1804}.
\newblock \DOIprefix\doi{10.1111/j.1365-2966.2009.14869.x},
  \href{http://arxiv.org/abs/0904.2565}{\tt arXiv:0904.2565}.
\bibitem[{{Kreidberg}(2015)}]{Kreidberg2015}
\bibinfo{author}{{Kreidberg}, L.}, \bibinfo{year}{2015}.
\newblock \bibinfo{title}{{batman: BAsic Transit Model cAlculatioN in Python}}.
\newblock \bibinfo{journal}{\pasp} \bibinfo{volume}{127},
  \bibinfo{pages}{1161}.
\newblock \DOIprefix\doi{10.1086/683602},
  \href{http://arxiv.org/abs/1507.08285}{\tt arXiv:1507.08285}.
\bibitem[{{Lietzow} et~al.(2021){Lietzow}, {Wolf} and
  {Brunngr{\"a}ber}}]{Lietzow2021}
\bibinfo{author}{{Lietzow}, M.}, \bibinfo{author}{{Wolf}, S.},
  \bibinfo{author}{{Brunngr{\"a}ber}, R.}, \bibinfo{year}{2021}.
\newblock \bibinfo{title}{{Three-dimensional continuum radiative transfer of
  polarized radiation in exoplanetary atmospheres}}.
\newblock \bibinfo{journal}{\aap} \bibinfo{volume}{645}, \bibinfo{pages}{A146}.
\newblock \DOIprefix\doi{10.1051/0004-6361/202038932},
  \href{http://arxiv.org/abs/2012.12992}{\tt arXiv:2012.12992}.
\bibitem[{{Lightkurve Collaboration} et~al.(2018){Lightkurve Collaboration},
  {Cardoso}, {Hedges}, {Gully-Santiago}, {Saunders}, {Cody}, {Barclay}, {Hall},
  {Sagear}, {Turtelboom}, {Zhang}, {Tzanidakis}, {Mighell}, {Coughlin}, {Bell},
  {Berta-Thompson}, {Williams}, {Dotson} and
  {Barentsen}}]{LightkurveCollaboration2018}
\bibinfo{author}{{Lightkurve Collaboration}}, \bibinfo{author}{{Cardoso},
  J.V.d.M.}, \bibinfo{author}{{Hedges}, C.}, \bibinfo{author}{{Gully-Santiago},
  M.}, \bibinfo{author}{{Saunders}, N.}, \bibinfo{author}{{Cody}, A.M.},
  \bibinfo{author}{{Barclay}, T.}, \bibinfo{author}{{Hall}, O.},
  \bibinfo{author}{{Sagear}, S.}, \bibinfo{author}{{Turtelboom}, E.},
  \bibinfo{author}{{Zhang}, J.}, \bibinfo{author}{{Tzanidakis}, A.},
  \bibinfo{author}{{Mighell}, K.}, \bibinfo{author}{{Coughlin}, J.},
  \bibinfo{author}{{Bell}, K.}, \bibinfo{author}{{Berta-Thompson}, Z.},
  \bibinfo{author}{{Williams}, P.}, \bibinfo{author}{{Dotson}, J.},
  \bibinfo{author}{{Barentsen}, G.}, \bibinfo{year}{2018}.
\newblock \bibinfo{title}{{Lightkurve: Kepler and TESS time series analysis in
  Python}}.
\newblock \bibinfo{howpublished}{Astrophysics Source Code Library, record
  ascl:1812.013}.
\newblock \href{http://arxiv.org/abs/1812.013}{\tt arXiv:1812.013}.
\bibitem[{{Lopez-Morales} et~al.(2019){Lopez-Morales}, {Currie}, {Teske},
  {Gaidos}, {Kempton}, {Males}, {Lewis}, {Rackham}, {Ben-Ami}, {Birkby},
  {Charbonneau}, {Close}, {Crane}, {Dressing}, {Froning}, {Hasegawa},
  {Konopacky}, {Kopparapu}, {Mawet}, {Mennesson}, {Ramirez}, {Stelter},
  {Szentgyorgyi}, {Wang}, {Alam}, {Collins}, {Dupree}, {Karovska}, {Kirk},
  {Levi}, {McGruder}, {Packman}, {Rugheimer} and {Rukdee}}]{Lopez-Morales2019}
\bibinfo{author}{{Lopez-Morales}, M.}, \bibinfo{author}{{Currie}, T.},
  \bibinfo{author}{{Teske}, J.}, \bibinfo{author}{{Gaidos}, E.},
  \bibinfo{author}{{Kempton}, E.}, \bibinfo{author}{{Males}, J.},
  \bibinfo{author}{{Lewis}, N.}, \bibinfo{author}{{Rackham}, B.V.},
  \bibinfo{author}{{Ben-Ami}, S.}, \bibinfo{author}{{Birkby}, J.},
  \bibinfo{author}{{Charbonneau}, D.}, \bibinfo{author}{{Close}, L.},
  \bibinfo{author}{{Crane}, J.}, \bibinfo{author}{{Dressing}, C.},
  \bibinfo{author}{{Froning}, C.}, \bibinfo{author}{{Hasegawa}, Y.},
  \bibinfo{author}{{Konopacky}, Q.}, \bibinfo{author}{{Kopparapu}, R.K.},
  \bibinfo{author}{{Mawet}, D.}, \bibinfo{author}{{Mennesson}, B.},
  \bibinfo{author}{{Ramirez}, R.}, \bibinfo{author}{{Stelter}, D.},
  \bibinfo{author}{{Szentgyorgyi}, A.}, \bibinfo{author}{{Wang}, J.},
  \bibinfo{author}{{Alam}, M.}, \bibinfo{author}{{Collins}, K.},
  \bibinfo{author}{{Dupree}, A.}, \bibinfo{author}{{Karovska}, M.},
  \bibinfo{author}{{Kirk}, J.}, \bibinfo{author}{{Levi}, A.},
  \bibinfo{author}{{McGruder}, C.}, \bibinfo{author}{{Packman}, C.},
  \bibinfo{author}{{Rugheimer}, S.}, \bibinfo{author}{{Rukdee}, S.},
  \bibinfo{year}{2019}.
\newblock \bibinfo{title}{{Detecting Earth-like Biosignatures on Rocky
  Exoplanets around Nearby Stars with Ground-based Extremely Large
  Telescopes}}.
\newblock \bibinfo{journal}{\baas} \bibinfo{volume}{51}, \bibinfo{pages}{162}.
\newblock \href{http://arxiv.org/abs/1903.09523}{\tt arXiv:1903.09523}.
\bibitem[{{Matthews} et~al.(2021){Matthews}, {Hinkley}, {Stapelfeldt}, {Vigan},
  {Mawet}, {Crossfield}, {David}, {Mamajek}, {Meshkat}, {Morales} and
  {Padgett}}]{Matthews2021}
\bibinfo{author}{{Matthews}, E.C.}, \bibinfo{author}{{Hinkley}, S.},
  \bibinfo{author}{{Stapelfeldt}, K.}, \bibinfo{author}{{Vigan}, A.},
  \bibinfo{author}{{Mawet}, D.}, \bibinfo{author}{{Crossfield}, I.J.M.},
  \bibinfo{author}{{David}, T.J.}, \bibinfo{author}{{Mamajek}, E.},
  \bibinfo{author}{{Meshkat}, T.}, \bibinfo{author}{{Morales}, F.},
  \bibinfo{author}{{Padgett}, D.}, \bibinfo{year}{2021}.
\newblock \bibinfo{title}{{Three New Late-type Stellar Companions to Very Dusty
  WISE Debris Disks Identified with SPHERE Imaging}}.
\newblock \bibinfo{journal}{\aj} \bibinfo{volume}{161}, \bibinfo{pages}{78}.
\newblock \DOIprefix\doi{10.3847/1538-3881/abcfca},
  \href{http://arxiv.org/abs/2012.03980}{\tt arXiv:2012.03980}.
\bibitem[{{Moskovitz} et~al.(2009){Moskovitz}, {Gaidos} and
  {Williams}}]{Moskovitz2009}
\bibinfo{author}{{Moskovitz}, N.A.}, \bibinfo{author}{{Gaidos}, E.},
  \bibinfo{author}{{Williams}, D.M.}, \bibinfo{year}{2009}.
\newblock \bibinfo{title}{{The Effect of Lunarlike Satellites on the Orbital
  Infrared Light Curves of Earth-Analog Planets}}.
\newblock \bibinfo{journal}{Astrobiology} \bibinfo{volume}{9},
  \bibinfo{pages}{269--277}.
\newblock \DOIprefix\doi{10.1089/ast.2007.0209},
  \href{http://arxiv.org/abs/0810.2069}{\tt arXiv:0810.2069}.
\bibitem[{{Oklopcic} et~al.(2019){Oklopcic}, {Hirata}, {Montero Camacho} and
  {Silva}}]{Oklopcic2019}
\bibinfo{author}{{Oklopcic}, A.}, \bibinfo{author}{{Hirata}, C.},
  \bibinfo{author}{{Montero Camacho}, P.}, \bibinfo{author}{{Silva}, M.},
  \bibinfo{year}{2019}.
\newblock \bibinfo{title}{{Detecting Magnetic Fields in Exoplanets with
  Spectropolarimetry in the Helium Line at 1083 nm}}, in:
  \bibinfo{booktitle}{AAS/Division for Extreme Solar Systems Abstracts}, p.
  \bibinfo{pages}{103.05}.
\bibitem[{{Perryman}(2018)}]{Perryman2018}
\bibinfo{author}{{Perryman}, M.}, \bibinfo{year}{2018}.
\newblock \bibinfo{title}{{The Exoplanet Handbook}}.
\bibitem[{{Piro} and {Vissapragada}(2020)}]{Piro2020}
\bibinfo{author}{{Piro}, A.L.}, \bibinfo{author}{{Vissapragada}, S.},
  \bibinfo{year}{2020}.
\newblock \bibinfo{title}{{Exploring Whether Super-puffs can be Explained as
  Ringed Exoplanets}}.
\newblock \bibinfo{journal}{\aj} \bibinfo{volume}{159}, \bibinfo{pages}{131}.
\newblock \DOIprefix\doi{10.3847/1538-3881/ab7192},
  \href{http://arxiv.org/abs/1911.09673}{\tt arXiv:1911.09673}.
\bibitem[{{Rein} and {Ofir}(2019)}]{Rein2019}
\bibinfo{author}{{Rein}, E.}, \bibinfo{author}{{Ofir}, A.},
  \bibinfo{year}{2019}.
\newblock \bibinfo{title}{{Fast and precise light-curve model for transiting
  exoplanets with rings}}.
\newblock \bibinfo{journal}{\mnras} \bibinfo{volume}{490},
  \bibinfo{pages}{1111--1119}.
\newblock \DOIprefix\doi{10.1093/mnras/stz2556},
  \href{http://arxiv.org/abs/1910.12099}{\tt arXiv:1910.12099}.
\bibitem[{{Russell}(1916)}]{Russell1916}
\bibinfo{author}{{Russell}, H.N.}, \bibinfo{year}{1916}.
\newblock \bibinfo{title}{{On the Albedo of the Planets and Their Satellites}}.
\newblock \bibinfo{journal}{\apj} \bibinfo{volume}{43},
  \bibinfo{pages}{173--196}.
\newblock \DOIprefix\doi{10.1086/142244}.
\bibitem[{{Sandford} and {Kipping}(2019)}]{Sandford2019}
\bibinfo{author}{{Sandford}, E.}, \bibinfo{author}{{Kipping}, D.},
  \bibinfo{year}{2019}.
\newblock \bibinfo{title}{{Shadow Imaging of Transiting Objects}}.
\newblock \bibinfo{journal}{\aj} \bibinfo{volume}{157}, \bibinfo{pages}{42}.
\newblock \DOIprefix\doi{10.3847/1538-3881/aaf565},
  \href{http://arxiv.org/abs/1812.01618}{\tt arXiv:1812.01618}.
\bibitem[{{Santos} et~al.(2015){Santos}, {Martins}, {Bou{\'e}}, {Correia},
  {Oshagh}, {Figueira}, {Santerne}, {Sousa}, {Melo}, {Montalto}, {Boisse},
  {Ehrenreich}, {Lovis}, {Pepe}, {Udry} and {Garcia Munoz}}]{Santos2015}
\bibinfo{author}{{Santos}, N.C.}, \bibinfo{author}{{Martins}, J.H.C.},
  \bibinfo{author}{{Bou{\'e}}, G.}, \bibinfo{author}{{Correia}, A.C.M.},
  \bibinfo{author}{{Oshagh}, M.}, \bibinfo{author}{{Figueira}, P.},
  \bibinfo{author}{{Santerne}, A.}, \bibinfo{author}{{Sousa}, S.G.},
  \bibinfo{author}{{Melo}, C.}, \bibinfo{author}{{Montalto}, M.},
  \bibinfo{author}{{Boisse}, I.}, \bibinfo{author}{{Ehrenreich}, D.},
  \bibinfo{author}{{Lovis}, C.}, \bibinfo{author}{{Pepe}, F.},
  \bibinfo{author}{{Udry}, S.}, \bibinfo{author}{{Garcia Munoz}, A.},
  \bibinfo{year}{2015}.
\newblock \bibinfo{title}{{Detecting ring systems around exoplanets using high
  resolution spectroscopy: the case of 51 Pegasi b}}.
\newblock \bibinfo{journal}{\aap} \bibinfo{volume}{583}, \bibinfo{pages}{A50}.
\newblock \DOIprefix\doi{10.1051/0004-6361/201526673},
  \href{http://arxiv.org/abs/1509.00723}{\tt arXiv:1509.00723}.
\bibitem[{{Schlichting} and {Chang}(2011)}]{Schlichting2011}
\bibinfo{author}{{Schlichting}, H.E.}, \bibinfo{author}{{Chang}, P.},
  \bibinfo{year}{2011}.
\newblock \bibinfo{title}{{Warm Saturns: On the Nature of Rings around
  Extrasolar Planets That Reside inside the Ice Line}}.
\newblock \bibinfo{journal}{\apj} \bibinfo{volume}{734}, \bibinfo{pages}{117}.
\newblock \DOIprefix\doi{10.1088/0004-637X/734/2/117},
  \href{http://arxiv.org/abs/1104.3863}{\tt arXiv:1104.3863}.
\bibitem[{{Seager} and {Mall{\'e}n-Ornelas}(2003)}]{Seager2003}
\bibinfo{author}{{Seager}, S.}, \bibinfo{author}{{Mall{\'e}n-Ornelas}, G.},
  \bibinfo{year}{2003}.
\newblock \bibinfo{title}{{A Unique Solution of Planet and Star Parameters from
  an Extrasolar Planet Transit Light Curve}}.
\newblock \bibinfo{journal}{\apj} \bibinfo{volume}{585},
  \bibinfo{pages}{1038--1055}.
\newblock \DOIprefix\doi{10.1086/346105},
  \href{http://arxiv.org/abs/astro-ph/0206228}{\tt arXiv:astro-ph/0206228}.
\bibitem[{{Seager} et~al.(2000){Seager}, {Whitney} and {Sasselov}}]{Seager2000}
\bibinfo{author}{{Seager}, S.}, \bibinfo{author}{{Whitney}, B.A.},
  \bibinfo{author}{{Sasselov}, D.D.}, \bibinfo{year}{2000}.
\newblock \bibinfo{title}{{Photometric Light Curves and Polarization of
  Close-in Extrasolar Giant Planets}}.
\newblock \bibinfo{journal}{\apj} \bibinfo{volume}{540},
  \bibinfo{pages}{504--520}.
\newblock \DOIprefix\doi{10.1086/309292},
  \href{http://arxiv.org/abs/astro-ph/0004001}{\tt arXiv:astro-ph/0004001}.
\bibitem[{{Serrano} et~al.(2018){Serrano}, {Barros}, {Oshagh}, {Santos},
  {Faria}, {Demangeon}, {Sousa} and {Lendl}}]{Serrano2018}
\bibinfo{author}{{Serrano}, L.M.}, \bibinfo{author}{{Barros}, S.C.C.},
  \bibinfo{author}{{Oshagh}, M.}, \bibinfo{author}{{Santos}, N.C.},
  \bibinfo{author}{{Faria}, J.P.}, \bibinfo{author}{{Demangeon}, O.},
  \bibinfo{author}{{Sousa}, S.G.}, \bibinfo{author}{{Lendl}, M.},
  \bibinfo{year}{2018}.
\newblock \bibinfo{title}{{Distinguishing the albedo of exoplanets from stellar
  activity}}.
\newblock \bibinfo{journal}{\aap} \bibinfo{volume}{611}, \bibinfo{pages}{A8}.
\newblock \DOIprefix\doi{10.1051/0004-6361/201731206},
  \href{http://arxiv.org/abs/1712.01805}{\tt arXiv:1712.01805}.
\bibitem[{{Sing}(2010)}]{Sing2010}
\bibinfo{author}{{Sing}, D.K.}, \bibinfo{year}{2010}.
\newblock \bibinfo{title}{{Stellar limb-darkening coefficients for CoRot and
  Kepler}}.
\newblock \bibinfo{journal}{\aap} \bibinfo{volume}{510}, \bibinfo{pages}{A21}.
\newblock \DOIprefix\doi{10.1051/0004-6361/200913675},
  \href{http://arxiv.org/abs/0912.2274}{\tt arXiv:0912.2274}.
\bibitem[{{Singh} et~al.(2022){Singh}, {Bonomo}, {Scandariato}, {Cibrario},
  {Barbato}, {Fossati}, {Pagano} and {Sozzetti}}]{Singh2022}
\bibinfo{author}{{Singh}, V.}, \bibinfo{author}{{Bonomo}, A.S.},
  \bibinfo{author}{{Scandariato}, G.}, \bibinfo{author}{{Cibrario}, N.},
  \bibinfo{author}{{Barbato}, D.}, \bibinfo{author}{{Fossati}, L.},
  \bibinfo{author}{{Pagano}, I.}, \bibinfo{author}{{Sozzetti}, A.},
  \bibinfo{year}{2022}.
\newblock \bibinfo{title}{{Probing Kepler's hottest small planets via
  homogeneous search and analysis of optical secondary eclipses and phase
  variations}}.
\newblock \bibinfo{journal}{\aap} \bibinfo{volume}{658}, \bibinfo{pages}{A132}.
\newblock \DOIprefix\doi{10.1051/0004-6361/202039037},
  \href{http://arxiv.org/abs/2111.05716}{\tt arXiv:2111.05716}.
\bibitem[{{Sobolev}(1975)}]{Sobolev1975}
\bibinfo{author}{{Sobolev}, V.V.}, \bibinfo{year}{1975}.
\newblock \bibinfo{title}{{Light scattering in planetary atmospheres}}.
\bibitem[{{Sucerquia} et~al.(2022){Sucerquia}, {Alvarado-Montes}, {Bayo},
  {Cuadra}, {Cuello}, {Giuppone}, {Montesinos}, {Olofsson}, {Schwab}, {Spitler}
  and {Zuluaga}}]{Sucerquia2022}
\bibinfo{author}{{Sucerquia}, M.}, \bibinfo{author}{{Alvarado-Montes}, J.A.},
  \bibinfo{author}{{Bayo}, A.}, \bibinfo{author}{{Cuadra}, J.},
  \bibinfo{author}{{Cuello}, N.}, \bibinfo{author}{{Giuppone}, C.A.},
  \bibinfo{author}{{Montesinos}, M.}, \bibinfo{author}{{Olofsson}, J.},
  \bibinfo{author}{{Schwab}, C.}, \bibinfo{author}{{Spitler}, L.},
  \bibinfo{author}{{Zuluaga}, J.I.}, \bibinfo{year}{2022}.
\newblock \bibinfo{title}{{Cronomoons: origin, dynamics, and light-curve
  features of ringed exomoons}}.
\newblock \bibinfo{journal}{\mnras} \bibinfo{volume}{512},
  \bibinfo{pages}{1032--1044}.
\newblock \DOIprefix\doi{10.1093/mnras/stab3531},
  \href{http://arxiv.org/abs/2112.02687}{\tt arXiv:2112.02687}.
\bibitem[{{Sucerquia} et~al.(2017){Sucerquia}, {Alvarado-Montes},
  {Ram{\'\i}rez} and {Zuluaga}}]{Sucerquia2017}
\bibinfo{author}{{Sucerquia}, M.}, \bibinfo{author}{{Alvarado-Montes}, J.A.},
  \bibinfo{author}{{Ram{\'\i}rez}, V.}, \bibinfo{author}{{Zuluaga}, J.I.},
  \bibinfo{year}{2017}.
\newblock \bibinfo{title}{{Anomalous light curves of young tilted exorings}}.
\newblock \bibinfo{journal}{\mnras} \bibinfo{volume}{472},
  \bibinfo{pages}{L120--L124}.
\newblock \DOIprefix\doi{10.1093/mnrasl/slx151},
  \href{http://arxiv.org/abs/1708.04600}{\tt arXiv:1708.04600}.
\bibitem[{{Sucerquia} et~al.(2019){Sucerquia}, {Alvarado-Montes}, {Zuluaga},
  {Cuello} and {Giuppone}}]{Sucerquia2019}
\bibinfo{author}{{Sucerquia}, M.}, \bibinfo{author}{{Alvarado-Montes}, J.A.},
  \bibinfo{author}{{Zuluaga}, J.I.}, \bibinfo{author}{{Cuello}, N.},
  \bibinfo{author}{{Giuppone}, C.}, \bibinfo{year}{2019}.
\newblock \bibinfo{title}{{Ploonets: formation, evolution, and detectability of
  tidally detached exomoons}}.
\newblock \bibinfo{journal}{\mnras} \bibinfo{volume}{489},
  \bibinfo{pages}{2313--2322}.
\newblock \DOIprefix\doi{10.1093/mnras/stz2110},
  \href{http://arxiv.org/abs/1906.11400}{\tt arXiv:1906.11400}.
\bibitem[{{Sucerquia} et~al.(2020a){Sucerquia}, {Alvarado-Montes}, {Zuluaga},
  {Montesinos} and {Bayo}}]{Sucerquia2020a}
\bibinfo{author}{{Sucerquia}, M.}, \bibinfo{author}{{Alvarado-Montes}, J.A.},
  \bibinfo{author}{{Zuluaga}, J.I.}, \bibinfo{author}{{Montesinos}, M.},
  \bibinfo{author}{{Bayo}, A.}, \bibinfo{year}{2020}a.
\newblock \bibinfo{title}{{Scattered light may reveal the existence of ringed
  exoplanets}}.
\newblock \bibinfo{journal}{\mnras} \bibinfo{volume}{496},
  \bibinfo{pages}{L85--L90}.
\newblock \DOIprefix\doi{10.1093/mnrasl/slaa080},
  \href{http://arxiv.org/abs/2004.14121}{\tt arXiv:2004.14121}.
\bibitem[{{Sucerquia} et~al.(2020b){Sucerquia}, {Ram{\'\i}rez},
  {Alvarado-Montes} and {Zuluaga}}]{Sucerquia2020b}
\bibinfo{author}{{Sucerquia}, M.}, \bibinfo{author}{{Ram{\'\i}rez}, V.},
  \bibinfo{author}{{Alvarado-Montes}, J.A.}, \bibinfo{author}{{Zuluaga}, J.I.},
  \bibinfo{year}{2020}b.
\newblock \bibinfo{title}{{Can close-in giant exoplanets preserve detectable
  moons?}}
\newblock \bibinfo{journal}{\mnras} \bibinfo{volume}{492},
  \bibinfo{pages}{3499--3508}.
\newblock \DOIprefix\doi{10.1093/mnras/stz3548},
  \href{http://arxiv.org/abs/1912.08049}{\tt arXiv:1912.08049}.
\bibitem[{{Szab{\'o}} et~al.(2006){Szab{\'o}}, {Szatm{\'a}ry}, {Div{\'e}ki} and
  {Simon}}]{Szabo2006}
\bibinfo{author}{{Szab{\'o}}, G.M.}, \bibinfo{author}{{Szatm{\'a}ry}, K.},
  \bibinfo{author}{{Div{\'e}ki}, Z.}, \bibinfo{author}{{Simon}, A.},
  \bibinfo{year}{2006}.
\newblock \bibinfo{title}{{Possibility of a photometric detection of
  ``exomoons''}}.
\newblock \bibinfo{journal}{\aap} \bibinfo{volume}{450},
  \bibinfo{pages}{395--398}.
\newblock \DOIprefix\doi{10.1051/0004-6361:20054555},
  \href{http://arxiv.org/abs/astro-ph/0601186}{\tt arXiv:astro-ph/0601186}.
\bibitem[{{Teachey} et~al.(2018){Teachey}, {Kipping} and
  {Schmitt}}]{Teachey2018}
\bibinfo{author}{{Teachey}, A.}, \bibinfo{author}{{Kipping}, D.M.},
  \bibinfo{author}{{Schmitt}, A.R.}, \bibinfo{year}{2018}.
\newblock \bibinfo{title}{{HEK. VI. On the Dearth of Galilean Analogs in
  Kepler, and the Exomoon Candidate Kepler-1625b I}}.
\newblock \bibinfo{journal}{\aj} \bibinfo{volume}{155}, \bibinfo{pages}{36}.
\newblock \DOIprefix\doi{10.3847/1538-3881/aa93f2},
  \href{http://arxiv.org/abs/1707.08563}{\tt arXiv:1707.08563}.
\bibitem[{{Tinetti} et~al.(2018){Tinetti}, {Drossart}, {Eccleston}, {Hartogh},
  {Heske}, {Leconte}, {Micela}, {Ollivier}, {Pilbratt}, {Puig} and
  et~al.}]{Tinetti2018}
\bibinfo{author}{{Tinetti}, G.}, \bibinfo{author}{{Drossart}, P.},
  \bibinfo{author}{{Eccleston}, P.}, \bibinfo{author}{{Hartogh}, P.},
  \bibinfo{author}{{Heske}, A.}, \bibinfo{author}{{Leconte}, J.},
  \bibinfo{author}{{Micela}, G.}, \bibinfo{author}{{Ollivier}, M.},
  \bibinfo{author}{{Pilbratt}, G.}, \bibinfo{author}{{Puig}, L.},
  \bibinfo{author}{et~al.}, \bibinfo{year}{2018}.
\newblock \bibinfo{title}{{A chemical survey of exoplanets with ARIEL}}.
\newblock \bibinfo{journal}{Experimental Astronomy} \bibinfo{volume}{46},
  \bibinfo{pages}{135--209}.
\newblock \DOIprefix\doi{10.1007/s10686-018-9598-x}.
\bibitem[{{van Dam} et~al.(2020){van Dam}, {Kenworthy}, {David}, {Mamajek},
  {Hillenbrand}, {Cody}, {Howard}, {Isaacson}, {Ciardi}, {Rebull}, {Stauffer},
  {Patel}, {Cameron + WASP Collaborators}, {Rodriguez}, {Pojma{\'n}ski},
  {Gonzales}, {Schlieder}, {Hambsch}, {Dufoer}, {Vanmunster}, {Dubois},
  {Vanaverbeke}, {Logie} and {Rau}}]{vanDam2020}
\bibinfo{author}{{van Dam}, D.M.}, \bibinfo{author}{{Kenworthy}, M.A.},
  \bibinfo{author}{{David}, T.J.}, \bibinfo{author}{{Mamajek}, E.E.},
  \bibinfo{author}{{Hillenbrand}, L.A.}, \bibinfo{author}{{Cody}, A.M.},
  \bibinfo{author}{{Howard}, A.W.}, \bibinfo{author}{{Isaacson}, H.},
  \bibinfo{author}{{Ciardi}, D.R.}, \bibinfo{author}{{Rebull}, L.M.},
  \bibinfo{author}{{Stauffer}, J.R.}, \bibinfo{author}{{Patel}, R.},
  \bibinfo{author}{{Cameron + WASP Collaborators}, A.C.},
  \bibinfo{author}{{Rodriguez}, J.E.}, \bibinfo{author}{{Pojma{\'n}ski}, G.},
  \bibinfo{author}{{Gonzales}, E.J.}, \bibinfo{author}{{Schlieder}, J.E.},
  \bibinfo{author}{{Hambsch}, F.J.}, \bibinfo{author}{{Dufoer}, S.},
  \bibinfo{author}{{Vanmunster}, T.}, \bibinfo{author}{{Dubois}, F.},
  \bibinfo{author}{{Vanaverbeke}, S.}, \bibinfo{author}{{Logie}, L.},
  \bibinfo{author}{{Rau}, S.}, \bibinfo{year}{2020}.
\newblock \bibinfo{title}{{An Asymmetric Eclipse Seen toward the
  Pre-main-sequence Binary System V928 Tau}}.
\newblock \bibinfo{journal}{\aj} \bibinfo{volume}{160}, \bibinfo{pages}{285}.
\newblock \DOIprefix\doi{10.3847/1538-3881/abc259},
  \href{http://arxiv.org/abs/2010.11199}{\tt arXiv:2010.11199}.
\bibitem[{{{\v{S}}ubjak} et~al.(2022){{\v{S}}ubjak}, {Endl}, {Chaturvedi},
  {Karjalainen}, {Cochran}, {Esposito}, {Gandolfi}, {Lam}, {Stassun},
  {{\v{Z}}{\'a}k}, {Lodieu}, {Boffin}, {MacQueen}, {Hatzes}, {Guenther},
  {Georgieva}, {Grziwa}, {Schmerling}, {Skarka}, {Bla{\v{z}}ek}, {Karjalainen},
  {{\v{S}}pokov{\'a}}, {Isaacson}, {Howard}, {Burke}, {Van Eylen}, {Falk},
  {Fridlund}, {Goffo}, {Jenkins}, {Korth}, {Lissauer}, {Livingston}, {Luque},
  {Muresan}, {Osborn}, {Pall{\'e}}, {Persson}, {Redfield}, {Ricker}, {Seager},
  {Serrano}, {Smith} and {Kab{\'a}th}}]{Subjak2022}
\bibinfo{author}{{{\v{S}}ubjak}, J.}, \bibinfo{author}{{Endl}, M.},
  \bibinfo{author}{{Chaturvedi}, P.}, \bibinfo{author}{{Karjalainen}, R.},
  \bibinfo{author}{{Cochran}, W.D.}, \bibinfo{author}{{Esposito}, M.},
  \bibinfo{author}{{Gandolfi}, D.}, \bibinfo{author}{{Lam}, K.W.F.},
  \bibinfo{author}{{Stassun}, K.}, \bibinfo{author}{{{\v{Z}}{\'a}k}, J.},
  \bibinfo{author}{{Lodieu}, N.}, \bibinfo{author}{{Boffin}, H.M.J.},
  \bibinfo{author}{{MacQueen}, P.J.}, \bibinfo{author}{{Hatzes}, A.},
  \bibinfo{author}{{Guenther}, E.W.}, \bibinfo{author}{{Georgieva}, I.},
  \bibinfo{author}{{Grziwa}, S.}, \bibinfo{author}{{Schmerling}, H.},
  \bibinfo{author}{{Skarka}, M.}, \bibinfo{author}{{Bla{\v{z}}ek}, M.},
  \bibinfo{author}{{Karjalainen}, M.}, \bibinfo{author}{{{\v{S}}pokov{\'a}},
  M.}, \bibinfo{author}{{Isaacson}, H.}, \bibinfo{author}{{Howard}, A.W.},
  \bibinfo{author}{{Burke}, C.J.}, \bibinfo{author}{{Van Eylen}, V.},
  \bibinfo{author}{{Falk}, B.}, \bibinfo{author}{{Fridlund}, M.},
  \bibinfo{author}{{Goffo}, E.}, \bibinfo{author}{{Jenkins}, J.M.},
  \bibinfo{author}{{Korth}, J.}, \bibinfo{author}{{Lissauer}, J.J.},
  \bibinfo{author}{{Livingston}, J.H.}, \bibinfo{author}{{Luque}, R.},
  \bibinfo{author}{{Muresan}, A.}, \bibinfo{author}{{Osborn}, H.P.},
  \bibinfo{author}{{Pall{\'e}}, E.}, \bibinfo{author}{{Persson}, C.M.},
  \bibinfo{author}{{Redfield}, S.}, \bibinfo{author}{{Ricker}, G.R.},
  \bibinfo{author}{{Seager}, S.}, \bibinfo{author}{{Serrano}, L.M.},
  \bibinfo{author}{{Smith}, A.M.S.}, \bibinfo{author}{{Kab{\'a}th}, P.},
  \bibinfo{year}{2022}.
\newblock \bibinfo{title}{{TOI-1268b: the youngest, hot, Saturn-mass transiting
  exoplanet}}.
\newblock \bibinfo{journal}{arXiv e-prints} ,
  \bibinfo{pages}{arXiv:2201.13341}\href{http://arxiv.org/abs/2201.13341}{\tt
  arXiv:2201.13341}.
\bibitem[{{Wakeford} et~al.(2013){Wakeford}, {Sing}, {Deming}, {Gibson},
  {Fortney}, {Burrows}, {Ballester}, {Nikolov}, {Aigrain}, {Henry}, {Knutson},
  {Lecavelier des Etangs}, {Pont}, {Showman}, {Vidal-Madjar} and
  {Zahnle}}]{Wakeford2013}
\bibinfo{author}{{Wakeford}, H.R.}, \bibinfo{author}{{Sing}, D.K.},
  \bibinfo{author}{{Deming}, D.}, \bibinfo{author}{{Gibson}, N.P.},
  \bibinfo{author}{{Fortney}, J.J.}, \bibinfo{author}{{Burrows}, A.S.},
  \bibinfo{author}{{Ballester}, G.}, \bibinfo{author}{{Nikolov}, N.},
  \bibinfo{author}{{Aigrain}, S.}, \bibinfo{author}{{Henry}, G.},
  \bibinfo{author}{{Knutson}, H.}, \bibinfo{author}{{Lecavelier des Etangs},
  A.}, \bibinfo{author}{{Pont}, F.}, \bibinfo{author}{{Showman}, A.P.},
  \bibinfo{author}{{Vidal-Madjar}, A.}, \bibinfo{author}{{Zahnle}, K.},
  \bibinfo{year}{2013}.
\newblock \bibinfo{title}{{HST hot Jupiter transmission spectral survey:
  detection of water in HAT-P-1b from WFC3 near-IR spatial scan observations}}.
\newblock \bibinfo{journal}{\mnras} \bibinfo{volume}{435},
  \bibinfo{pages}{3481--3493}.
\newblock \DOIprefix\doi{10.1093/mnras/stt1536},
  \href{http://arxiv.org/abs/1308.2106}{\tt arXiv:1308.2106}.
\bibitem[{{Wittenmyer} et~al.(2020){Wittenmyer}, {Wang}, {Horner}, {Butler},
  {Tinney}, {Carter}, {Wright}, {Jones}, {Bailey}, {O'Toole} and
  {Johns}}]{Wittenmyer2020}
\bibinfo{author}{{Wittenmyer}, R.A.}, \bibinfo{author}{{Wang}, S.},
  \bibinfo{author}{{Horner}, J.}, \bibinfo{author}{{Butler}, R.P.},
  \bibinfo{author}{{Tinney}, C.G.}, \bibinfo{author}{{Carter}, B.D.},
  \bibinfo{author}{{Wright}, D.J.}, \bibinfo{author}{{Jones}, H.R.A.},
  \bibinfo{author}{{Bailey}, J.}, \bibinfo{author}{{O'Toole}, S.J.},
  \bibinfo{author}{{Johns}, D.}, \bibinfo{year}{2020}.
\newblock \bibinfo{title}{{Cool Jupiters greatly outnumber their toasty
  siblings: occurrence rates from the Anglo-Australian Planet Search}}.
\newblock \bibinfo{journal}{\mnras} \bibinfo{volume}{492},
  \bibinfo{pages}{377--383}.
\newblock \DOIprefix\doi{10.1093/mnras/stz3436},
  \href{http://arxiv.org/abs/1912.01821}{\tt arXiv:1912.01821}.
\bibitem[{{Wong} et~al.(2021){Wong}, {Shporer}, {Zhou}, {Kitzmann}, {Komacek},
  {Tan}, {Tronsgaard}, {Buchhave}, {Vissapragada}, {Greklek-McKeon},
  {Rodriguez}, {Ahlers}, {Quinn}, {Furlan}, {Howell}, {Bieryla}, {Heng},
  {Knutson}, {Collins}, {McLeod}, {Berlind}, {Brown}, {Calkins}, {de Leon},
  {Esparza-Borges}, {Esquerdo}, {Fukui}, {Gan}, {Girardin}, {Gnilka}, {Ikoma},
  {Jensen}, {Kielkopf}, {Kodama}, {Kurita}, {Lester}, {Lewin}, {Marino},
  {Murgas}, {Narita}, {Pall{\'e}}, {Schwarz}, {Stassun}, {Tamura}, {Watanabe},
  {Benneke}, {Ricker}, {Latham}, {Vanderspek}, {Seager}, {Winn}, {Jenkins},
  {Caldwell}, {Fong}, {Huang}, {Mireles}, {Schlieder}, {Shiao} and {Noel
  Villase{\~n}or}}]{Wong2021}
\bibinfo{author}{{Wong}, I.}, \bibinfo{author}{{Shporer}, A.},
  \bibinfo{author}{{Zhou}, G.}, \bibinfo{author}{{Kitzmann}, D.},
  \bibinfo{author}{{Komacek}, T.D.}, \bibinfo{author}{{Tan}, X.},
  \bibinfo{author}{{Tronsgaard}, R.}, \bibinfo{author}{{Buchhave}, L.A.},
  \bibinfo{author}{{Vissapragada}, S.}, \bibinfo{author}{{Greklek-McKeon}, M.},
  \bibinfo{author}{{Rodriguez}, J.E.}, \bibinfo{author}{{Ahlers}, J.P.},
  \bibinfo{author}{{Quinn}, S.N.}, \bibinfo{author}{{Furlan}, E.},
  \bibinfo{author}{{Howell}, S.B.}, \bibinfo{author}{{Bieryla}, A.},
  \bibinfo{author}{{Heng}, K.}, \bibinfo{author}{{Knutson}, H.A.},
  \bibinfo{author}{{Collins}, K.A.}, \bibinfo{author}{{McLeod}, K.K.},
  \bibinfo{author}{{Berlind}, P.}, \bibinfo{author}{{Brown}, P.},
  \bibinfo{author}{{Calkins}, M.L.}, \bibinfo{author}{{de Leon}, J.P.},
  \bibinfo{author}{{Esparza-Borges}, E.}, \bibinfo{author}{{Esquerdo}, G.A.},
  \bibinfo{author}{{Fukui}, A.}, \bibinfo{author}{{Gan}, T.},
  \bibinfo{author}{{Girardin}, E.}, \bibinfo{author}{{Gnilka}, C.L.},
  \bibinfo{author}{{Ikoma}, M.}, \bibinfo{author}{{Jensen}, E.L.N.},
  \bibinfo{author}{{Kielkopf}, J.}, \bibinfo{author}{{Kodama}, T.},
  \bibinfo{author}{{Kurita}, S.}, \bibinfo{author}{{Lester}, K.V.},
  \bibinfo{author}{{Lewin}, P.}, \bibinfo{author}{{Marino}, G.},
  \bibinfo{author}{{Murgas}, F.}, \bibinfo{author}{{Narita}, N.},
  \bibinfo{author}{{Pall{\'e}}, E.}, \bibinfo{author}{{Schwarz}, R.P.},
  \bibinfo{author}{{Stassun}, K.G.}, \bibinfo{author}{{Tamura}, M.},
  \bibinfo{author}{{Watanabe}, N.}, \bibinfo{author}{{Benneke}, B.},
  \bibinfo{author}{{Ricker}, G.R.}, \bibinfo{author}{{Latham}, D.W.},
  \bibinfo{author}{{Vanderspek}, R.}, \bibinfo{author}{{Seager}, S.},
  \bibinfo{author}{{Winn}, J.N.}, \bibinfo{author}{{Jenkins}, J.M.},
  \bibinfo{author}{{Caldwell}, D.A.}, \bibinfo{author}{{Fong}, W.},
  \bibinfo{author}{{Huang}, C.X.}, \bibinfo{author}{{Mireles}, I.},
  \bibinfo{author}{{Schlieder}, J.E.}, \bibinfo{author}{{Shiao}, B.},
  \bibinfo{author}{{Noel Villase{\~n}or}, J.}, \bibinfo{year}{2021}.
\newblock \bibinfo{title}{{TOI-2109: An Ultrahot Gas Giant on a 16 hr Orbit}}.
\newblock \bibinfo{journal}{\aj} \bibinfo{volume}{162}, \bibinfo{pages}{256}.
\newblock \DOIprefix\doi{10.3847/1538-3881/ac26bd},
  \href{http://arxiv.org/abs/2111.12074}{\tt arXiv:2111.12074}.
\bibitem[{{Zellem} et~al.(2014){Zellem}, {Lewis}, {Knutson}, {Griffith},
  {Showman}, {Fortney}, {Cowan}, {Agol}, {Burrows}, {Charbonneau}, {Deming},
  {Laughlin} and {Langton}}]{Zellem2014}
\bibinfo{author}{{Zellem}, R.T.}, \bibinfo{author}{{Lewis}, N.K.},
  \bibinfo{author}{{Knutson}, H.A.}, \bibinfo{author}{{Griffith}, C.A.},
  \bibinfo{author}{{Showman}, A.P.}, \bibinfo{author}{{Fortney}, J.J.},
  \bibinfo{author}{{Cowan}, N.B.}, \bibinfo{author}{{Agol}, E.},
  \bibinfo{author}{{Burrows}, A.}, \bibinfo{author}{{Charbonneau}, D.},
  \bibinfo{author}{{Deming}, D.}, \bibinfo{author}{{Laughlin}, G.},
  \bibinfo{author}{{Langton}, J.}, \bibinfo{year}{2014}.
\newblock \bibinfo{title}{{The 4.5 {\ensuremath{\mu}}m Full-orbit Phase Curve
  of the Hot Jupiter HD 209458b}}.
\newblock \bibinfo{journal}{\apj} \bibinfo{volume}{790}, \bibinfo{pages}{53}.
\newblock \DOIprefix\doi{10.1088/0004-637X/790/1/53},
  \href{http://arxiv.org/abs/1405.5923}{\tt arXiv:1405.5923}.
\bibitem[{{Zugger} et~al.(2010){Zugger}, {Kasting}, {Williams}, {Kane} and
  {Philbrick}}]{Zugger2010}
\bibinfo{author}{{Zugger}, M.E.}, \bibinfo{author}{{Kasting}, J.F.},
  \bibinfo{author}{{Williams}, D.M.}, \bibinfo{author}{{Kane}, T.J.},
  \bibinfo{author}{{Philbrick}, C.R.}, \bibinfo{year}{2010}.
\newblock \bibinfo{title}{{Light Scattering from Exoplanet Oceans and
  Atmospheres}}.
\newblock \bibinfo{journal}{\apj} \bibinfo{volume}{723},
  \bibinfo{pages}{1168--1179}.
\newblock \DOIprefix\doi{10.1088/0004-637X/723/2/1168},
  \href{http://arxiv.org/abs/1006.3525}{\tt arXiv:1006.3525}.
\bibitem[{{Zuluaga} et~al.(2015){Zuluaga}, {Kipping}, {Sucerquia} and
  {Alvarado}}]{Zuluaga2015}
\bibinfo{author}{{Zuluaga}, J.I.}, \bibinfo{author}{{Kipping}, D.M.},
  \bibinfo{author}{{Sucerquia}, M.}, \bibinfo{author}{{Alvarado}, J.A.},
  \bibinfo{year}{2015}.
\newblock \bibinfo{title}{{A Novel Method for Identifying Exoplanetary Rings}}.
\newblock \bibinfo{journal}{\apjl} \bibinfo{volume}{803}, \bibinfo{pages}{L14}.
\newblock \DOIprefix\doi{10.1088/2041-8205/803/1/L14},
  \href{http://arxiv.org/abs/1502.07818}{\tt arXiv:1502.07818}.

\end{thebibliography}
\end{document}